\documentclass[aps,twocolumn,pr*,amsfonts,showpacs,superscriptaddress,longbibliography]{revtex4-2}
\usepackage[utf8]{inputenc}
\usepackage{verbatim,epsfig,amsmath,amssymb,bm,epsf,graphicx,psfrag,bbold,amsthm,amsfonts}
\usepackage[bottom]{footmisc}
\usepackage{hyperref}
\usepackage{cleveref} 
\usepackage{enumitem}
\usepackage{framed}
\usepackage{mathrsfs}
\usepackage{esint}
\setlist{nosep}
\usepackage{placeins}
\usepackage[all]{xy}
\usepackage{color}
\usepackage[utf8]{inputenc}
\usepackage{float}
\usepackage{natbib}
\usepackage{tikz-cd}
\usepackage{verbatim}
\usepackage{leftidx}
\usepackage{physics}
\usepackage{ulem}

\usepackage{graphicx} 
\usepackage{makecell}

\def\brcurs{{\mbox{$\resizebox{.09in}{.08in}{\includegraphics[trim= 1em 0 14em 0,clip]{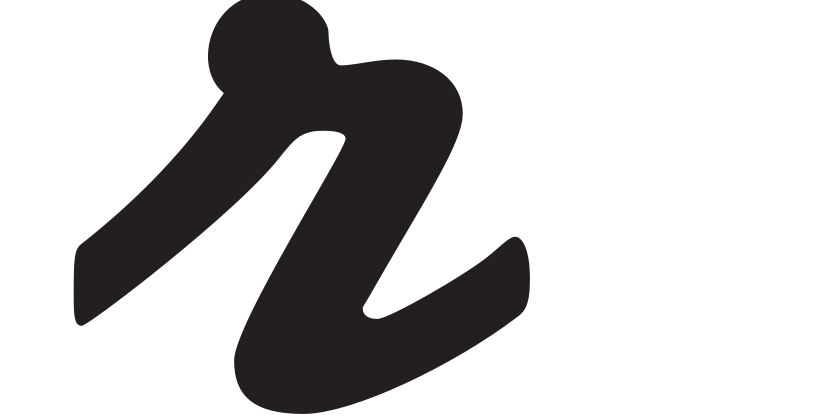}}$}}}

\newcommand{\br}{\mathbf{r}}
\newcommand{\bk}{\mathbf{k}}
\newcommand{\bq}{\mathbf{q}}

\newcommand{\bG}{\mathbf{G}}

\newcommand{\bR}{\mathbf{R}}
\newcommand{\bp}{\mathbf{p}}
\newcommand{\bL}{\mathbf{L}}

\title{Higher Chern in Helical Trilayer TMDs (draft)}

\date{\today}

\begin{document}
\title{Higher Chern bands in helical homotrilayer transition metal dichalcogenides }

\author{Jungho Daniel Choi}
\affiliation{Center for Computational Quantum Physics, Flatiron Institute, 162 5th Avenue, New York, NY 10010, USA}
\affiliation{Center for Quantum Phenomena, Department of Physics, New York University, 726 Broadway, New York, NY 10003, USA}

\author{Nicol\'as Morales-Dur\'an}
\affiliation{Center for Computational Quantum Physics, Flatiron Institute, 162 5th Avenue, New York, NY 10010, USA}

\author{Yves H. Kwan}
\affiliation{Princeton Center for Theoretical Science, Princeton University, Princeton, NJ 08544}

\author{Andrew J. Millis}
\affiliation{Center for Computational Quantum Physics, Flatiron Institute, 162 5th Avenue, New York, NY 10010, USA}
\affiliation{Department of Physics, Columbia University, 538 West 120th Street, New York, NY 10027, USA}

\author{Nicolas Regnault}
\affiliation{Center for Computational Quantum Physics, Flatiron Institute, 162 5th Avenue, New York, NY 10010, USA}
\affiliation{Department of Physics, Princeton University, Princeton, New Jersey 08544, USA}
\affiliation{Laboratoire de Physique de l’Ecole normale sup\'erieure, ENS, Universit\'e PSL, CNRS, Sorbonne Universit\'e, Universit\'e Paris-Diderot, Sorbonne Paris, Cit\'e, 75005 Paris, France}

\author{Daniele Guerci}
\affiliation{Department of Physics, Massachusetts Institute of Technology, Cambridge, MA-02139,USA}

\begin{abstract}
    We propose helically twisted homotrilayer transition metal dichalcogenides as a platform for realizing correlated topological phases of matter with higher and tunable Chern numbers. 
    We show that a clear separation of scales emerges for small twist angles, allowing us to derive a low-energy continuum model that captures the physics within moiré-scale domains. 
    We identify regimes of twist angle and displacement field for which the highest-lying hole band is isolated from other bands and is topological with $K$-valley Chern number $C=-2$. 
    We demonstrate that varying the displacement field can induce a transition from $C=-2$ to $C=-1$, as well as from a topologically trivial band to a $C=-1$ band. 
    We derive an effective tight-binding description for a high-symmetry stacking domain which is valid for a wide range of twist angles, and we show that the $C=-2$ band can remain stable at filling fraction $\nu=-1$ in the presence of interactions in Hartree-Fock calculations. 
\end{abstract}

\maketitle

\section{Introduction}
Moiré materials have emerged as a versatile experimental platform in which the interplay between topology and electronic correlations gives rise to exotic phases of matter. 
Recent studies of transition metal dichalcogenide (TMD)-based moiré heterostructures have reported a plethora of correlated phases, including various types of correlated insulators \cite{tang2020simulation,wang2020correlated,xu2020correlated,regan2020mott,jin2021stripe,xu2022tunable,anderson2023programming}, heavy fermion physics \cite{zhao2023gate,zhao2025emergencechernmetalmoire,han2025evidencetopologicalkondoinsulating}, the integer \cite{li2021quantum,zhao2024realization,tao2024valley} and fractional quantum anomalous Hall effects \cite{FCI_Experiment1,FCI_Experiment2,FCI_Transport1,FCI_Transport2,xu2025signatures}, and superconductivity \cite{xia2025superconductivity,guo2025superconductivity,xu2025signatures}.
A very active area of research relates to the effect on correlation physics of the non-trivial quantum geometry and band topology occurring in some moir\'e materials. 
Band topology may be characterized by a Chern number, $C$; situations with $|C|=1$ have been studied but less is known both theoretically and experimentally about the physics of bands with higher Chern numbers. 
Situations with $|C|>1$ are of interest because they provide an avenue to investigate topological phases that fall outside the conventional quantum Hall paradigm \cite{kwan2024fractional,PhysRevB.110.075417,niu2025quantum}. 
Theoretical proposals~\cite{zhang2019nearlyflat,liu2019multilayer,koshino2019doublebilayer,chebrolu2019doublebi,lee2019theory,rademaker2020monobi,ma2021topological,park2020monobi,ledwith2022family,wang2022hierarchy,liang2022doubletwisted,park2023C4,phong2025coulombinteractionstabilizedisolatednarrow} and experimental signatures~\cite{Chen2020,zhou2021half,Polshyn_2020,Wu2021,polshyn2022topological,Choi_2025} of topological bands with Chern numbers greater than one have been reported in certain multilayer graphene heterostructures. 
One example is helically twisted trilayer graphene \cite{Xia2025_TrilayerGraphene,yuncheng_2023,guerci2024chern,Nakatsuji2023,Devakul_2023,Popov2023,popov2023magicanglebutterflytwisted,Foo2024,yang_prb_2024,PhysRevB.109.125141}, which is predicted to host a narrow Chern band with $\lvert C \rvert =2$ near charge neutrality.

\begin{figure*}
    \centering
    \includegraphics[width=0.98\linewidth]{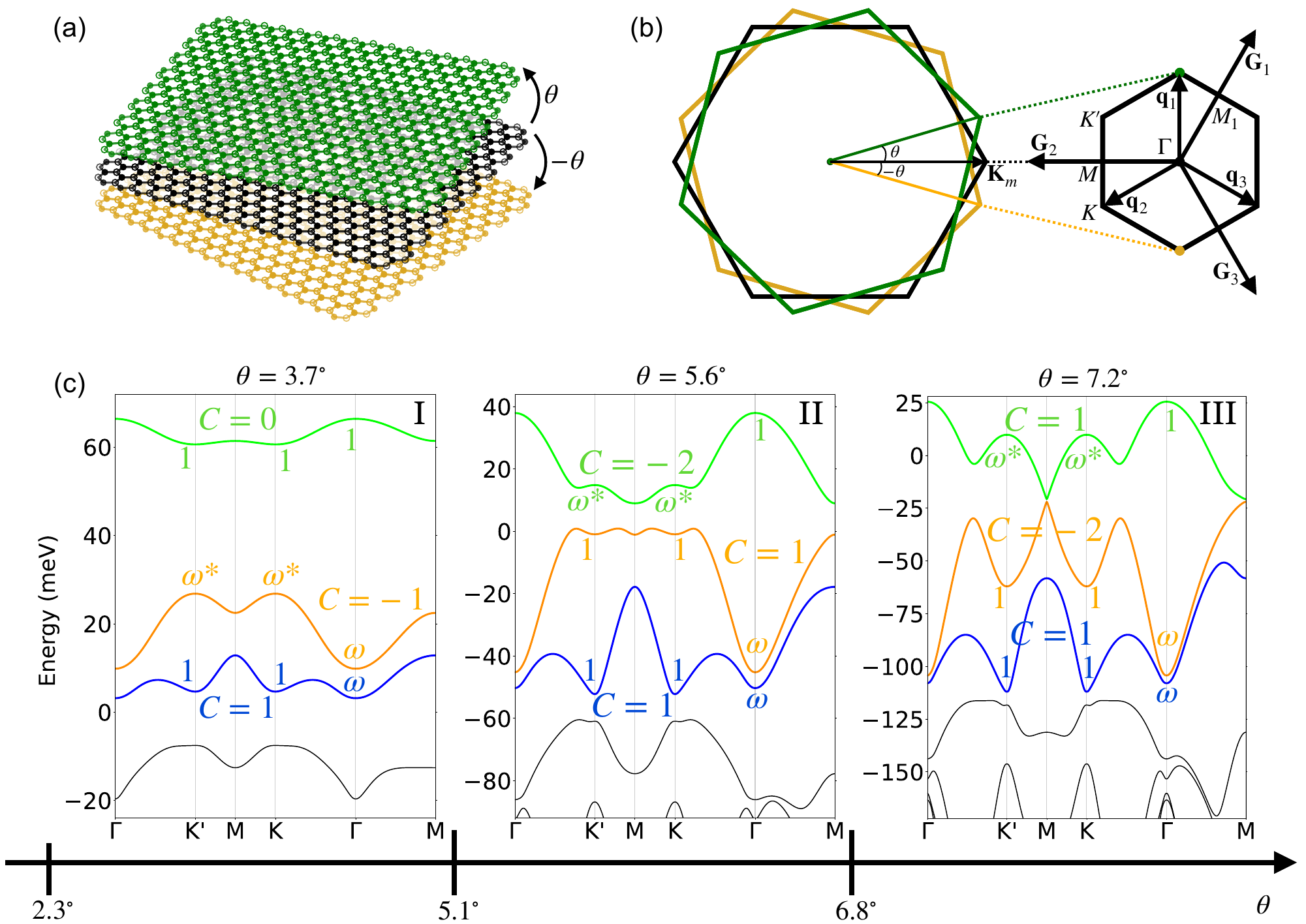}
    \caption{\textbf{(a)} Schematic of the helical trilayer TMD in real space depicting the relative twist angles between the top (green), middle (black), and bottom (yellow) layers. 
    Filled circles indicate a metal ($M$) atom, while open circles indicate a chalcogen ($X$) atom. 
    \textbf{(b)} Overlapping hexagons on the left depict the rotated Brillouin zones of the three layers, with $\mathbf{K}_m$ marking the momentum of the middle layer's $K$ valley. 
    Right zoom-in illustrates the emergent moir\'e Brillouin zone (mBZ) arising from valley $K$ of the monolayer Brillouin zone. 
    \textbf{(c)} Phase diagram for the continuum model for the $XMX$-stacked helical trilayer TMD, with $(V, \psi, w, m^*) = (16.5\text{ meV}, -100^\circ, -18.8\text{ meV}, 0.60m_e)$. 
    For twist angles $\theta \gtrsim 2.3^\circ$, the top three valence bands are gapped from the other remote valence bands. 
    There are three regimes (denoted $I, \,II, \, III$) of Chern sequences. When $5.1^\circ \lesssim \theta \lesssim 6.8^\circ$, the top valence band (in green) has $K$-valley Chern number $C=-2$. 
    The representative band structures are computed for twist angles $\theta = 3.7^\circ, 5.6^\circ, 7.2^\circ$. 
    Each band structure is labeled with the Chern numbers and $\mathcal{C}_{3z}$ eigenvalues at the high-symmetry points of the three topmost valence bands. }
    \label{fig:helicalgeometry}
\end{figure*}

A natural question is whether twisted semiconductor multilayers could also display Chern bands with $\lvert C \rvert > 1$ near charge neutrality. Recent theoretical studies have considered TMD homomultilayers \cite{Vogl_PhysRevB2023,liang2024moireflatbandsalternating} in the alternating stacking configuration and TMD heterotrilayers~\cite{he2024topologicalinsulatortwistedtransition}. 
They have found rich phase diagrams with topologically non-trivial valence bands close to charge neutrality, which can become flat. 
However, the topmost bands in these systems possess either $C=0$ or $\lvert C \rvert=1$. 
In this work, we focus on helically twisted $K$-valley TMD homotrilayers, which consist of three layers rotated with equal consecutive twist angles, as depicted in Fig.~\ref{fig:helicalgeometry}(a). 
In this helically stacked geometry, two distinct length scales emerge (moiré and moiré-of-moiré) in the small angle limit. 
We identify high-symmetry stacking configurations that form moir\'e-scale domains, and a clear separation of scales allows us to derive a local continuum model on the moir\'e scale, in which the spin and valley degrees of freedom are locked. 
We study the single-particle phase diagram of a moir\'e-scale domain that emerges in the helical homotrilayer, using continuum model parameters appropriate for MoTe$_2$. 
We find that for twist angles $\theta\gtrsim  2.3^\circ$, the topmost three moiré valence bands are isolated from the remote bands, and we identify three regimes with distinct Chern sequences that emerge as the twist angle is increased, as indicated in Fig.~\ref{fig:helicalgeometry}(c). 
A topmost $|C|=2$ valence band emerges for twist angles $5.1^\circ \lesssim \theta \lesssim 6.8^\circ$. 
We explain the values for the Chern numbers of the three topmost continuum bands by an analysis of symmetry indicators, and derive an effective three-orbital tight-binding model for the system. 
We find that a displacement field of sufficient strength can lead to a transition from a trivial band of $C=0$ to a topological band with $|C|=1$, in contrast to the MoTe$_2$ homobilayer where no such transition occurs, and that the Chern number of the topological $C = -2$ band may be tuned to $C = -1$. 
Finally, we perform a Hartree-Fock (HF) analysis of our continuum model and present mean-field phase diagrams at filling factor $\nu=-1$ as a function of twist angle and displacement field. 
We identify parameter regimes in which the $|C|=2$ band remains stable to adding gate-screened Coulomb interactions.

Our results establish helical homotrilayer TMDs as a promising candidate to experimentally explore interacting physics in bands with Chern numbers greater than one. In contrast to the higher Chern bands of helical trilayer graphene, we find that the homotrilayer TMD topmost band is dispersive and isolated, although the gap to the remote bands is small. 

The paper is organized as follows. In Sec.~\ref{sect2:ctm_model}, we introduce the continuum model description of helical TMD homotrilayers. In Sec.~\ref{sect3:CtmModPhaseDiag} we obtain the single-particle phase diagram as a function of twist angle and displacement field. In Sec.~\ref{sect4:tightbinding} we derive an effective three-orbital tight-binding description of the continuum model. A Hartree-Fock study of the continuum model is presented in Sec.~\ref{sect5:hf_phasediag}. Finally, in Sec.~\ref{sect6:Discussion}, we discuss implications of our results and future research directions. 

\section{Continuum Model of Helical TMDs}
\label{sect2:ctm_model}

\subsection{Helical Stacking Geometry}
In this section, we present a continuum model for three layers of $K$-valley TMDs in the helical configuration with equal twist angle $\theta$ between each pair of adjacent layers shown in Fig.~\ref{fig:helicalgeometry}. 
We define $\mathbf{K}_m$ as the high-symmetry momentum (corresponding to the $K$ valley) of the middle layer's Brillouin zone, as illustrated in Fig.~\ref{fig:helicalgeometry}(b).
The top ($t$) and middle ($m$) layers generate a moir\'e pattern characterized by wave vector $\bq^{tm}_1=R_{\theta}\mathbf K_m-\mathbf K_m$, while the middle and bottom ($b$) layers generate a moir\'e pattern characterized by $\bq^{mb}_1=\mathbf K_m-R_{-\theta}\mathbf K_m = R_{-\theta} \bq^{tm}_1$, where $R_{\theta}$ is the counterclockwise rotation matrix by $\theta$. 
The two moir\'e patterns have a relative rotation of $\theta$ and are generally incommensurate.
These two wave vectors can be decomposed into a moir\'e wave vector $\bq_1=(\bq^{tm}_1+\bq^{mb}_1)/2 \simeq \theta (\hat{\mathbf z}\times \mathbf K_m)$ and a moir\'e-of-moir\'e (or supermoir\'e) wave vector $\delta\mathbf q_1=\bq^{tm}_1-\bq^{mb}_1 \simeq \theta^2(\hat{\mathbf z}\times\hat{\mathbf z}\times\mathbf K_m)$, where we have used the small angle approximation. 
The corresponding moir\'e and supermoir\'e lattices are both triangular, and their principal axes are rotated by $90^\circ$ with respect to each other in the equal twist angle configuration~\cite{yuncheng_2023}.

In the small twist angle regime, the ratio between the periods of the moiré and supermoiré lattices becomes large (e.g., for $\theta\le7^\circ$, $1/\theta\ge8.2$), leading to a clear separation between the two length scales.
This enables an adiabatic approach, where the physics on the moir\'e length scale is captured by a moir\'e-periodic Hamiltonian whose parameters vary slowly over the larger supermoir\'e length scale~\cite{yuncheng_2023,guerci2024chern}. 
As explained in the following subsections, this moir\'e-periodic Hamiltonian is parametrized by the local relative stacking of the $tm$ and $mb$ moir\'e patterns. 
This adiabatic approach is further justified by pronounced lattice relaxation observed in graphene-based supermoir\'e systems~\cite{hoke2024imagingsupermoirerelaxationconductive,Nakatsuji2023,Devakul_2023,yang_prb_2024,Foo2024}, a phenomenon we expect to play a crucial role in $K$-valley helical trilayer TMDs as well~\cite{nakatsuji2025moirebandengineeringtwisted}, especially as TMDs are known to be more elastic than graphene~\cite{sun2019elastic,lee2012young}. 
Such relaxation leads to domains of high-symmetry stacking configurations, which extend over large areas of the moir\'e-of-moir\'e unit cell. 

\subsection{High-Symmetry Stacking Configurations}

We identify three high-symmetry stackings within the moir\'e-of-moir\'e unit cell, each located at a point invariant under the three-fold rotational symmetry $\mathcal{C}_{3z}$.
These stacking configurations may be characterized by relative shifts between the moir\'e patterns formed by the top–middle ($tm$) and middle–bottom ($mb$) layer pairs, and are defined by the local stacking arrangement of the three layers relative to one another. 

\begin{figure*}[ht]
    \centering
    \includegraphics[width=0.95\linewidth]{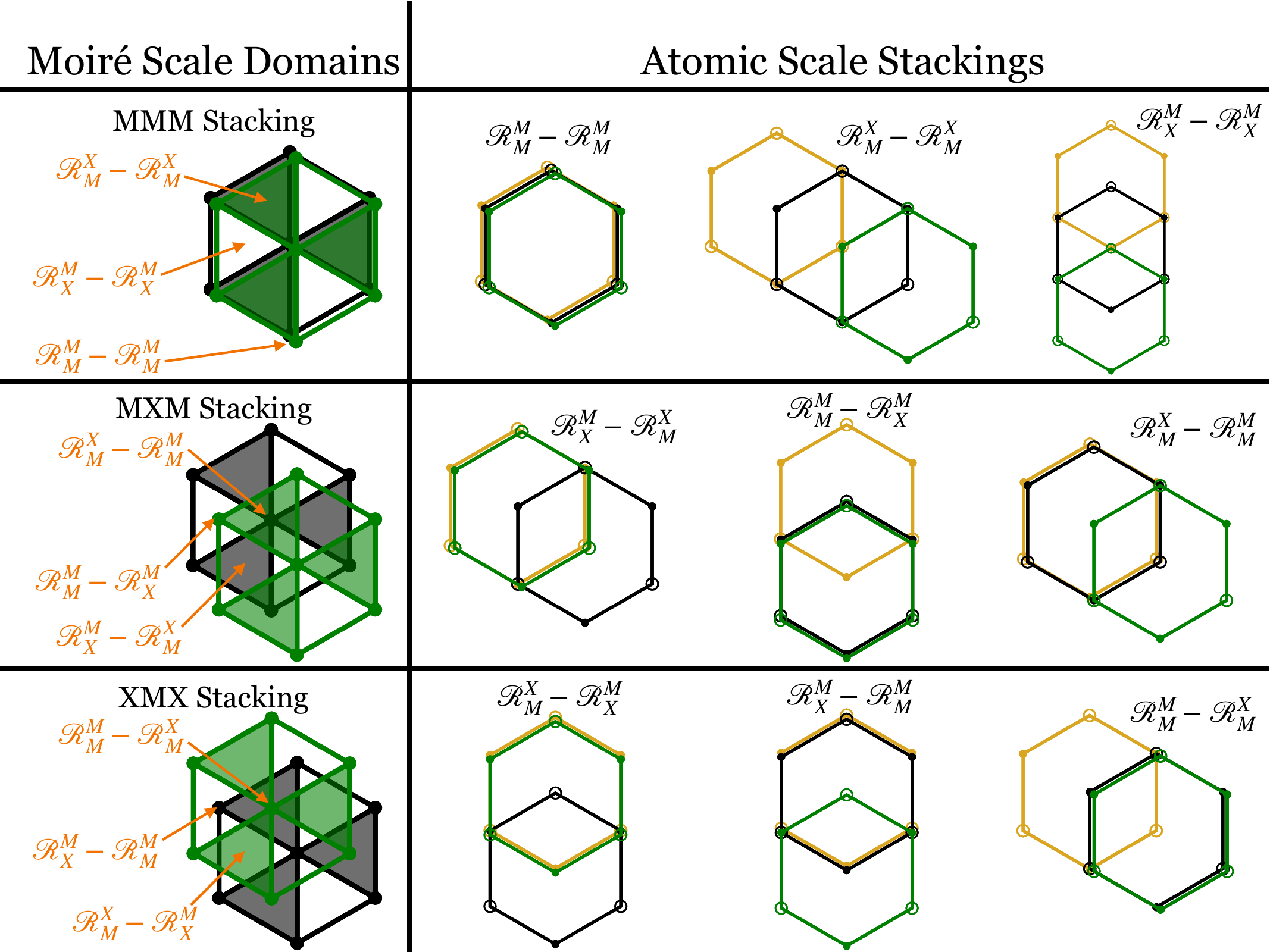}
    \caption{\textbf{Left column:} Local moir\'e-scale stacking configurations created by the $tm$ (green) and $mb$ (black) moir\'e patterns. For each moir\'e pattern, the solid dots at the vertices indicate $\mathscr{R}^M_M$ stacking, the shaded triangles indicate areas of $\mathscr{R}^X_M$ stacking, and the unshaded triangles indicate areas of $\mathscr{R}^M_X$ stacking. The rows correspond to the three high-symmetry stacking configurations $MMM$, $MXM$, and $XMX$. \textbf{Right column:} Corresponding atomic-scale high-symmetry configurations created by the top (green), middle (black), and bottom (yellow) layers. The filled circles represent a transition metal ($M$) atom, while the open circles represent a chalcogen ($X$) atom. }
    \label{fig:stackingconfigs}
\end{figure*}

The configuration displayed in the first row of Fig.~\ref{fig:stackingconfigs}, which we denote $MMM$, is the one in which the $tm$ and $mb$ moir\'e patterns are aligned without any relative shift. 
At the atomic scale, denoting the case in which an atom $\alpha$ is stacked above an atom $\beta$ as $\mathscr{R}^\alpha_\beta$, the $MMM$ domain adiabatically modulates between three atomic-scale high-symmetry configurations: $\mathscr{R}^M_M - \mathscr{R}^M_M, \mathscr{R}^X_M-\mathscr{R}^X_M, \mathscr{R}^M_X-\mathscr{R}^M_X$, where the atomic-scale stackings are written in the order $tm-mb$.
On the other hand, the $MXM$ domain contains the high-symmetry stackings $\mathscr{R}^M_M-\mathscr{R}^M_X, \mathscr{R}^X_M-\mathscr{R}^M_M, \mathscr{R}^M_X-\mathscr{R}^X_M$, and the $XMX$ domain contains $\mathscr{R}^M_M-\mathscr{R}^X_M, \mathscr{R}^X_M-\mathscr{R}^M_X, \mathscr{R}^M_X-\mathscr{R}^M_M$, as shown in the middle and bottom rows of Fig.~\ref{fig:stackingconfigs}. 

\bgroup
\def\arraystretch{1.6}
\begin{table}[t]
    \centering
    \begin{tabular}{|c|c|c|c|c|} \hline  
         Domain & $\boldsymbol{\phi}$ & \multicolumn{3}{|c|}{High Symmetry Stackings} \\ \hline  
         $MMM$ & $[0,0,0]$ & $\mathscr{R}^M_M - \mathscr{R}^M_M$ &  $\mathscr{R}^X_M-\mathscr{R}^X_M$  & $\mathscr{R}^M_X-\mathscr{R}^M_X$\\ \hline  
         $MXM$ & $\left[0,-\frac{2\pi}{3},\frac{2\pi}{3}\right]$ & $\mathscr{R}^M_M-\mathscr{R}^M_X$ & $\mathscr{R}^X_M-\mathscr{R}^M_M$ & $\mathscr{R}^M_X-\mathscr{R}^X_M$\\ \hline  
         $XMX$ &  $\left[0,\frac{2\pi}{3},-\frac{2\pi}{3}\right]$ & $\mathscr{R}^M_M-\mathscr{R}^X_M$& $\mathscr{R}^X_M-\mathscr{R}^M_X$& $\mathscr{R}^M_X-\mathscr{R}^M_M$ \\ \hline \end{tabular}
    \caption{Each high-symmetry stacking configuration on the moir\'e scale corresponds to three high-symmetry stackings on the atomic scale. 
    These are listed with the notation that $\mathscr{R}^\alpha_\beta - \mathscr{R}^\gamma_\delta$ denotes that the $\alpha$ atom of the top layer is aligned with the $\beta$ atom of the middle layer, and that the $\gamma$ atom of the middle layer is aligned with the $\delta$ atom on the bottom layer. 
    The local stacking configuration is parametrized in the Hamiltonian, Eq.~\eqref{eq:Hamiltonian_1}, by a vector $\boldsymbol{\phi}$ of phases, which are derived in Appendix~\ref{sect:stacking_phases}. }
    \label{tab:stackingconfigs}
\end{table}
\egroup

These latter two moir\'e-scale stacking configurations have a relative shift of $\mathbf{d}^{MXM/XMX} = \pm (\bL_1 - 2 \bL_2)/3$ between the $tm$ and $mb$ moir\'e structures, where the real space moir\'e lattice vectors $\bL_i$ are defined by $\bL_i \cdot \bG_j = 2\pi \delta_{ij}$ in relation to the moir\'e reciprocal lattice vectors $\bG_j$ in Fig.~\ref{fig:helicalgeometry}(b). 
For convenience, the three high-symmetry stacking configurations are summarized in Table~\ref{tab:stackingconfigs}. 

\subsection{Continuum Model Hamiltonian}
\label{sect:ctm_mod_hamiltonian}

We follow Refs.~\cite{yuncheng_2023,guerci2024chern} to derive the local continuum Hamiltonian that describes the physics on the moir\'e length scale (see Appendix~\ref{sect:CtmModDetails} for details). 
We restrict ourselves to the first harmonic approximation of the moir\'e potential and only allow couplings between adjacent layers. 
Within these assumptions, the local $K$-valley Hamiltonian  reads
\begin{equation}
\label{eq:Hamiltonian_1}
    H_K(\br, \boldsymbol{\phi}) =-\frac{\hbar^2 \bk^2}{2m^*}\mathbb 1+ \begin{pmatrix}
    V_t(\br,\boldsymbol{\phi}) &  t_{\boldsymbol{\phi}}(\br)  & 0 \\ 
    t_{\boldsymbol{\phi}}^*(\br) &V_m(\br,\boldsymbol{\phi}) &  t_{-{\boldsymbol{\phi}}}(\br)  \\ 
    0 & t^*_{-\boldsymbol{\phi}}(\br) &V_b(\br,\boldsymbol{\phi})
    \end{pmatrix},
\end{equation}
where the first term describes the monolayer valence band dispersions, modeled by parabolas with effective mass $m^*$. 
Here, the vector $\boldsymbol{\phi} = [\phi_1, \phi_2, \phi_3]$ parametrizes the stacking configurations at the moir\'e scale. 
We note that this is equivalent to treating the effect of the supermoir\'e pattern adiabatically (see Appendix~\ref{sect:adiabatic_approximation} for details). 
As shown in Appendix~\ref{sect:stacking_phases}, the $MMM$ stacking configuration corresponds to the case in which all $\phi_j = 0$, the $MXM$ configuration is represented by $\boldsymbol{\phi}^{MXM} = [0, -2\pi/3, 2\pi/3]$, and the $XMX$ configuration is represented by $\boldsymbol{\phi}^{XMX} = [0, 2\pi/3, -2\pi/3]$. 
The $\boldsymbol{\phi}$ vectors corresponding to each high-symmetry stacking configuration are also listed in Table~\ref{tab:stackingconfigs}. 

Note that Eq.~\eqref{eq:Hamiltonian_1} is expressed with a layer-dependent momentum boost such that $\mathbf{k}=0$ corresponds to the maximum of each monolayer valence band dispersion. 
The elements of Eq.~\eqref{eq:Hamiltonian_1} are dependent on a set of continuum model parameters that can be \emph{partially} constrained via the rotational symmetries $\mathcal{C}_{3z}$ and $\mathcal{C}_{2y}$ of the helical trilayer tTMD system. 
The details of these constraints are provided in Appendix~\ref{sect:ctm_mod_prms}.
The diagonal terms represent the intralayer moir\'e potentials 
\begin{equation}
    \begin{split}
        V_t(\br,\boldsymbol{\phi}) = 2V &\sum_{j=1}^{3}\cos(\bG_j\cdot \br+\psi+\Delta \phi_j),\\
        V_m(\br,\boldsymbol{\phi}) = 2V &\sum_{j=1}^{3} \left[ \cos(\bG_j\cdot \br-\psi+\Delta\phi_j) \right. \\
        &\left.+\cos(\bG_j\cdot \br+\psi-\Delta\phi_j) \right], \\
        V_b(\br,\boldsymbol{\phi}) = 2V &\sum_{j=1}^{3}\cos(\bG_j\cdot \br-\psi-\Delta \phi_j),
    \end{split}
    \label{eq:intralayerpotentials}
\end{equation}
where $\Delta \phi_j = \phi_j - \phi_{j+1}$, and the off-diagonal terms describe the tunneling between adjacent layers with amplitude $w$,
\begin{equation}
    t_{\bm\phi}(\br)=w\sum_{j=1}^{3}e^{-i \bq_j\cdot \br} e^{-i\phi_j}. 
    \label{eq:interlayer_tunnelling}
\end{equation}
The moir\'e reciprocal lattice vectors $\bG_j$ and momentum transfer vectors $\bq_j$ are illustrated in Fig.~\ref{fig:helicalgeometry}(b). 
The continuum model parameters $(V, \psi, w, m^*)$ are analogous to those in the continuum model of the TMD homobilayer~\cite{wu2019topological}. 

The Hamiltonian of the other valley is simply obtained via time-reversal symmetry, $H_{K'}(\boldsymbol{\phi})=H_K^*(\boldsymbol{\phi})$. 
We note that the $MXM$ and $XMX$ stacking configurations are related to each other by the symmetry $\psi \rightarrow - \psi, \, \br  \rightarrow -\br$. 
In addition, the Hamiltonian in Eq.~\eqref{eq:Hamiltonian_1} is invariant under the symmetries $\mathcal{C}_{3z}$ and $\mathcal{C}_{2y}\mathcal{T}$. 
There is also a three-dimensional pseudoinversion symmetry $\mathcal{I}$ that is defined by swapping the $t$ and $b$ layers, which acts as $D(\mathcal{I}) H_K(\br)D^\dagger(\mathcal{I}) = H_K(-\br)$. 
As in the TMD homobilayer case, the $\mathcal{I}$ symmetry is only present in our model as we restrict the moir\'e potential to first harmonic terms~\cite{Jia-PhysRevB.109.205121}. $\mathcal{I}$ is not a symmetry of the actual system, and the addition of higher harmonics would generically remove it from the model. 
A more expansive analysis of the symmetries of the continuum model for each stacking configuration is provided in Appendix~\ref{sect:localHam_symm}.

\section{Single-Particle Phase Diagram}
\label{sect3:CtmModPhaseDiag}

In this section, we study the single-particle properties of the model corresponding to Eq.~\eqref{eq:Hamiltonian_1} by exploring its phase space of parameters.

\subsection{Chern Numbers in the Continuum Model}

\begin{figure}[h]
    \centering
    \includegraphics[width=0.98\linewidth]{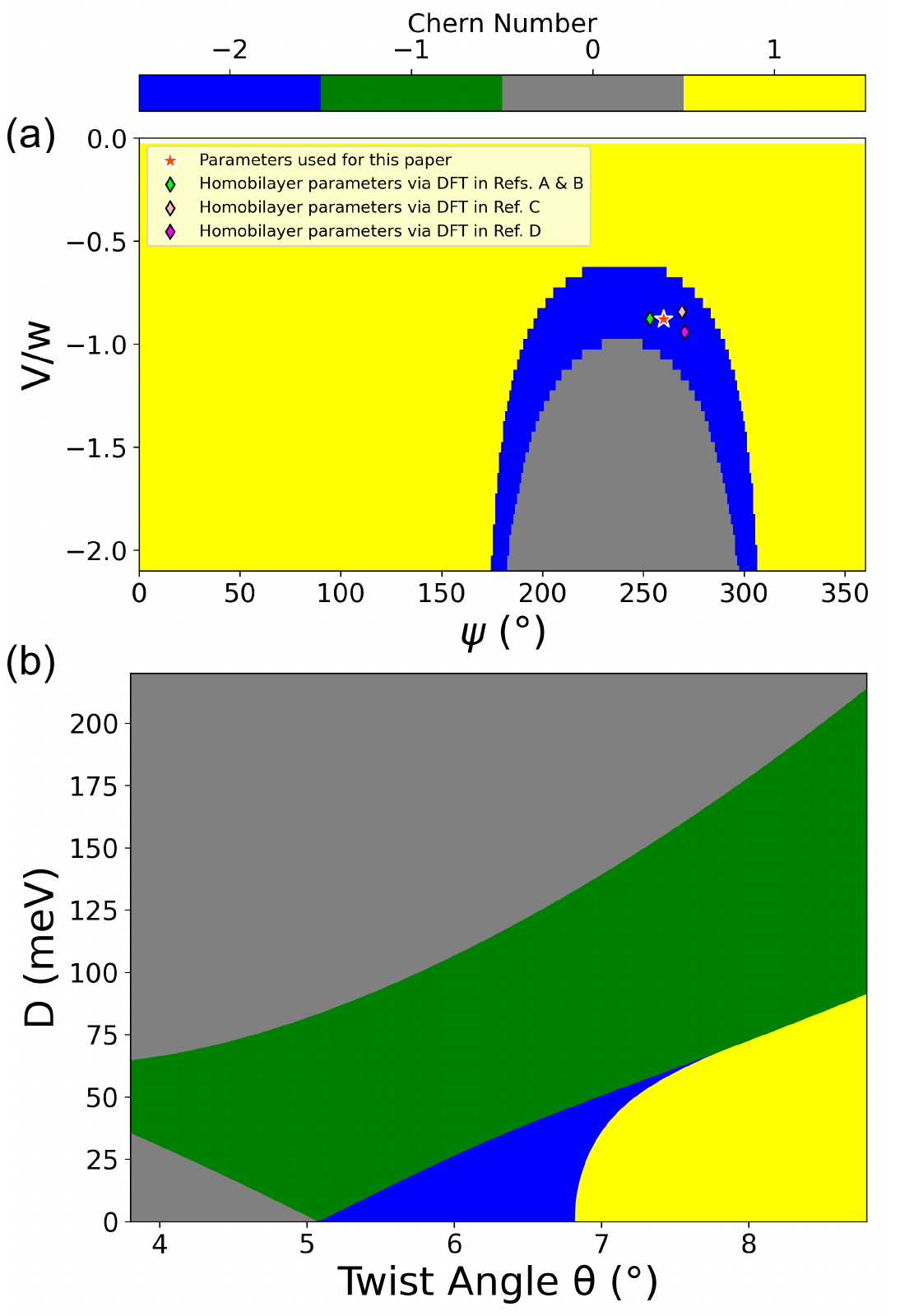}
    \caption{\textbf{(a)} Chern number phase diagram for the topmost valence band of the $XMX$-stacked helical TMD homotrilayer at twist angle $\theta = 5.6^\circ$ as a function of varying continuum model parameters $V/w$ vs.~$\psi$ with fixed $w = -18.8\text{ meV}$ and $m^* = 0.60m_e$. Diamond markers indicate the parameter values obtained via DFT for homobilayer tMoTe$_2$ at $\theta = 3.89^\circ$ in Refs.~\cite{Jia-PhysRevB.109.205121,wang2024fractional} (denoted Refs.~A~\&~B) with $m^* = 0.60m_e$, and $\theta = 4.4^\circ$ in Ref.~\cite{reddy2023} (denoted Ref.~C) with $m^* = 0.62m_e$, as well as for three high-symmetry displacements for the untwisted MoTe$_2$ homobilayer in Ref.~\cite{wu2019topological} (denoted Ref.~D) with $m^* = 0.62m_e$. 
    The orange star indicates the values we use throughout this paper: $(V, \psi, w, m^*) = (16.5\text{ meV}, -100^\circ, -18.8\text{ meV}, 0.60m_e)$. Each of these sets of parameters lie within the $C = -2$ region. 
    \textbf{(b)} $K$-valley Chern number phase diagram of the topmost valence band for $XMX$ stacking as a function of varying twist angle and applied displacement field $D$. As $D$ increases, there is a region of $C = -1$ that emerges for a range of twist angles in which the topmost band has $C = 0$ or $-2$ when $D = 0$. 
    When $D$ is sufficiently large, the topmost band is trivialized at all $\theta$. }
    \label{fig:chernphases}
\end{figure}

We consider each high-symmetry stacking configuration on the moir\'e scale and investigate the band structure of the corresponding $K$-valley Hamiltonian given in Eq.~\eqref{eq:Hamiltonian_1}. 
Following Ref.~\cite{wu2019topological}, we diagonalize the Hamiltonian in a plane wave basis, using a cut-off of $6$ shells of reciprocal lattice vectors, which is sufficient to guarantee convergence of the resulting band structure. 
We study the properties of the top few valence bands as a function of continuum model parameters. 

For the $MMM$ stacking domain, the topmost moir\'e valence band is topologically trivial for a wide range of continuum model parameters, including the continuum model parameters for the MoTe$_2$ homobilayer found using DFT calculations~\cite{Jia-PhysRevB.109.205121,wang2024fractional,reddy2023,wu2019topological}. 
Meanwhile, in the area of the phase space where the topmost valence band is topologically non-trivial, sometimes exhibiting a Chern number $\lvert C \rvert = 3$ and sometimes $\lvert C \rvert = 1$, the gap to the nearest remote band is small ($\lesssim 2\text{ meV}$). 
We provide more details in Appendix~\ref{sect:mmm-stacked_numerics}. 
In addition, lattice relaxation effects will likely cause the relative areas of the $MMM$-stacked domain to be very small relative to those of the domains of $MXM/XMX$ stacking, as demonstrated for WSe$_2$ in Ref.~\cite{nakatsuji2025moirebandengineeringtwisted}. 
Since in this work we are mostly interested in bands with $|C|>1$ that may be experimentally accessible, we focus on the phase diagram of the $XMX$ stacking in the following discussion, with the understanding that the phase diagram of the $MXM$ stacking is related by the symmetry mentioned in Sec.~\ref{sect:ctm_mod_hamiltonian}. 

For the $XMX$ stacking, our numerical computations reveal that there is a range of parameters for which the topmost valence band has a $K$-valley Chern number $C=-2$ (as the $K, K'$ valleys are time-reversal symmetry pairs, the $K'$-valley Hamiltonian will have an equivalent region with Chern number $C=+2$).
In the $V/w$ vs.~$\psi$ plane, this $C=-2$ region of the phase diagram shifts as the twist angle is varied, and we present a representative phase diagram for $\theta = 5.6^\circ$ in Fig.~\ref{fig:chernphases}(a) using $w = -18.8$ meV and $m^*=0.60\,m_e$. 
The diamond markers on Fig.~\ref{fig:chernphases}(a) refer to the various DFT-derived continuum model parameters for twisted homobilayer MoTe$_2$ at twist angles $\theta = 3.89^\circ$ (with $m^* = 0.60 m_e$)~\cite{Jia-PhysRevB.109.205121,wang2024fractional} and $\theta=4.4^\circ$ (with $m^*=0.62m_e$)~\cite{reddy2023}, as well as an untwisted AA-stacked MoTe$_2$ homobilayer (with $m^* = 0.62m_e$)~\cite{wu2019topological}. 
We note that the fitted parameters of Refs.~\cite{Jia-PhysRevB.109.205121,wang2024fractional} still provide a reliable approximation of the topmost valence bands for the bilayer case around $\theta=5.09^\circ$, as shown in Appendix~\ref{sect:dft_prms_bilayer}. 

As an analogous DFT calculation for the helical trilayer supermoir\'e structure would be computationally intractable, we take the parameters for the homobilayer case to be a reasonable approximation for our trilayer system, given that our Hamiltonian, Eq.~\eqref{eq:Hamiltonian_1}, only directly couples adjacent layers together. 
As there is a small variance in these DFT-obtained values, we use the parameters $V = 16.5$ meV, $\psi = -100^\circ$, $w = -18.8$ meV, and $m^* = 0.60\,m_e$ (denoted by the orange star in Fig.~\ref{fig:chernphases}(a)) for all the results presented in the remainder of the main text.

For this choice of continuum model parameters, the topmost three valence bands are isolated from the other remote bands when $\theta \gtrsim 2.3^\circ$. 
As long as they are isolated, there are three primary regimes of interest, as depicted in Fig.~\ref{fig:helicalgeometry}(c). 
In Regime $I$ ($2.3^\circ \lesssim \theta \lesssim 5.1^\circ$), the Chern sequence of the topmost valence bands is $[0, -1, +1]$, where the Chern numbers are listed starting from the valence band with the highest energy. 
In Regime $II$ ($5.1^\circ \lesssim \theta \lesssim 6.8^\circ$), the Chern sequence of the topmost bands is $[-2, +1, +1]$. 
We note that the $\lvert C \rvert = 2$ band is robust to a change of the middle layer's intralayer parameters (that are not enforced by symmetry), as shown in Appendix~\ref{sect:addl_ctm_prms}. 
Finally, in Regime $III$ ($\theta \gtrsim 6.8^\circ$), the Chern sequence of the topmost bands is $[+1, -2, +1]$. 
This last Chern sequence persists into the large angle limit. 
Note that for the $MXM$ stacking, keeping the same continuum model parameters, we do not observe a $|C|=2$ regime but only a $|C|=1$ regime (see Appendix~\ref{sect:addl_mxm}).

\subsection{Symmetry Indicators and the Large Angle Limit}
\label{sect:SymmAtLargeAngle}

With increasing twist angle $\theta$, we find that the transition from a $C = 0$ to $C = -2$ topmost valence band occurs due to a gap closure and reopening between the first two bands at $K, K'$, while a band inversion at $M$ marks the transition from $C = -2$ to $C = +1$. 
To explain how this gives rise to the three regimes of Chern number sequences, we analyze high-symmetry indicators by examining the eigenvalues of the threefold rotation operator $\mathcal{C}_{3z}$ and the 3D pseudoinversion operator $\mathcal{I}$ at high-symmetry points of the mBZ. 

The Chern number of the topmost band in Fig.~\ref{fig:helicalgeometry}(c) can be inferred (modulo $3$) from the $\mathcal{C}_{3z}$ operator, the representation of which is given by 
\begin{equation}
    D^{XMX}(\mathcal{C}_{3z})=\text{diag}[\omega^*,1,\omega^*]
    \label{eq:c3xmx}
\end{equation}
for the XMX stacking, where $\omega = e^{2\pi i/3}$ (see Appendix~\ref{sect:localHam_symm}). Denoting the $\mathcal{C}_{3z}$ eigenvalues at $\mathbf{k} = {K, K', \Gamma}$ by $\xi_{\mathbf{k}}$, we use the relation~\cite{fang_2012,Po_2017,PhysRevB.110.205124,reddy2023,crepel_2025}
\begin{equation}\label{eq:mod_3}
    e^{2\pi i C/3} = \xi_K \xi_{K'} \xi_{\Gamma},
\end{equation} 
to infer that the Chern number of the topmost valence band must satisfy $C = 1 \mod 3$ in the large angle limit. 

As our model retains only the first harmonic of the moir\'e potential, it features an emergent pseudoinversion symmetry with representation 
\begin{equation}
    D(\mathcal{I}) = \gamma^x = \begin{pmatrix}
    0 & 0 & 1\\
    0 & 1 & 0\\
    1 & 0 & 0
    \end{pmatrix}, 
    \label{eq:pseudoinv_oper}
\end{equation}
which constrains the Chern number mod 6 according to
\begin{equation}\label{eq:mod_6}
    e^{2\pi i C/6} = \xi_K\xi_\Gamma \gamma_M,
\end{equation}
where $\gamma_M$ is the pseudoinversion eigenvalue of the topmost eigenstate at $M$. 

The above allows us to distinguish between e.g.~$C=1$ and $C=-2$ in the large angle limit.
The main idea is only outlined here; a full derivation is provided in Appendix~\ref{sect:pseudoinv_eigs}. 
We employ perturbation theory in the large angle limit, where the kinetic energy is much larger than the interlayer potential. We find that the avoided crossing at $M_1$ (see rightmost panel in Fig.~\ref{fig:helicalgeometry}(b)) involving the two plane waves originating from $K$ and $K'$ in the moir\'e Brillouin zone, which we denote as $\lvert \psi_K(M_1) \rangle$ and $\lvert \psi_{K'}(M_1) \rangle$, is lifted by a matrix element of the form
\begin{equation}
\label{eq:perturbative_hamiltonian}
    \left[\delta H_{w}\right]_{ij} ={2w^2}\left[1-{\sigma^x}/{2}\right]_{ij}/{E_{\rm gap}}.
\end{equation}
Here, $E_{\rm gap}$ denotes the nonzero energy separation between the topmost plane-wave state and the remote bands at $M_1$. 
The perturbation given in Eq.~\eqref{eq:perturbative_hamiltonian} lifts the degeneracy and selects the antisymmetric linear combination $\ket{-} = (\ket{\psi_K(M_1)} - \ket{\psi_{K'}(M_1)})/\sqrt{2}$ as the highest-energy eigenstate. 
This state is odd under inversion, which corresponds to $\gamma_M = -1$.
Using Eq.~\eqref{eq:mod_3}, we find that $C = 1 \mod{6}$, which is consistent with our numerical finding that the topmost band has $C = +1$ when $\theta \gtrsim 6.8^\circ$. 

For smaller twist angles, the perturbative analysis is no longer valid, so we compute the symmetry eigenvalues numerically to track the evolution of the topmost valence band’s Chern number. 
We present these results in Table~\ref{tab:symm_eigvals}. 
\bgroup
\def\arraystretch{1.6}
\begin{table}[]
    \centering
    \begin{tabular}{|c|c|c|c|}
        \hline
        & \textbf{Regime $I$} & \textbf{Regime $II$} & \textbf{Regime $III$} \\
        \hline
        \textbf{Twist Angles} & $2.3^\circ \lesssim \theta \lesssim 5.1^\circ$ & $5.1^\circ \lesssim \theta \lesssim 6.8^\circ$ & $\theta \gtrsim 6.8^\circ$\\
        \hline
        \makecell{\textbf{Chern} \\ \textbf{Sequences}} & $[0,-1,-1]$ & $[-2,1,1]$ & $[1,-2,1]$\\
        \hline
        \makecell{$\mathcal{C}_{3z}$\\ \textbf{Eigenvalues}\\ \textbf{
        at }$\mathbf{[K', K, \Gamma]}$} & $[1,1,1]$& $[\omega^*, \omega^*, 1]$ & $[\omega^*, \omega^*, 1]$ \\
        \hline
        \makecell{\textbf{$\mathcal{I}$ Eigenvalues}\\ \textbf{at M}} & +1 & +1 & -1 \\
        \hline
    \end{tabular}
    \caption{Chern sequences (ordered from the highest band) for top three valence bands, with $\mathcal{C}_{3z}$ and $\mathcal{I}$ eigenvalues at high-symmetry points for the topmost band in the absence of an applied displacement field. 
    The symmetry eigenvalues determine the Chern number $C\,\text{mod}\,6$ of the topmost band.}
    \label{tab:symm_eigvals}
\end{table}
\egroup
Having confirmed that $C = 1 \mod{6}$ in Regime $III$, we next observe that the pseudoinversion symmetry eigenvalue at $M$ changes sign when moving from Regime $III$ to Regime $II$. 
Eq.~\eqref{eq:mod_6} then implies that the topmost band has $C = -2 \mod{6}$ in the intermediate regime, as expected. 
Finally, across the transition from Regime $II$ to Regime $I$, the pseudoinversion eigenvalue remains the same, but the $\mathcal{C}_{3z}$ eigenvalues at $K$ and $K'$ are exchanged, yielding a topmost band with $C = 0 \mod{6}$. 

\subsection{Displacement Field}
\begin{figure*}
    \centering
    \includegraphics[width=0.98\linewidth]{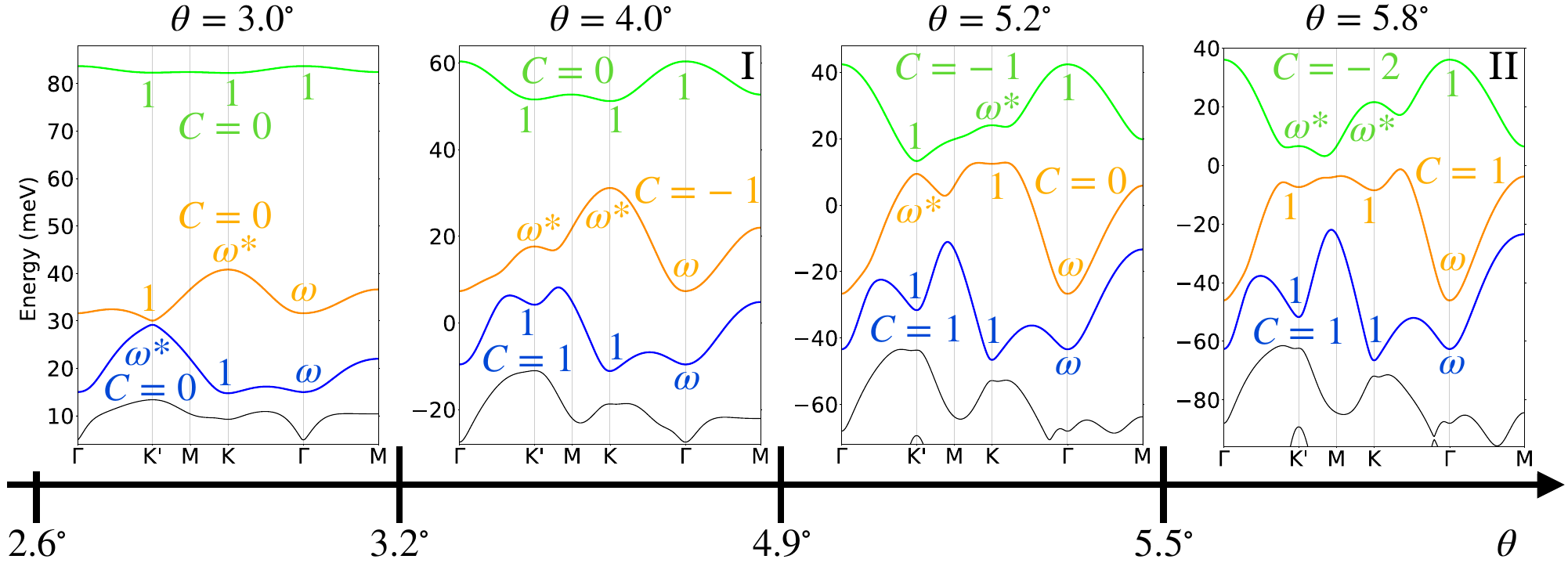}
    \caption{Representative band structures labeled with Chern numbers and $\mathcal{C}_{3z}$ eigenvalues at the high-symmetry points after the application of a displacement field of $D = 8\,\text{meV}$. 
    The band inversion at $K$ occurs at a lower angle ($\theta \simeq 4.9^\circ$) than the inversion at $K'$ ($\theta \simeq 5.5^\circ$).
    Labels $I$ and $II$ indicate band structures with identical Chern number sequences as in Regimes $I$ and $II$ in Fig.~\ref{fig:helicalgeometry}(c).}
    \label{fig:dispfield_c3eigs}
\end{figure*}
To incorporate the effect of an out-of-plane displacement field $D$, we add to the continuum model (Eq.~\eqref{eq:Hamiltonian_1}) the term
\begin{equation}
    H_K^\text{disp} = \begin{pmatrix}
        +D & 0 & 0\\
        0 & 0 & 0\\
        0 & 0 & -D
    \end{pmatrix} 
    \label{eq:displHamil},
\end{equation}
which breaks the pseudoinversion symmetry and thus the degeneracy between the $K, K'$ points in the mBZ. 
We illustrate the Chern number phase diagram of the topmost valence band as a function of $D$ vs. $\theta$ in Fig.~\ref{fig:chernphases}(b). 
We restrict ourselves to $D \geq 0$ since the phase diagram is invariant under $D \rightarrow -D$ (see Appendix~\ref{sect:pseudoinv_at_Dneq0}). 

The phase diagram shows that a nonzero $D > 0$ induces a $C = -1$ state at twist angles that would otherwise have $C = 0$ or $-2$. 
At large twist angles of $\theta \gtrsim 6.8^\circ$, the $C = -1$ band transitions into a $C = +1$ state when a strong $D$ is applied, though there is a small intermediate phase of $C = +2$ for $6.8^\circ \lesssim \theta \lesssim 8.3^\circ$. 
As expected, in the large $D$ limit, the topmost band is topologically trivial regardless of twist angle, as the displacement field eventually completely layer-polarizes the TMD trilayer, preventing hybridization between the electronic dispersions centered on each layer. 

In twisted homobilayer MoTe$_2$, it was previously found that the application of an external displacement field can drive a topological phase transition from a topologically non-trivial $C=1$ band to a trivial $C=0$ one~\cite{FCI_Experiment1,FCI_Experiment2}.
Unlike the bilayer case, however, we find that the application of a sufficiently strong displacement field in the helical MoTe$_2$ trilayer induces a transition in the topmost valence band from a trivial $C=0$ band to a topological $C=-1$ band in Regime $I$. 
In addition, for a different range of twist angles, a sufficiently strong displacement field drives a transition from a $C=-2$ phase to a $C=-1$ one in the topmost valence band. 

An explanation for these displacement field-induced phase transitions is given in Fig.~\ref{fig:dispfield_c3eigs}, which depicts the critical values of $\theta$ and representative band structures of each phase with labeled $\mathcal{C}_{3z}$ eigenvalues for $D = 8\,\text{meV}$.  
A positive $D$ pushes the energy of the topmost band up at $K$ and down at $K'$. 
As a result, the topological transition from Regime $I$ to $II$ that is depicted in Fig.~\ref{fig:helicalgeometry}(c) for $D = 0$ happens in two steps when $D > 0$, and the values of $\theta$ at which the topological phase transitions occur are altered. 
Starting from Regime $I$, as $\theta$ is increased, there is a band crossing between the highest two valence bands that occurs at $K$, in which the $\mathcal{C}_{3z}$ eigenvalues at $K$ are exchanged. 
Thus, there is a transition from a trivial topmost valence band to a $C=-1$ phase. 
Then, as $\theta$ is further increased, a band crossing at $K'$ occurs, and so the $\mathcal{C}_{3z}$ eigenvalue exchange at $K'$ causes a transition from the $C=-1$ to $C=-2$ phase in the topmost valence band. 

\section{Effective Lattice Description}
\label{sect4:tightbinding}

In this section, we present an effective tight-binding model obtained via Wannierization of the topmost three continuum bands. 
The Wannier orbitals used in the construction of this model form a triangular and a honeycomb lattice. 
Using this tight-binding model, we further explain the Chern sequences numerically obtained in Section~\ref{sect3:CtmModPhaseDiag} for each regime in terms of the hoppings within and between these lattices. 

\subsection{Wannier Orbital Construction}
As our continuum model results above indicate (see Figs.~\ref{fig:helicalgeometry}(c) and~\ref{fig:dispfield_c3eigs}), if we use parameters similar to those obtained in Refs.~\cite{Jia-PhysRevB.109.205121,reddy2023,wang2024fractional,wu2019topological} for the MoTe$_2$ homobilayer, the three topmost bands are isolated from the remote bands and have Chern numbers that add up to zero for a wide range of twist angles $\theta \gtrsim 2.3^\circ$ (or at a larger critical angle if $D \neq 0$).  

\begin{figure}
    \centering
    \includegraphics[width=0.98\linewidth,]{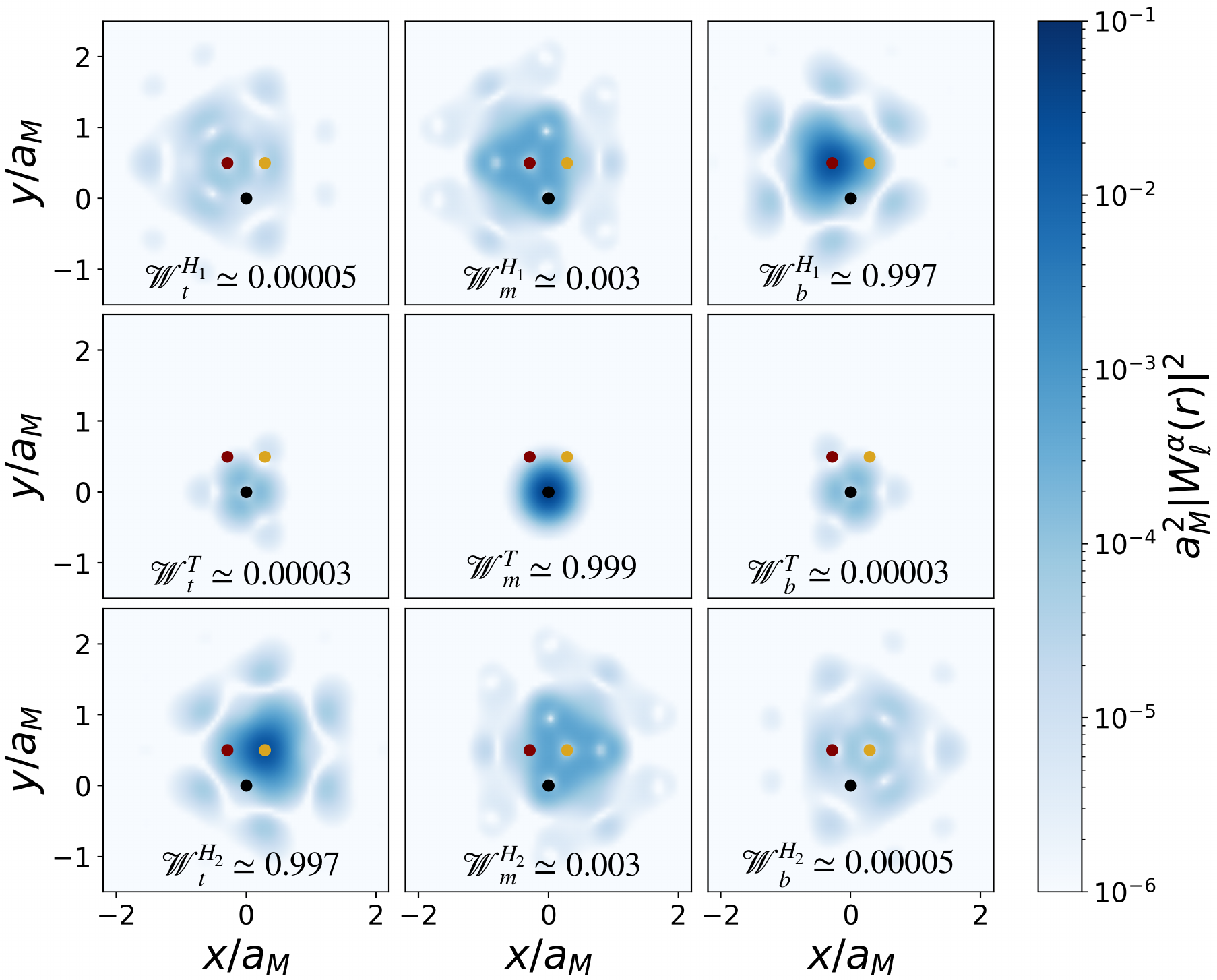}
    \caption{Squared modulus of Wannier orbitals centered at $\alpha$, $W^\alpha(\br)$, for each layer $\ell \in \{t,m,b\}$ at $\theta = 3^\circ$ plotted on a logarithmic color scale in real space, with a cut-off of $10^{-6}$. 
    Each plot provides the weight $\mathcal{W}_\ell^\alpha$ as defined in Eq.~\eqref{eq:wannierorbital_weights}. 
    The maroon, black, and yellow dots mark the $\alpha \in \{H_1, T, H_2 \}$ sites, respectively. 
    Real space coordinates are measured in units of the moir\'e period $a_M$. }
    \label{fig:Wannier_theta3}
\end{figure}

Motivated by this observation, we Wannierize~\cite{PhysRevB.56.12847, RevModPhys.84.1419, PhysRevLett.98.046402} the three topmost bands of the continuum Hamiltonian, Eq.~\eqref{eq:Hamiltonian_1}, of the $XMX$-stacked domain. 
Following Refs.~\cite{devakul2021magic,Wu_PhysRevX.13.041026,Crepel_PhysRevResearch.6.033127}, we then use the Wannier orbitals to construct a minimal tight-binding model comprising two orbitals centered at honeycomb sites and one orbital centered at triangular sites. 
The honeycomb sites are located at atomic stackings $\mathscr R^M_X - \mathscr R^M_M$ and $\mathscr R^M_M - \mathscr R^X_M$, which we denote as $H_1$ and $H_2$ (hexagonal), respectively. 
The triangular sites are located at the atomic stacking $\mathscr R^X_M - \mathscr R^M_X$, which we denote as $T$ (triangular). 
We start from the Bloch states of the three continuum model bands 
\begin{align}
    \psi_{\bm k}^{n}({\bm r})=u^n_{{\bm k}}({\bm r})e^{i{\bm k}\cdot {\bm r}}=\sum_{\bm G}z^{n}_{{\bm k},{\bm G}}\,e^{i{({\bm k}+ {\bm G})}\cdot {\bm r}},
    \label{eq:Bloch_States}
\end{align}
where $n \in \{ 1,2,3 \}$ is the band index, $u^n_{{\bm k}}({\bm r})$ is the periodic component of the Bloch state, and we have used its Fourier expansion in the second equality. 
The numerical coefficients $z^{n}_{{\bm k},{\bm G}}$ are obtained by diagonalizing the continuum model. 
We apply a unitary transformation $U_\Pi({\bm k})$ to the states in Eq.~\eqref{eq:Bloch_States} to obtain three linear combinations labeled using the orbital index $\alpha \in \{H_1,T, H_2\}$ that are mostly polarized in the top, middle, and bottom layers, respectively. The resulting quasi-Bloch states are written as
\begin{align}
    \widetilde{\psi}_{\bm k}^{\alpha}({\bm r})=\sum_{m}U^{m\alpha}_{\Pi}({\bm {k}})\,\psi_{\bm k}^m({\bm r}). 
    \label{eq:quasiBloch}
\end{align}
The construction of the transformation matrix $U_\Pi({\bm k})$ and other details of the Wannierization procedure are given in Appendix~\ref{sect:Wannierisation_method}.
In order to fix the gauge of the quasi-Bloch states given in Eq.~\eqref{eq:quasiBloch}, we require that the top component of $\widetilde{\psi}_{\bm k}^{H_1}$, the middle component of $\widetilde{\psi}_{\bm k}^{T}$, and the bottom component of $\widetilde{\psi}_{\bm k}^{H_2}$ are all real and positive at the position where each orbital is centered. This can be achieved by applying a unitary transformation $U_\varphi(\bk)$ to the quasi-Bloch states for each $\alpha$
\begin{align}
    \label{eq:gauge_quasiBloch}
    \varphi_{\bm k}^{\alpha'}({\bm r})=\sum_{\alpha} U^{\alpha'\alpha}_{\varphi}({\bm k})\,\widetilde{\psi}_{\bm k}^{\alpha}({\bm r}).
\end{align}
Explicit expressions for $U_{\varphi}({\bm k})$ are provided in Appendix~\ref{sect:Wannierisation_method}.
The Wannier function centered at site $\alpha$ is then expressed as 
\begin{align}
    W^{\alpha}({\bm r})=\frac{1}{\sqrt{N}}\sum_{\bm k}\varphi_{{\bm k}}^{\alpha}(\bm r), 
    \label{eq:Wannier_orbitals}
\end{align}
where $N$ is the number of moiré unit cells. 
The Wannier function can be written in terms of its layer components $W^\alpha(\br)=(W^\alpha_t(\br), W^\alpha_m (\br), W^\alpha_b (\br))^T$, and we define the weight of the orbital on a given layer via 
\begin{equation}
    \mathcal{W}^\alpha_\ell = \int \lvert W^\alpha_\ell (\br) \rvert^2 \, d^2\br. 
    \label{eq:wannierorbital_weights}
\end{equation}
A representative example of the resulting Wannier functions for the topmost three continuum model bands in the case $\theta = 3.0^\circ$ is shown in Fig.~\ref{fig:Wannier_theta3} using a logarithmic scale. 
In Appendix~\ref{sect:wannier_localisation}, we present the equivalent figure for $\theta = 5.6^\circ$, and show line cuts across the Wannier functions, demonstrating that they are exponentially localized. 
The Wannier orbitals respect the $\mathcal{C}_{2y}$ and $\mathcal{C}_{3z}$ symmetries of the system about the high-symmetry points at which they are centered. 
The $\mathcal{C}_{2y}$ symmetry relates the orbitals at $H_1$ and $H_2$ to each other and the orbital at $T$ to itself.

\subsection{Tight-Binding Model}
\label{sect:tight-binding_model}

\begin{figure}
    \centering
    \includegraphics[width=\linewidth]{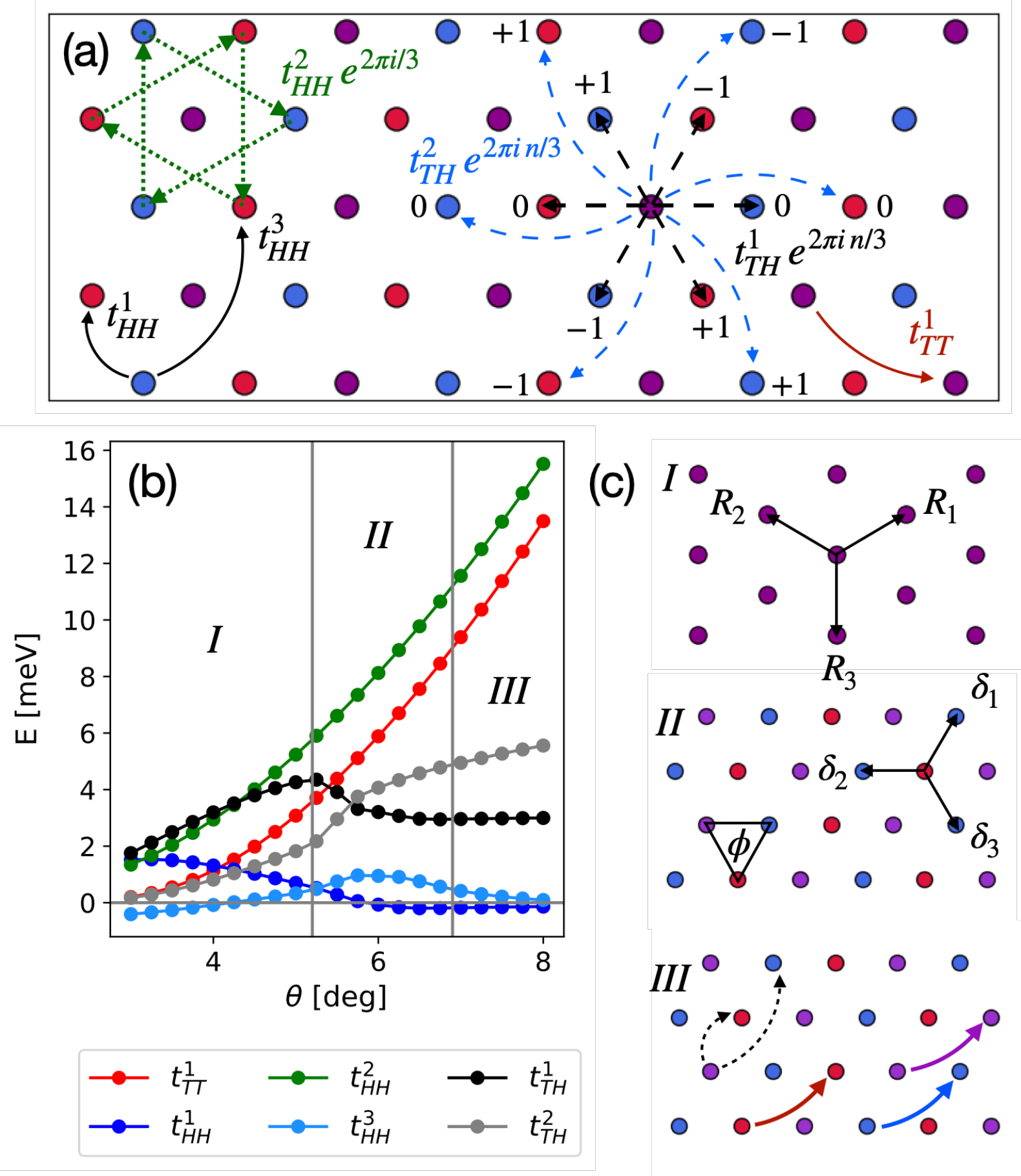}
    \caption{\textbf{(a)} Schematic of the effective tight-binding model with the triangular lattice of $T$ sites displayed in purple and the honeycomb lattice of $H_1$ and $H_2$ sites shown in blue and red, respectively. 
    Solid lines correspond to real hopping elements while dashed lines indicate complex hoppings. 
    The phase $e^{\frac{2\pi ni}{3}}$ is determined by the integer next to each site. 
    \textbf{(b)} Behavior of the hoppings as a function of twist angle. 
    The regimes $I$, $II$, and $III$ correspond to those indicated in Fig. \ref{fig:helicalgeometry}(c). 
    \textbf{(c)} Effective minimal model for the top band in each regime. 
    $I$ corresponds to a triangular lattice, $II$ is a triangular lattice with a smaller unit cell and complex hoppings, such that the flux enclosed in each plaquette is $\phi=2\pi/3$, and $III$ is the same triangular lattice as in $II$, but the hoppings between orbitals of the same kind (solid, colored lines) dominate and the hoppings $t^j_{TH}$ (dashed, black lines) are taken as a perturbation.}
    \label{fig:Hoppings}
\end{figure}

Using the Wannier states, Eq.~\eqref{eq:Wannier_orbitals}, we can construct a tight-binding effective description of the trilayer by computing the hopping elements, which are given by
\begin{align}
    t^{i,j}_{\alpha,\beta}=\braket{W^{\alpha}({\bm R_i})|H_K}{W^{\beta}({\bm R_j})},
    \label{eq:tb_hoppings}
\end{align}
where ${\bm R}_i$ and ${\bm R}_j$ label the unit cell positions of the orbitals and $\alpha, \beta$ are orbital indices. 
Using the translational symmetry, we can define every hopping with respect to the unit cell at the origin, i.e. $t^{i,j}_{\alpha,\beta} \equiv t^{\lvert i - j \rvert,0}_{\alpha\beta}$. 
We also use the fact that the two hexagonal orbitals $H_1$ and $H_2$ are equivalent in terms of their hoppings to simplify the notation of the hopping parameters, which we denote as $t^j_{HH}$, $t^j_{TT}$, and $t^j_{TH}$, where $j$ indicates the $j^\text{th}$ nearest neighbor, as depicted in Fig.~\ref{fig:Hoppings}(a). 
The evolution of the main hopping parameters as a function of twist angle is shown in Fig.~\ref{fig:Hoppings}(b). 

The effective tight-binding Hamiltonian is written as
\begin{align}
    H({\bm k})=\sum_{\alpha,\beta}c^{\dagger}_{\alpha,{\bm k}}\,h^{\alpha\beta}_{\bm k}\,c_{\beta,{\bm k}}~,
\end{align}
where $c^{\dagger}_{\alpha,{\bm k}}$ creates a hole with momentum ${\bm k}$ at orbital $\alpha$. 
The hoppings determine the matrix $h^{\alpha,\beta}_{\bm k}$. 
The diagonal elements are given by
\begin{align}
    h^{H_1H_1/H_2H_2}_{\bm k}=&~t^{2}_{\text{HH}}\sum_{j=1}^3e^{\pm2\pi i/3}\left( e^{i{\bm k}\cdot {\bm R}_j}+e^{-i{\bm k}\cdot {\bm R}_j}\right),\label{eq:tb_diags_HH}\\
    h^{TT}_{\bm k}=&~(t^0_{\text{TT}}-t^0_{\text{HH}})+t^{1}_{\text{TT}}\sum_{j=1}^32\cos({\bm k}\cdot {\bm R}_j), \label{eq:tb_diags_TT}
\end{align}
while the off-diagonal elements are given by 
\begin{align}
    h_{\bm k}^{H_1H_2}=&~t^{1}_{\text{HH}}\sum_{j=1}^3e^{i{\bm k}\cdot {\bm \delta}_j}+t^{3}_{\text{HH}}\sum_{j=1}^3e^{-2i{\bm k}\cdot {\bm \delta}_j},\label{eq:tb_offdiags_HH}\\
    h_{\bm k}^{H_1T/H_2T}=&~t^{1}_{\text{TH}}\sum_{j=1}^3e^{2\pi(j-2)i/3} e^{\pm i{\bm k}\cdot {\bm \delta}_j}\nonumber\\
    &+t^{2}_{\text{TH}}\sum_{j=1}^3e^{2\pi(j-2)i/3}e^{\mp 2i{\bm k}\cdot {\bm \delta}_j}.
    \label{eq:tb_offdiags_HT}
\end{align}
The vectors ${\bm R}_j$ and ${\bm \delta}_j$ are indicated in Fig. \ref{fig:Hoppings}(c). 
In Eqs.~\eqref{eq:tb_diags_HH}--\eqref{eq:tb_offdiags_HT} above, we use a cut-off of $\lvert \br \rvert \leq 2\lvert \boldsymbol{\delta}_j\rvert$ for the hoppings. 
In Appendix~\ref{sect:wannier_vs_ctm}, we show a comparison between the band structures obtained from the continuum model and the corresponding band structures obtained from our lattice description with this cut-off, as well as those resulting from increasing the cut-off to $\lvert \br \rvert \leq 2 a_M$, where $a_M$ is the moir\'e period. 
As the twist angle $\theta$ is increased, long-range hoppings become non-negligible, and a larger cut-off in hopping distance is required to accurately reproduce the energetics of the continuum model. 
This is expected as the Wannier functions become more extended for larger values of the twist angle, as we show in Appendix~\ref{sect:wannier_localisation}.

\subsection{Behavior in Different Twist Angle Regimes}
The tight-binding model that we derived reproduces the Chern sequences for the three topmost bands of Eq. \eqref{eq:Hamiltonian_1} and allows us to explain the three regimes displayed in Fig. \ref{fig:helicalgeometry} in terms of simplified lattice descriptions. 
\subsubsection{Large angle limit (Regime $III$)}
From the behavior of the hoppings, we see that the dominant terms in the regime $\theta\ge 6.9^{\circ}$ (region $III$ in Fig. \ref{fig:Hoppings}) are $t_{TT}^1$ and $t_{HH}^2$, which correspond to first-neighbor hoppings between the three triangular sublattices formed by $T$, $H_1$, and $H_2$ sites. 
If we only kept those terms, the model would correspond to three decoupled triangular lattices, although with slightly different onsite energies. 
The reason for these terms to be dominant in the large angle limit is that hoppings between the honeycomb and triangular sites ($t_{TH}^j$) correspond to tunneling processes from the top to the bottom layer, which are not directly coupled in the continuum Hamiltonian, and hence are second-order tunnelings. 
Therefore, the terms $t_{TH}^1$ and $t_{TH}^2$ (which are nonzero but much smaller in magnitude) can be taken as perturbative; they will induce couplings between the three triangular sublattices (comprising orbitals centered at the $T,~H_1,~H_2$ sites) and open gaps at the high-symmetry points in the mBZ, giving rise to topological bands. 
In particular $t_{HH}^1$ opens gaps at $\Gamma$ and $M$, while $t_{TH}^1$ opens gaps at $M$. 

\subsubsection{Small angle limit (Regime $I$)} 
For $\theta \le 5.2^{\circ}$ (region $I$ in Fig. \ref{fig:Hoppings}), the triangular (purple) and honeycomb  (red and blue) orbitals are separated by a large energy gap, such that the main hopping terms are $t^{i}_{TT}$ and $t^{i}_{HH}$. 
All triangular-triangular hoppings $t^{i}_{TT}$ are real, corresponding to a topologically trivial band in a triangular lattice formed by the orbitals at the $T$ sites. 
The hexagonal-hexagonal hoppings $t^{1,3}_{HH}$ are real, while the $t^{2}_{HH}$ hoppings are complex with phases that are multiples of $2\pi/3$, as indicated in Fig. \ref{fig:Hoppings}(a). 
This corresponds to a Haldane model in the honeycomb lattice formed by orbitals located at the $H_1$ and $H_2$ sites, which explains the Chern numbers $C=\pm 1$ of the second and third bands. 
Therefore, in the small angle limit the model corresponds to a topologically trivial triangular model weakly coupled to a higher energy Haldane model. 
The physics of low-energy holes will thus be dictated by the triangular band, which is highest in energy. 
Furthermore, in this limit, the honeycomb sites can be integrated out to yield an effective triangular lattice model with a modified first-neighbor hopping $\tilde{t}_{TT}^1=t_{TT}^1+|t_{TH}^1|^2/(t^0_{\text{TT}}-t^0_{\text{HH}})$. 
At half-filling of the topmost band, we expect a Mott insulating state~\cite{devakul2021magic,Jiawei2021,NicolasPRL2022}. By tuning the displacement field and the twist angle, which together reduce the gap to the remote band, the model provides a platform for heavy Fermi liquid physics with topologically non-trivial interlayer hybridization~\cite{Guerci2023Kondo,GuerciTKI2024,Valderrama2024,xie2025topologicalkondosemimetalsemulated}.

\subsubsection{Intermediate angle regime (Regime $II$)}

In this regime the low-energy physics of holes is described by an isolated $C=-2$ Chern band (as depicted in Fig.~\ref{fig:helicalgeometry}(c)). 
We can understand the lattice description as a $2\pi/3$\textendash{Hofstadter model} on a triangular lattice formed by the $H_1$, $T$, and $H_2$ sites \cite{Crepel_PhysRevResearch.6.033127}, whose Chern sequence is $[-2,1,1]$. 
This can be confirmed by computing the flux enclosed by each triangular plaquette to be $\pm2\pi/3$, as indicated in Fig. \ref{fig:Hoppings}. 
The symmetry indicator analysis described in Section~\ref{sect3:CtmModPhaseDiag} also explains how the Chern number of the topmost band is determined. 
This intermediate angle regime corresponds to a possible physical realization of tight-binding models discussed previously in which fractional Chern insulators (FCIs)~\cite{PhysRevLett.106.236804,Sheng2011,FCI-PhysRevX.1.021014,PhysRevLett.106.236802,checkerboard,parameswaran_fractional_2013,bergholtz_topological_2013} in a $|C|=2$ band were numerically studied~\cite{Sheng_PhysRevB.86.201101,Liu-PhysRevLett.109.186805,Sterdyniak-PhysRevB.87.205137,Wu-PhysRevLett.110.106802}. 
We note that our continuum model does not yield a set of hoppings such that the resulting $C=-2$ band is flat, which potentially disfavors topologically-ordered phases at fractional fillings, but we still expect interesting correlated topological phases, such as Chern insulators at integer fillings and (topological) charge density waves at fractional fillings.

\section{Hartree-Fock Phase Diagram}
\label{sect5:hf_phasediag}

\begin{figure}
    \centering
    \includegraphics[width=0.8\linewidth]{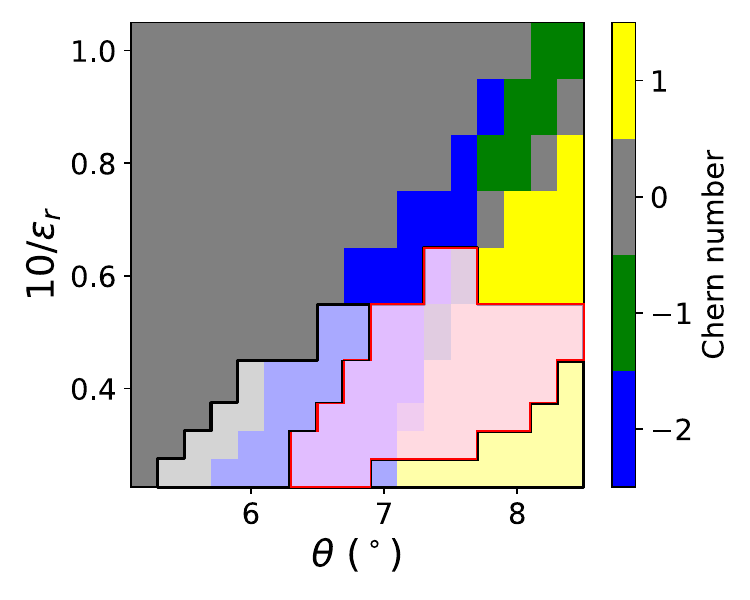}
    \caption{HF phase diagram in the $XMX$-stacked configuration at $\nu=-1$ as a function of interaction strength $10/\epsilon_r$ and twist angle $\theta$. The Chern number of the lowest-energy fully valley-polarized solution is depicted. White (pink) highlighted region with black (red) border indicates where the ground state has vanishing (partial) valley polarization. We use the continuum model parameters $(V, \psi, w, m^*) = (16.5\text{ meV}, -100^\circ, -18.8\text{ meV},0.60m_e)$. System size is $18\times 18$, and three valence bands are kept.}
    \label{fig:u_theta_HF_phase}
\end{figure}

To investigate the interacting physics of the top valence band in the $XMX$-stacked configuration, we perform self-consistent HF calculations at filling $\nu=-1$ on the continuum model. 
The goal is to understand how interactions affect the topology, effective band dispersion, and the stability of valley ferromagnetism within mean-field theory. 
We incorporate density-density interactions with a dual gate-screened potential 
\begin{equation}
    V(q)=\frac{e^2}{2\epsilon_0\epsilon_r q}\tanh(qd_{\text{sc}}), 
\end{equation}
where $d_{\text{sc}}=20$\,nm is the gate distance and $\epsilon_r$ is the relative dielectric constant. 
The physical value of $\epsilon_r$ is influenced by the intrinsic values~\cite{laturia2018dielectric} in non-twisted MoTe$_2$ ($\epsilon_r\simeq15$) and the encapsulating hBN substrate ($\epsilon_r\simeq5$), as well as details of screening from remote moir\'e bands. 
Note that previous theoretical studies found that a dielectric constant in the range $0.5<10/\epsilon_r<1.0$ reproduces the experimental phenomenology of the magnetism and fractional Chern insulators in twisted bilayer MoTe$_2$~\cite{yu2024nonmagnetic}. 
In the following, we treat $\epsilon_r$ as a tunable parameter to capture the uncertainty in the appropriate dielectric parameters for helical trilayer MoTe$_2$. 

We restrict to HF states that preserve moir\'e translation and valley-$U(1)$ conservation symmetry. This means that the one-body density matrix can be expressed as
\begin{equation}
    P_\sigma^{mn}(\bm{k})=\langle d^\dagger_{\bm{k},\sigma,m}d_{\bm{k},\sigma,n}\rangle,
\end{equation}
where $d^\dagger_{\bm{k},\sigma,m}$ is the electron creation operator for mBZ momentum $\bm{k}$, spin-valley sector $\sigma$, and continuum band $m$. We define the form factors $M^{nn'}_{\bm{k},\bm{q},\sigma} = \braket{u^n_{\bm{k}+\bm{q}, \sigma}}{u^{n'}_{\bm{k}, \sigma}}$, where $\ket{u^n_{\bm{k}, \sigma}}$ is the moir\'e-cell--periodic Bloch function. In terms of these, the Hartree and Fock potentials are
\begin{align}
\begin{split}
    [H^\text{H}(\bm{k})]_\sigma^{mn}=&\sum_{\bm{G}}\frac{V(\bm{G})}{\Omega_\text{tot}}M_{\bm{k},\bm{G},\sigma}^{mn}\\
    &\times \sum_{\bm{k}',m',n',\sigma'}\left[M^{m'n'}_{\bm{k}',\bm{G},\sigma'}\right]^* \delta P^{n'm'}_{\sigma'}(\bm{k}')
\end{split}
\end{align}
\begin{align}
\begin{split}
    [H^\text{F}(\bm{k})]_\sigma^{mn}=-&\sum_{\bm{q},\bm{G},m',n'}\frac{V(\bm{q}+\bm{G})}{\Omega_\text{tot}}M_{\bm{k},\bm{q}+\bm{G},\sigma}^{n'n}\\
    &\times \left[M_{\bm{k},\bm{q}+\bm{G},\sigma}^{m'm }\right]^*\delta P_{\sigma}^{n'm'}(\bm{k}+\bm{q}),\label{eq:Fock_Ham}
\end{split}
\end{align}
where $\Omega_\text{tot}$ is the total system area, $\bm{q}$ runs over mBZ momenta, and $\bm{G}$ runs over moir\'e reciprocal lattice vectors. The appearance of $\delta P_\sigma^{mn}(\bm{k})=P_\sigma^{mn}(\bm{k})-\delta^{mn}$ reflects the fact that interactions are measured relative to charge neutrality. The valley polarization is defined as $\bar{\sigma}\equiv\frac{1}{N_\text{M}}\sum_{n,\bm{k},\sigma}\sigma P_{\sigma}^{n,n} (\bk)$, where $N_\text{M}$ is the number of moir\'e unit cells in the calculation.

\begin{figure}
    \centering
    \includegraphics[width=0.8\linewidth]{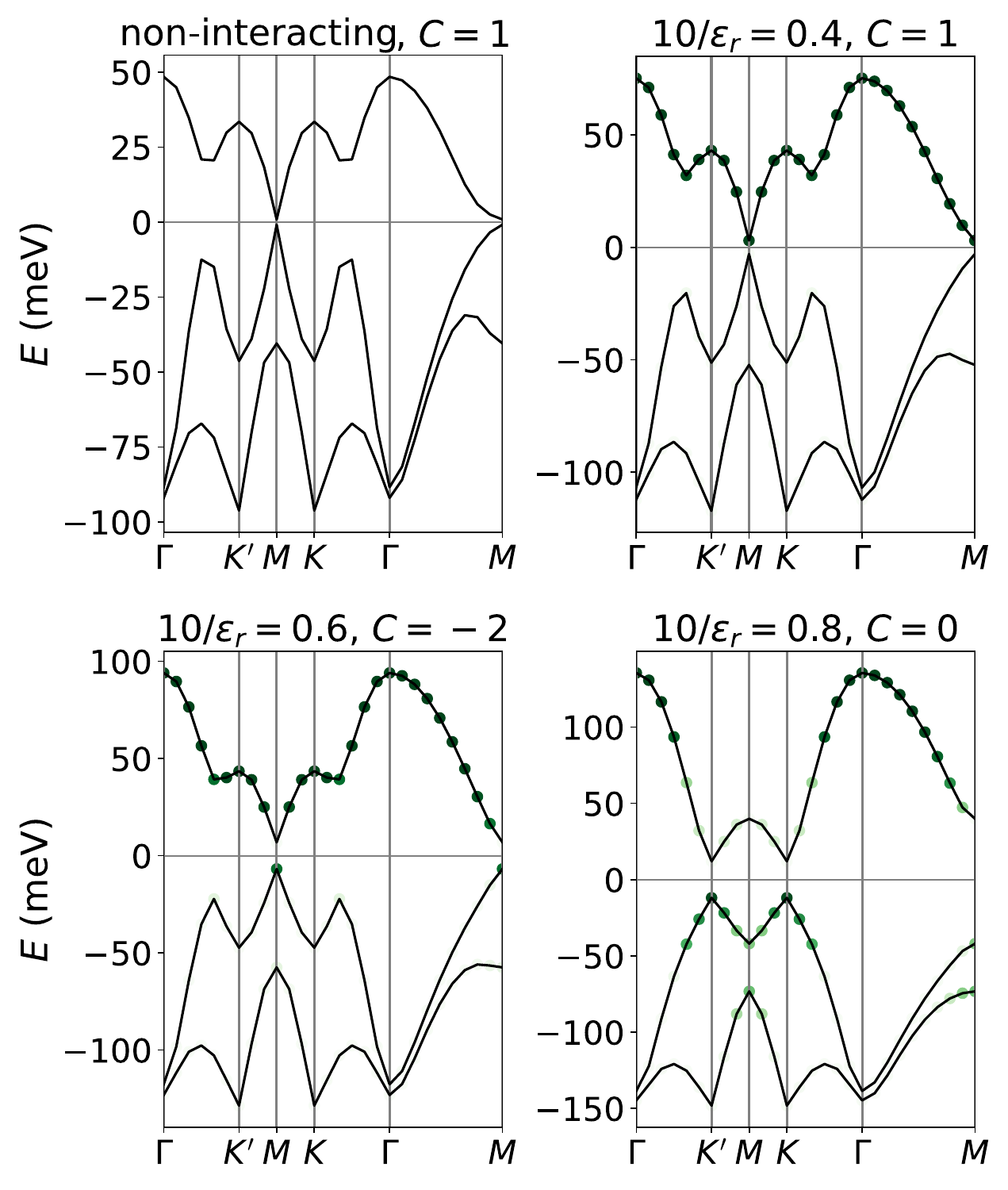}
    \caption{HF band structures of fully valley-polarized states in the $XMX$-stacked configuration at $\nu=-1$ and twist angle $\theta=7.4^\circ$. We only show the band structure in the hole-doped valley sector. For the interacting calculations, the intensity of the green dots indicates the overlap of the HF wave function with the highest non-interacting valence band. We use the continuum model parameters $(V, \psi, w, m^*) = (16.5\text{ meV}, -100^\circ, -18.8\text{ meV},0.60m_e)$. System size is $18\times 18$, and three valence bands are kept.}
    \label{fig:u_HF_transition}
\end{figure}

Fig.~\ref{fig:u_theta_HF_phase} shows the Chern number of the unoccupied valence band as a function of interaction strength and twist angle, assuming a fully valley-polarized state (i.e.~$|\bar{\sigma}|=1$). 
For the weakest interactions considered ($\epsilon_r=40$), the topology of the HF band closely follows the non-interacting case (Fig.~\ref{fig:helicalgeometry}). 
As the interaction strength is increased, the boundaries between the $C=0,-2,1$ phases all shift to higher $\theta$. 
Hence, it is possible to obtain a $C=-2$ HF ground state for twist angles where the top non-interacting valence band has $C=1$. 
The $C=-2$ region narrows for stronger interactions. 
Between the $C=-2$ and $C=1$ phases, we also find small regions of $C=0,-1$ states. 
We further indicate the regions where the fully polarized valley ferromagnet is undercut by states with only partial (pink shading) or vanishing valley polarization (white shading). 
As expected, full polarization is only sustained for sufficiently strong interactions $10/\epsilon_r\gtrsim 0.5$, which lies within the physically reasonable values of $\epsilon_r$ expected for twisted MoTe$_2$. We caution though that mean-field theory is known to overestimate the stability of valley-polarized phases.

\begin{figure}
    \centering
    \includegraphics[width=1\linewidth]{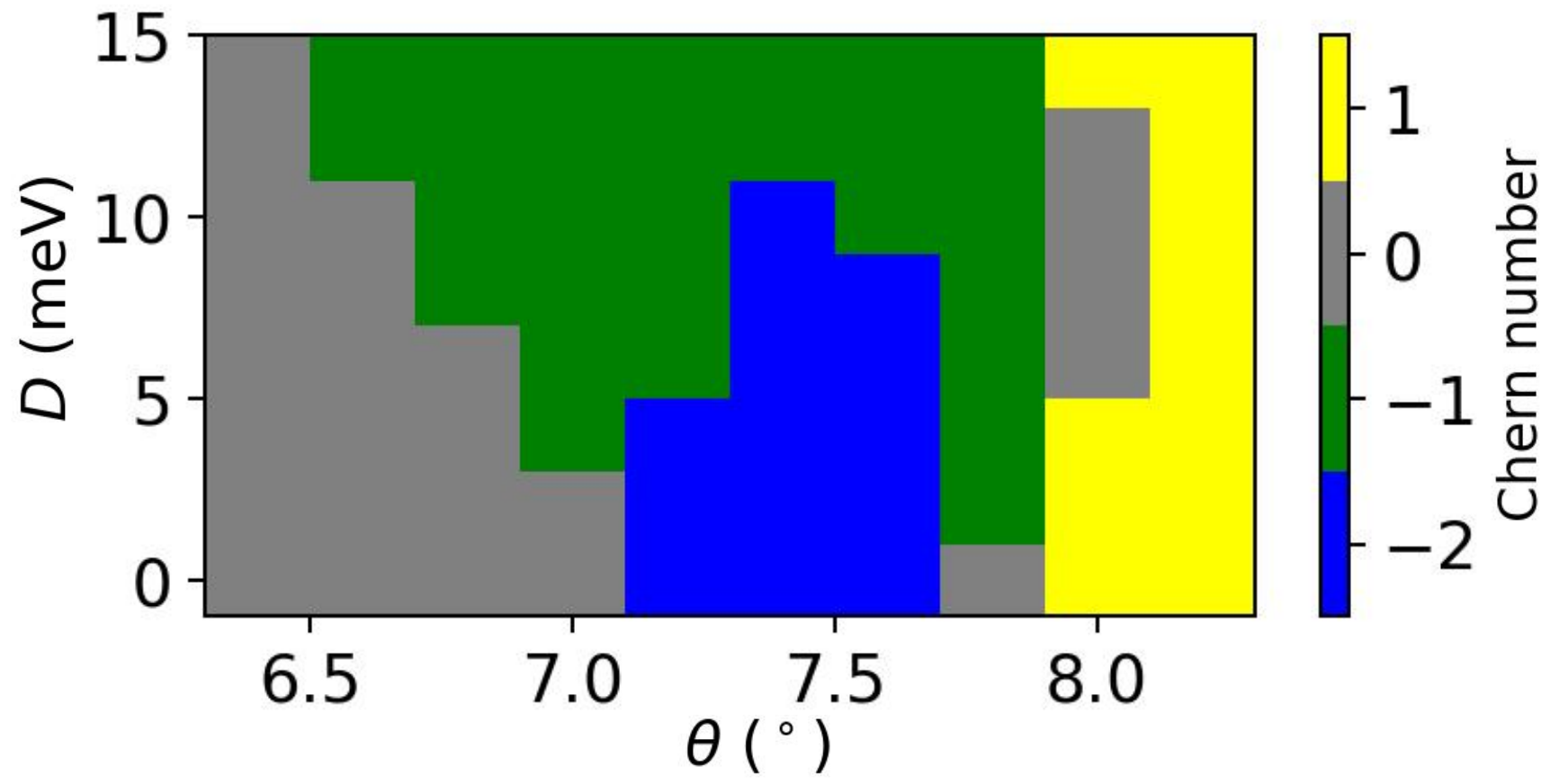}
    \caption{HF phase diagram of the $XMX$-stacked configuration at $\nu=-1$ and interaction strength $10/\epsilon_r=0.7$ as a function of displacement field $D$ and twist angle $\theta$. We use the continuum model parameters $(V, \psi, w,m^*) = (16.5\text{ meV}, -100^\circ, -18.8\text{ meV},0.60m_e)$. System size is $18\times 18$, and top three valence bands are kept.}
    \label{fig:D_theta_epsr14.29_HF_phase}
\end{figure}

To explain the phase boundaries of the dominant $C=0,-2,1$ phases, we consider the HF band structure in Fig.~\ref{fig:u_HF_transition} for $\theta=7.4^\circ$, where the top non-interacting band has $C=1$ with a small direct gap. 
We find that interactions consistently enhance the bandwidth of the HF valence band. 
This observation can be motivated by considering the properties of the single-particle Bloch functions. 
In a large region around the $\Gamma$ point, the Bloch functions of the top valence band are mostly localized on the middle layer and do not change appreciably with $\bm{k}$, as encoded by the Fubini-Study metric. 
Around the mBZ boundary where the single-particle gap to the adjacent band is reduced, the Bloch functions vary more rapidly with $\bm{k}$, especially near the $M$ points (see Appendix~\ref{app:QG_results}). As a result, for such $\bm{k}$, the magnitudes of the form factors $M_{\bm{k},\bm{q},\sigma}^{nn}$ are reduced even for small momentum transfer $\bm{q}$. This suppresses the Fock self-energy at these points (see Eq.~\ref{eq:Fock_Ham} and Ref.~\cite{abouelkomsan2023metric}), which relatively raises the HF energy at $\Gamma$ point upon hole-doping.
This enhances the existing kinetic dispersion.

The topological transition between the $C=1$ state  and the $C=-2$ state is mediated by gap closures at the $M$ points. 
In particular, interactions favor the Bloch state with pseudoinversion eigenvalue $\mathcal{I}=+1$  (see Table~\ref{tab:symm_eigvals}). 
Further increasing the interaction strength leads to band inversions at the $K,K'$ points, where the Bloch states with $\mathcal{C}_{3z}=1$ are favored (see Table~\ref{tab:symm_eigvals}). 
This results in the $C=0$ phase for the smallest values of $\epsilon_r$.

In Fig.~\ref{fig:D_theta_epsr14.29_HF_phase}, we also consider the HF phase diagram as a function of displacement field $D$ and $\theta$. 
For the interaction strength $10/\epsilon_r=0.7$, which lies within the range physically expected for MoTe$_2$,
the HF ground state is always fully spin-polarized. 
The structure of the HF phase diagram is similar to the non-interacting case in Fig.~\ref{fig:chernphases}. 
In particular, applying a displacement field to either the $C=0$ or $C=-2$ phases can yield a transition to a $C=-1$ phase.

\section{Discussion and Outlook}
\label{sect6:Discussion}
In this work, we have proposed helical homotrilayer TMDs as candidate platforms to explore topological physics beyond the quantum Hall paradigm, by identifying a range of twist angles and displacement fields where the topmost moiré miniband is isolated and has Chern number $|C|=2$. 
We have constructed both continuum and lattice descriptions, which are applicable at twist angles where there is a clear separation between the moiré and supermoiré length scales. 

We demonstrate that the Chern number of the topmost valence band is tunable by an external displacement field. 
Interestingly, for certain twist angles (i.e.~$\theta \lesssim 4.9^\circ$ using parameters appropriate for MoTe$_2$), application of a displacement field can drive a transition from a topologically trivial band to a Chern band. 
This is in stark contrast with the homobilayer case, where it is observed that the displacement field-tuned transition in the topmost band always results in $C=0$ as the displacement field layer-polarizes the system. 
In the limit of large $D$, the MoTe$_2$ homotrilayer will similarly become completely layer-polarized, resulting in a trivial topmost band. 
However, unlike the homobilayer, there is a wide range of $D$ (dependent on twist angle) for which $\lvert C \rvert = 1$ prior to this trivialization, indicating that the topological state of the topmost band is much more robust to the application of an external displacement field. 
In addition, for twist angles where the topmost band has $C=-2$ when $D=0$, the application of a nonzero $D > 0$ can drive a transition to a $C=-1$ band, demonstrating the gate-tunability between different Chern numbers.

Moreover, we have studied the interacting physics at integer filling $\nu=-1$, and showed that the $\lvert C \rvert = 2$ band can be stable to interactions at the Hartree-Fock level. 
We expect a rich variety of interacting physics to be realized in helical homotrilayer TMD systems. 
For instance, translation symmetry-breaking states may arise at partial fillings, such as topological charge density waves at $\nu=-1/2$~\cite{polshyn2022topological,kumar2014generalizing,wilhelm2023noncoplanar,dong2023decomposition}. Various magnetic orders beyond the ferromagnetic states considered in this work are also possible. 
Additionally, the tight-binding model that we derived allows for the use of beyond--mean-field methods, such as dynamical mean-field theory~\cite{zang2022dynamical,ryee2025sitepolarizedmottphasescompeting} and lattice quantum Monte Carlo, to investigate the correlated phases at various filling fractions. 

We conclude by commenting on the possible experimental realization of the physics predicted here. 
In bulk TMDs, the top of the valence band is at the $\Gamma$ point of the microscopic Brillouin zone~\cite{manzeli20172d}. 
Although this remains the case all the way down to the bilayer level in some systems, it is expected that for MoTe$_2$, the valence band edge lies at $K$ and $K'$ for up to three layers, which justifies our theoretical modeling. 
The technology to fabricate and probe twisted homobilayer TMDs has proven successful and should be extendable to the trilayer case as well. 
For the bilayer case, as moiré TMDs show more robust quantum anomalous Hall physics than twisted bilayer graphene, this trend suggests that it could be possible to observe interesting transport signatures in helical TMD homotrilayers in the absence of magnetic fields as well. 
Note that because a sample is expected to form stacking domains at the supermoir\'e scale~\cite{hoke2024imagingsupermoirerelaxationconductive}, transport measurements may observe an anomalous Hall effect which is not necessarily quantized~\cite{Xia2025_TrilayerGraphene}. 
Interestingly, these domains may give rise to a Chern mosaic pattern of anomalous Hall bulk insulators with diverse Chern numbers hosting chiral modes at their edges. 
A detailed analysis of structural relaxation and the interplay between correlated anomalous Hall insulators is left for future investigations. 
We expect scanning tunneling microscopy probes to be able to unveil the topology of the topmost band for these domains~\cite{Zhang2025_STM,Thompson2025_STM}. 

{\emph{Note added}:} During the completion of this manuscript, several experimental observations of insulating phases related to higher Chern numbers have been reported~\cite{dong2025observationintegerfractionalchern,wang2025moir,liu2025diverse} in various configurations of twisted rhombohedral multilayer graphene.

\acknowledgments

We acknowledge V. Cr\'epel for his help in the Wannierization of the continuum model bands. We are grateful to Ting Cao and Di Xiao for helpful discussions. D.G. acknowledges Y. Mao and C. Mora for discussions and past collaborations on related topics. N.R. and Y.K. acknowledge B.A. Bernevig, J. Herzog-Arbeitman and J. Yu for collaborations on related topics. The Flatiron Institute is a division of the Simons Foundation.

\bibliography{ref}

\clearpage

\appendix

\section{Derivation and Symmetries of the Continuum Model}
\label{sect:CtmModDetails}

\subsection{Definition of Lattice Geometries}

\begin{figure}
    \centering
    \includegraphics[width=0.98\linewidth]{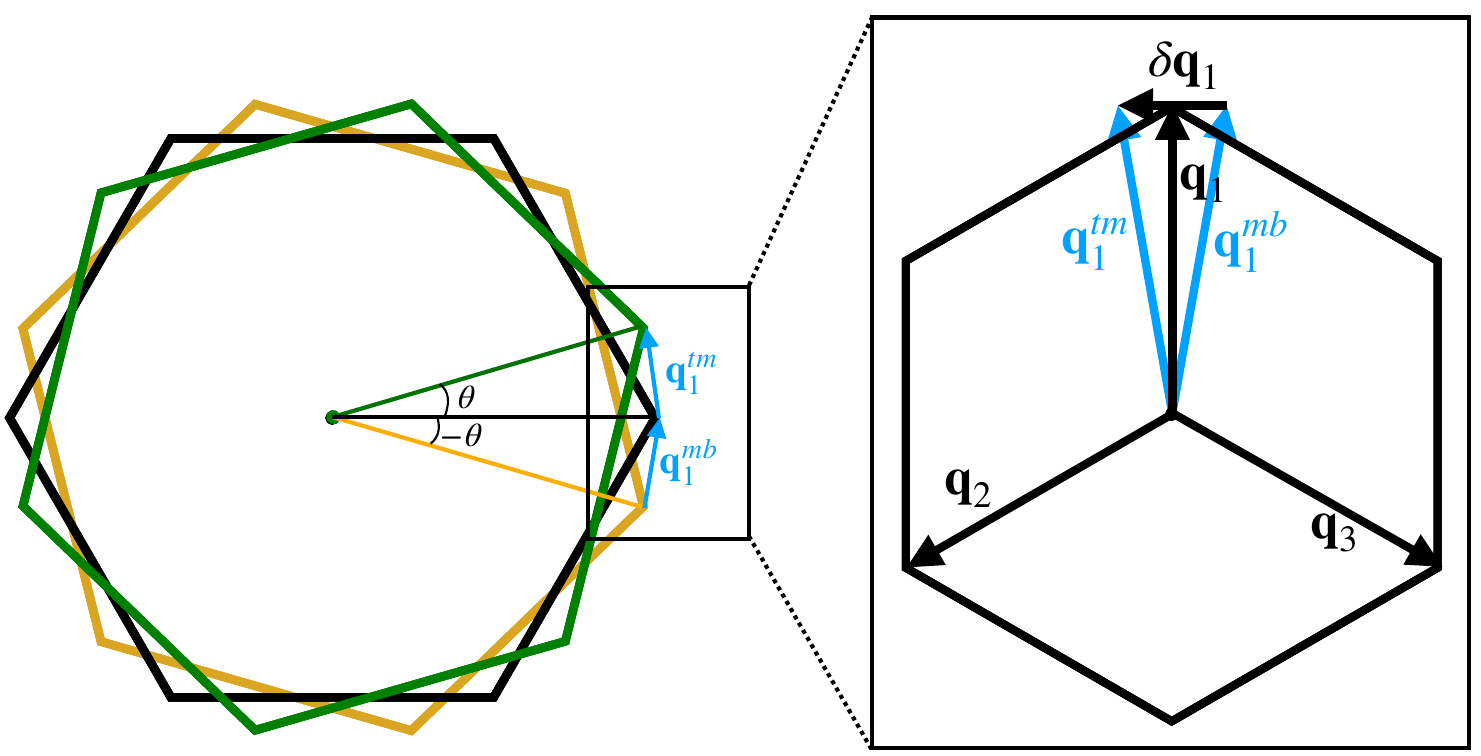}
    \caption{On the left, the Brillouin zones of the top (green), middle (black), and bottom (yellow) layers give rise to the momentum transfer vectors of each moir\'e pattern (sky blue), $\bq_1^{tm}$ and $\bq_1^{mb}$, which are superimposed on a diagram of the moir\'e Brillouin zone (mBZ) on the right. The mBZ diagram also depicts the center-of-mass moir\'e modulation $\bq_1$ and the small deviation $\delta\bq_1$. }
    \label{fig:appendix_mBZ}
\end{figure}

To fix notation, we list real and reciprocal lattice vectors for the moir\'e and moir\'e-of-moir\'e lattices, where we express lengths in units of the monolayer TMD lattice constant $a_0=1$ (or in units of $1/a_0$ for the momenta). 
For a given twist angle $\theta$, we define $k_\theta = (8\pi /3) \sin{(\theta/2)}$, where $k_\theta$ is the magnitude of each of the moir\'e momentum modulation vectors $\bq_j$, depicted in Fig.~\ref{fig:appendix_mBZ}.  
Consider a TMD monolayer with real space lattice vectors $\mathbf{a}_1 = (-1/2, \sqrt{3}/2), \mathbf{a}_2 = (-1, 0)$, and reciprocal lattice vectors $\mathbf{b}_1 = 2\pi(0, 2/\sqrt{3})$, $\mathbf{b}_2 = 2\pi (-1, -1/\sqrt{3})$. 
The position of the middle layer $K$ valley in momentum space is $\mathbf{K}_m = -\frac{1}{3} (\mathbf{b}_1 + 2 \mathbf{b}_2) = \frac{4\pi}{3} (1,0)$. 

As in the main text, we denote the layers as $t, m, b$, and use integer indices to label vectors in reciprocal and real space. 
We consider helical stacking, where we have equal twist angles between adjacent layers: $\theta_{tm}=\theta_{mb}=\theta$ and the top and bottom layers are twisted in opposite directions with respect to the middle one. 
These relative twists give rise to momentum transfer vectors between each pair of adjacent layers. 
For valley $K$, these are 
\begin{equation}\label{periodicities}
    \mathbf{q}^{tm}_1 = R_\theta \mathbf{K}_m- \mathbf{K}_m,\quad \mathbf{q}^{mb}_1= \mathbf{K}_m - R_{-\theta} \mathbf{K}_m,
\end{equation}
where $R_\theta$ denotes a counterclockwise rotation by $\theta$. 
The two wavevectors in Eq.~\eqref{periodicities} can be re-expressed in terms of the center-of-mass moir\'e modulation ${\bf q}_1$ and the small deviation $\delta{\bf q}_1$.
As we have equal twist angles, we find that 
\begin{equation}
   \mathbf{q}_1=\frac{\mathbf{q}_1^{tm}+\mathbf{q}_1^{mb}}{2},\quad \delta \mathbf{q}_1=\mathbf{q}_1^{tm}-\mathbf{q}_{1}^{mb},
\end{equation}
where in the regime of small twist angles we have 
\begin{equation}
\begin{aligned}
    \mathbf{q}_1 &=2\sin{(\theta/2)} \, (\mathbf{\hat{z}}\times \mathbf{K}_m),\\
    \delta \mathbf{q}_1 &=(2\sin{(\theta/2)})^2 \, \mathbf{\hat{z}}\times( \mathbf{\hat{z}}\times \mathbf{K}_m),
    \label{eq:modulationdecomp_appendix}
\end{aligned}
\end{equation}
with $\mathbf{q}_j=R_{2\pi (j-1)/3} \mathbf{q}_1$ and $\delta \mathbf{q}_j=R_{2\pi (j-1)/3}\delta \mathbf{q}_1$. 

Given the monolayer geometry conventions established above, the explicit values of the moir\'e momentum transfer vectors are $\mathbf{q}_1 =  k_\theta (0, 1)$, $\mathbf{q}_2 = R_{\frac{2\pi}{3}}\mathbf{q}_1 = k_\theta (-\sqrt{3}/2, -1/2)$, and $\mathbf{q}_3 = R_{\frac{2\pi}{3}}\mathbf{q}_2  = k_\theta (\sqrt{3}/2, -1/2)$. 
The moir\'e reciprocal lattice vectors are $\mathbf{G}_1 = \mathbf{q}_1 - \mathbf{q}_2, \mathbf{G}_2 = \mathbf{q}_2 - \mathbf{q}_3, \mathbf{G}_3 = \mathbf{q}_3 - \mathbf{q}_1$, i.e. $\bG_j = R_{2\pi (j-1)/3} \bG_1$.
We denote the moir\'e-scale real space lattice vectors $\bL_1, \bL_2$ such that 
\begin{equation}
    \bL_i \cdot \bG_j = 2\pi \delta_{ij}. 
    \label{eq:appendix_moirelattvec}
\end{equation}
The magnitude of each moir\'e lattice vector is the moir\'e lattice constant $a_M = 1/(2 \, \sin(\theta/2))$. 
A diagram of how the moir\'e Brillouin zone (mBZ) emerges from the helical stacking of the monolayer reciprocal lattices is provided in Fig.~\ref{fig:appendix_mBZ}. 

Finally, the moir\'e-of-moir\'e momentum transfer vectors are $\delta\bq_1 = \frac{3k^2_\theta}{4\pi}(-1,0)$, $\delta\bq_2 = R_{\frac{2\pi}{3}} \delta\bq_1=\frac{3k^2_\theta}{4\pi} (1/2, -\sqrt{3}/2)$ and $\delta\bq_3 = R_{\frac{2\pi}{3}} \delta\bq_2 =\frac{3k^2_\theta}{4\pi}(1/2, \sqrt{3}/2)$.  
The moir\'e and moir\'e-of-moir\'e lattices are both triangular, but their principal axes are rotated by $90^\circ$ with respect to each other. 
The moir\'e-of-moir\'e real and reciprocal lattice vectors are denoted $\delta \bL_i$ and $\delta \bG_i$, respectively. The supermoir\'e reciprocal lattice vectors are defined as 
\begin{equation}
    \begin{aligned}
        \delta \mathbf{G}_1 &=\delta \mathbf{q}_1-\delta \mathbf{q}_2\\
        \delta \mathbf{G}_2 &=\delta \mathbf{q}_2-\delta \mathbf{q}_3\\
        \delta \bG_3 &= \delta \bq_3 - \delta\bq_1, 
    \end{aligned}
\end{equation}
and the corresponding real space lattice vectors are defined such that 
\begin{equation}
    \delta\bL_i \cdot \delta \bG_j = 2\pi \delta_{ij}. 
\end{equation}

\subsection{Continuum Model}
\label{sect:ctm_mod_prms}
We now derive the helical twisted trilayer Hamiltonian following Ref.~\cite{yuncheng_2023}. 
We begin by constructing the full Hamiltonian for the helical trilayer system, following the continuum model for the TMD homobilayer in Ref.~\cite{wu2019topological}. 
Assuming that only adjacent layers are coupled, the Hamiltonian for valley $K$ at position $\brcurs$ reads as 
\begin{equation}
    \label{eq:appendix_trilayer_Hamiltonian}
    H_K(\brcurs) = -\frac{\hbar^2 \bk^2}{2m^*} \mathbb{1} + \begin{pmatrix}
        V_t(\brcurs) & t_{tm}(\brcurs) & 0\\
        t_{tm}^*(\brcurs) & V_m(\brcurs) & t_{mb}(\brcurs)\\
        0 & t_{mb}^*(\brcurs) & V_b(\brcurs)
    \end{pmatrix}. 
\end{equation}
The most general form for the intralayer moir\'e potential upon taking the first harmonic approximation with respect to the moir\'e-scale wave vectors $\bq^{\ell \ell'}_j$ (see Fig.~\ref{fig:appendix_mBZ}) is given by 
\begin{equation}
    \begin{aligned}
        V_t(\brcurs) = 2 \sum\limits_{j=1}^3 &V_j^t \cos{(\bG_j^{tm} \cdot \brcurs + \psi^t_j)}\\
        V_m (\brcurs) = 2\sum\limits_{j=1}^3 &\bigg[ V_j^m \cos{(\bG_j^{mb} \cdot \brcurs + \psi_j^{m})} \\
        &+ V_j^{m'} \cos{(\bG^{mb}_j \cdot \brcurs + \psi_j^{m'})} \bigg]\\
        V_b(\brcurs) = 2\sum\limits_{j=1}^3 &V_j^b \cos{(\bG^{mb}_j \cdot \brcurs + \psi_j^{b})}, 
    \end{aligned}
\end{equation}
where $\bG_j^{\ell \ell'} = \bq_j^{\ell \ell'} - \bq_{j+1}^{\ell \ell'}$ for any pair of adjacent layers $\ell, \ell'$, with the index defined modulo 3, i.e., $\bq_{j+1} \equiv \bq_{(j+1)\text{ mod 3}}$.

We now use the symmetries $\mathcal{C}_{3z}$ and $\mathcal{C}_{2y}$ to derive constraints on the parameters $V_j^\ell$ and $\psi_j^{\ell}$. 
$\mathcal{C}_{3z}$ is the three-fold rotational symmetry about the $\mathbf{\hat{z}}$-axis, which constrains the intralayer moir\'e potential of each layer as 
\begin{equation}
    V_\ell(\mathcal{C}_{3z} \brcurs) = V_\ell (\brcurs).
    \label{eq:appendix_c3z_constraint}
\end{equation}

Fig.~\ref{fig:appendix_mBZ} shows that the $\mathcal{C}_{3z}$ symmetry acts on the moir\'e wave vectors as $\mathcal{C}_{3z} \bq_j^{\ell \ell'} = \bq_{j+1}^{\ell \ell'}$, which implies that $\mathcal{C}_{3z} \bG_j^{\ell \ell'} = \bG_{j+1}^{\ell \ell'}$. 
As a result, we find that 
\begin{equation}
    \begin{aligned}
        \left( \mathcal{C}_{3z} \bG_j^{\ell \ell'} \right) \cdot \mathcal{C}_{3z} \brcurs &= \bG_j^{\ell \ell'} \cdot \brcurs\\
        \bG^{\ell \ell'}_{j+1} \cdot \mathcal{C}_{3z} \brcurs &= \bG_j^{\ell \ell'} \cdot \brcurs\\
        \implies \bG_j^{\ell \ell'} \cdot \mathcal{C}_{3z} \brcurs &= \bG_{j-1}^{\ell \ell'} \cdot \brcurs. 
    \end{aligned}
\end{equation}
Using this and Eq.~\eqref{eq:appendix_c3z_constraint}, we then find that 
\begin{equation}
    \begin{aligned}
        V_{t/b}(\mathcal{C}_{3z} \brcurs) &= 2\sum\limits_{j=1}^3 V_j^{t/b} \cos{\left[ \bG_j^{tm/mb} \cdot (\mathcal{C}_{3z} \brcurs) + \psi_j^{t/b} \right]} \\
        &= 2 \sum\limits_{j=1}^3 V_j^{t/b} \cos{\left( \bG_{j-1}^{tm/mb} \cdot \brcurs + \psi_j^{t/b}\right)} \\
        V_{t/b}(\brcurs) &= 2 \sum\limits_{j=1}^3 V_j^{t/b} \cos{\left( \bG_j^{tm/mb} \cdot \brcurs + \psi_j^{t/b} \right)}. 
    \end{aligned}
\end{equation}
Similarly, for the middle layer, 
\begin{equation}
    \begin{aligned}
        V_m(\mathcal{C}_{3z} \brcurs) = 2 &\sum\limits_{j=1}^3 \bigg[ V_j^m \cos{(\bG_{j-1}^{tm} \cdot \brcurs + \psi_j^m)} \\
        &+ V_j^{m'} \cos{(\bG_{j-1}^{mb} \cdot \brcurs + \psi_j^{m'})} \bigg]\\
        V_m(\brcurs) = 2 &\sum\limits_{j=1}^3 \bigg[ V_j^m \cos{(\bG_{j}^{tm} \cdot \brcurs + \psi_j^m)} \\
        &+ V_j^{m'} \cos{(\bG_{j}^{mb} \cdot \brcurs + \psi_j^{m'})} \bigg]. 
    \end{aligned}
\end{equation}
As a result, invariance under the $\mathcal{C}_{3z}$ symmetry requires that each of the parameters $V^\ell_j$ and $\psi^\ell_j$ be equal for all values of $j$, and so we remove the $j$ dependence and denote the parameters by $V_0^\ell$ and $\psi_0^\ell$ , respectively. 
This yields a simplified form for the intralayer potentials, which may be written as
\begin{equation}
    \begin{aligned}
        V_t(\brcurs) = 2V^t_0 &\sum\limits_{j=1}^3 \cos{(\bG_j^{tm} \cdot \brcurs + \psi_0^t)}\\
        V_m(\brcurs) = 2V^m_0 &\sum\limits_{j=1}^3 \cos{(\bG_j^{tm} \cdot \brcurs + \psi_0^m)} \\
        &+ 2V_0^{m'} \sum\limits_{j=1}^3 \cos{(\bG^{mb}_j \cdot \brcurs + \psi_0^{m'})}\\
        V_b(\brcurs) = 2V^b_0 &\sum\limits_{j=1}^3 \cos{(\bG^{mb}_j \cdot \brcurs + \psi_0^{b})}. 
    \end{aligned}
\end{equation}

We now constrain the model parameters further via the $\mathcal{C}_{2y}$ symmetry, which relates the intralayer potentials according to 
\begin{equation}
    V_t(\mathcal{C}_{2y} \brcurs) = V_b(\brcurs), \quad V_m(\mathcal{C}_{2y}\brcurs) = V_m(\brcurs). 
    \label{eq:appendix_c2y_layer}
\end{equation}
Though the full symmetry of the trilayer system is actually $\mathcal{C}_{2y}\mathcal{T}$, the effect of the time-reversal symmetry $\mathcal{T}$ is to relate the two different valleys $K$ and $K'$ such that $V_{t,K}(\brcurs) = V^*_{b, K'}(\brcurs)$. 
As the $\mathcal{C}_{2y}$ symmetry also relates $K$ and $K'$, the $\mathcal{C}_{2y}\mathcal{T}$ symmetry of the trilayer system leaves the $K$ valley invariant. 
However, since these intralayer potential terms lie on the diagonals of the Hamiltonian, and so the Hermitian property yields that $V_\ell(\brcurs)$ must be entirely real, and so Eq.~\eqref{eq:appendix_c2y_layer} holds. 

Then, to derive the constraint from the $\mathcal{C}_{2y}$ symmetry, we observe that from the mBZ geometry depicted in Fig.~\ref{fig:appendix_mBZ}, $\mathcal{C}_{2y}$ acts on the moir\'e wave vectors as $\mathcal{C}_{2y}\bq_1^{tm} = \bq_1^{mb}$ and $\mathcal{C}_{2y} \bq_2^{tm} = \bq_3^{mb}$. 
This in turn establishes the action of the $\mathcal{C}_{2y}$ symmetry on the moir\'e reciprocal lattice vectors as $\mathcal{C}_{2y} \bG_1^{tm} = -\bG_3^{mb}$, $\mathcal{C}_{2y} \bG_2^{tm} = -\bG_2^{mb}$, and $\mathcal{C}_{2y} \bG_3^{tm} = -\bG_1^{mb}$. 

As a result, we find that 
\begin{equation}
    \begin{aligned}
        V_{t/b}(\mathcal{C}_{2y} \brcurs) &= 2V^{t/b}_0\sum\limits_{j=1}^3 \cos{(\bG_j^{tm/mb} \cdot \mathcal{C}_{2y} \brcurs + \psi_0^{t/b})}\\
        &= 2V_0^{t/b} \sum\limits_{j=1}^3 \cos{(-\bG_{4-j}^{mb/tm} \cdot \brcurs + \psi_0^{t/b})}\\
        &= 2V_0^{t/b} \sum\limits_{j=1}^3 \cos{(\bG_{4-j}^{mb/tm} \cdot \brcurs - \psi_0^{t/b})}\\
        V_{b/t}(\brcurs) &= 2V_0^{b/t} \sum\limits_{j=1}^3 \cos{(\bG_j^{mb/tm} \cdot \brcurs + \psi_0^{b/t})}, 
    \end{aligned}
\end{equation}
which yields constraints $V_0^t = V_0^b$ and $\psi_0^t = -\psi_0^b$. 

Similarly, for the middle layer, 
\begin{equation}
    \begin{aligned}
        V_m(\mathcal{C}_{2y} \brcurs) =\; &2V_0^m \sum\limits_{j=1}^3 \cos{(\bG_j^{tm} \cdot \mathcal{C}_{2y} \brcurs + \psi_0^{m})}\\
        &+ 2V_0^{m'} \sum\limits_{j=1}^3 \cos{(\bG_j^{mb} \cdot \mathcal{C}_{2y}\brcurs + \psi_0^{m'})}\\
        = \; &2V_0^m \sum\limits_{j=1}^3 \cos{(-\bG_{4-j}^{mb} \cdot \brcurs + \psi_0^m)}\\
        &+ 2V_0^{m'} \sum\limits_{j=1}^3 \cos{(-\bG_{4-j}^{tm} \cdot \brcurs + \psi_0^{m'})}\\
        = \; &2V_0^m \sum\limits_{j=1}^3 \cos{(\bG_{4-j}^{mb} \cdot \brcurs - \psi_0^m)}\\
        &+ 2V_0^{m'} \sum\limits_{j=1}^3 \cos{(\bG_{4-j}^{tm} \cdot \brcurs - \psi_0^{m'})}\\
        V_m(\brcurs) = \; &2V_0^m \sum\limits_{j=1}^3 \cos{(\bG_j^{tm} \cdot \brcurs + \psi_0^{m})}\\
        &+ 2V_0^{m'} \sum\limits_{j=1}^3 \cos{(\bG_j^{mb} \cdot \brcurs + \psi_0^{m'})}, 
    \end{aligned}
\end{equation}
which yields constraints $V_0^m = V_0^{m'}$ and $\psi_0^{m'} = -\psi_0^m$. 

Thus, our model has now been reduced to 
\begin{equation}
    \begin{aligned}
        V_t(\brcurs) = 2V_o \sum\limits_{j=1}^3 \; &\cos{(\bG_j^{tm} \cdot \brcurs + \psi_o)}\\
        V_m(\brcurs) = 2V_m \sum\limits_{j=1}^3 \; &\bigg[ \cos{(\bG_j^{tm} \cdot \brcurs - \psi_m)}\\
        &+ \cos{(\bG_j^{mb} \cdot \brcurs + \psi_m)} \bigg] \\
        V_b(\brcurs) = 2V_o \sum\limits_{j=1}^3 \; &\cos{(\bG_j^{mb} \cdot \brcurs - \psi_o)}, 
    \end{aligned}
\end{equation}
where we use $V_o, \psi_o$ to denote the parameters of the outer layers. 
By analogy with the bilayer model, we will only focus on the case that $V_o = V_m = V$ and $\psi_o = \psi_m = \psi$, and so the final form for the intralayer moir\'e potentials of our trilayer tTMD Hamiltonian, Eq.~\eqref{eq:appendix_trilayer_Hamiltonian} is 
\begin{equation}
    \begin{aligned}
        V_t(\brcurs) = 2V \sum\limits_{j=1}^3 \; &\cos{(\bG_j^{tm} \cdot \brcurs + \psi)}\\
        V_m(\brcurs) = 2V \sum\limits_{j=1}^3 \; &\bigg[ \cos{(\bG_j^{tm} \cdot \brcurs - \psi)}\\
        &+ \cos{(\bG_j^{mb} \cdot \brcurs + \psi)} \bigg] \\
        V_b(\brcurs) = 2V \sum\limits_{j=1}^3 \; &\cos{(\bG_j^{mb} \cdot \brcurs - \psi)}. 
    \end{aligned}
    \label{eq:appendix_trilayer_intralayer}
\end{equation}

Because our Hamiltonian only couples adjacent layers, the interlayer tunneling terms are equivalent to the bilayer case, which is given in Ref.~\cite{wu2019topological}, 
\begin{equation}
    t_{\ell\ell'} (\br) = w \sum\limits_{j=1}^3 e^{-i\bq_j^{\ell\ell'} \cdot \brcurs}. 
    \label{appeq:trilayer_interlayer}
\end{equation}

In Appendix~\ref{sect:addl_ctm_prms}, we provide a brief study of the non-interacting phase diagram when $V_o \neq V_m$ and $\psi_o \neq \psi_m$. 

\subsection{Adiabatic Approximation}
\label{sect:adiabatic_approximation}

\begin{figure}
    \centering
    \includegraphics[width=0.96\linewidth]{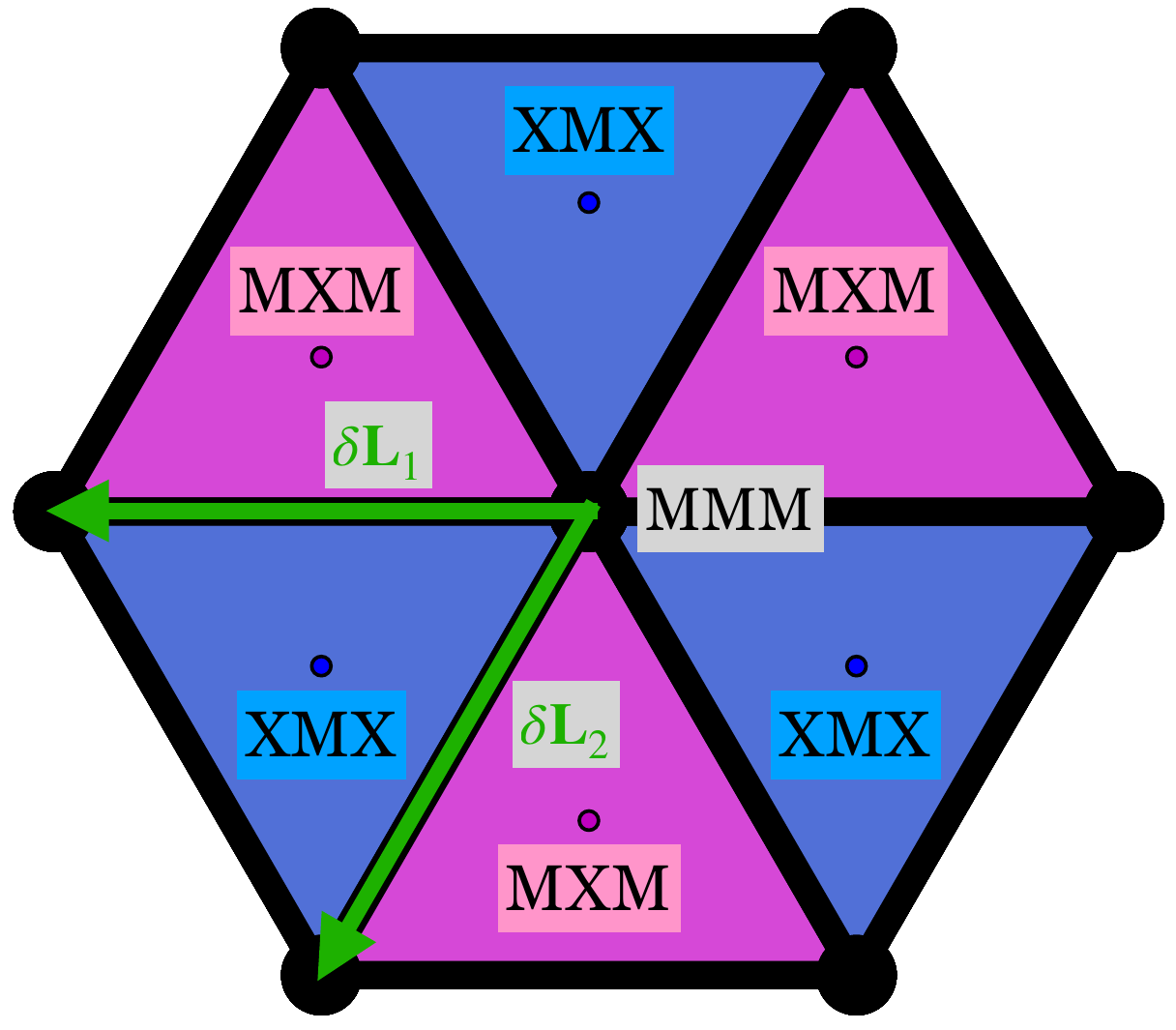}
    \caption{Moir\'e-scale domains within a supermoir\'e cell in real space formed by lattice relaxation effects that expand the energetically favorable areas of $MXM$ and $XMX$ stacking. 
    The high-symmetry points occur at the $MMM$ stacking sites and the centers of the $MXM$- and $XMX$-stacked domains. 
    The slow coordinate $\bR$ is measured from the origin (at $MMM$ stacking) to a high-symmetry point, while $\br$ is measured from said high-symmetry point.}
    \label{fig:appendix_supermoire}
\end{figure}

Fig.~\ref{fig:appendix_supermoire} depicts the moir\'e-scale domains of $MXM$ and $XMX$ stacking that form on the supermoir\'e scale due to lattice relaxation effects. 
These domains are created as the $MXM/XMX$ stacking configurations are energetically favorable relative to the $MMM$ stacking~\cite{nakatsuji2025moirebandengineeringtwisted}. 

In order to construct a local Hamiltonian describing a moir\'e-scale domain, we use Eq.~\eqref{eq:modulationdecomp_appendix} to re-express the terms of Eq.~\eqref{eq:appendix_trilayer_Hamiltonian} given in Eqs.~\eqref{eq:appendix_trilayer_intralayer} and~\eqref{appeq:trilayer_interlayer}. 
The intralayer potential terms become 
\begin{equation}
\begin{aligned}
    V_t(\brcurs) = 2V&\sum\limits_{j=1}^3 \cos{[(\bG_j + \delta\bG_j/2)\cdot \brcurs + \psi]}\\
    V_m(\brcurs) = 2V &\sum\limits_{j=1}^3 \bigg[ \cos{[(\bG_j + \delta\bG_j/2)\cdot \brcurs - \psi]} \\
    &+ \cos{[(\bG_j - \delta\bG_j/2)\cdot \brcurs + \psi]}\bigg] \\
    V_b(\brcurs) = 2V &\sum\limits_{j=1}^3 \cos{[(\bG_j - \delta\bG_j/2)\cdot \brcurs - \psi]}, 
\end{aligned}
\label{appeq:intralayer_trilayer_decomp}
\end{equation}
while the interlayer tunneling terms read 
\begin{equation}
    t_{tm/mb}(\brcurs)= w \sum_{j=1}^{3}e^{-i(\mathbf{q}^{}_{j}\pm\delta \mathbf{q}_j/2)\cdot \brcurs}. 
    \label{eq:tunnelling_appendix}
\end{equation}

Due to the clear separation of the moir\'e and supermoir\'e scales at small twist angles, $|\delta{\bf q}_1|/|{\bf q}_1|\sim \theta\ll 1$, we treat the slow periodicity $\delta \mathbf{q}_1$ adiabatically. 
This is done by introducing the slow variable $\mathbf{R}$, which we define to be a point on the moir\'e lattice, i.e. 
\begin{equation}
    \bR = n_1 \bL_1 + n_2 \bL_2, \quad n_1, n_2 \in \mathbb{Z}, 
    \label{eq:app_slow_coord_lincomb}
\end{equation}
and which changes with a characteristic rate of the supermoir\'e scale. 
Then, we decompose our position coordinate $\brcurs$ into $\bR$ that varies on the supermoir\'e scale and $\br$ that varies on the moir\'e scale: $\brcurs = \bR + \br$. 

We then select a value of the slow coordinate $\bR$ that is at or near the center of one of these moir\'e-scale domains in order to construct the local Hamiltonian of that domain. 
The center of each domain is a local high-symmetry point with respect to the $\mathcal{C}_{3z}$ symmetry, as are the points of $MMM$ stacking at the vertices of the domains. 
So we define the slow coordinate $\bR$ to go from the origin to one of these high-symmetry points, and the local position coordinate $\br$ has its origin at the high-symmetry point. 

Because we wish to construct a local Hamiltonian that is dependent on only $\br$ explicitly, we write the interlayer tunneling terms as 
\begin{equation}
    \begin{aligned}
        t_{tm/mb}(\brcurs) &= w \sum_{j=1}^{3} e^{-i(\mathbf{q}_{j}\pm\delta \mathbf{q}_j/2)\cdot (\bR + \mathbf{r})}\\
        &= w \sum\limits_{j=1}^3 e^{-i \bq_j \cdot \bR} e^{-i\bq_j \cdot \br} e^{\mp i \delta\bq_j \cdot \bR/2} e^{\mp i \delta\bq_j \cdot \br/2}. 
    \end{aligned}
    \label{eq:appendix_interlayer_deriv}
\end{equation}
We find that we may neglect the last term in the product, as $\delta \bq_j \cdot \br/2$ is of order $\theta \ll 1$ (in the small angle limit). 

We explicitly evaluate $\bq_j \cdot \bR$ using Eq.~\eqref{eq:app_slow_coord_lincomb} to find 
\begin{equation}
    \begin{aligned}
        \bq_1 \cdot \bR &= \bq_1 \cdot (n_1 \bL_1 + n_2 \bL_2) = \frac{4\pi n_1}{3} + \frac{2\pi n_2}{3}\\
        \bq_2 \cdot \bR &= \bq_2 \cdot (n_1 \bL_1 + n_2 \bL_2) = -\frac{2\pi n_1}{3} + \frac{2\pi n_2}{3}\\
        \bq_3 \cdot \bR &= \bq_3 \cdot (n_1 \bL_1 + n_2 \bL_2) = -\frac{2\pi n_1}{3} - \frac{4\pi n_2}{3}. 
    \end{aligned}
    \label{appeq:qj_dot_R}
\end{equation}
As a result, we can see that the term $e^{-i\bq_j \cdot \bR}$ is independent of $j$, and will thus contribute a global phase to the interlayer tunneling that may thus be gauged out. 
Then, in order to ensure that the Hamiltonian only has a parametric dependence on the slow coordinate $\bR$, we introduce the phases 
\begin{equation}
    \phi_j=\delta \bq_j\cdot \bR/2,  
    \label{eq:phiphase_appendix}
\end{equation}
that vary on the supermoir\'e scale. 
We collect these phase parameters into a vector $\boldsymbol{\phi} = [\phi_1, \phi_2, \phi_3]$, and thus find that the expression for the interlayer tunneling parameters defined in Eq.~\eqref{eq:tunnelling_appendix} becomes moir\'e-periodic and depend on the slow coordinate scale only parametrically as
\begin{equation}
    t_{tm/mb}(\br) = t_{\pm \boldsymbol{\phi}}(\br) = w \sum\limits_{j=1}^3 e^{-i\bq \cdot \br} e^{\mp i \phi_j}, 
    \label{appeq:interlayer_tunnelling}
\end{equation}
where 
\begin{equation}
    t_{\boldsymbol{\phi}}^*(-\br) = t_{-\boldsymbol{\phi}}(\br).     \label{appeq:interlayer_minusr_identity}
\end{equation}

Using $\bG_j = \bq_j - \bq_{j+1}$, we also define 
\begin{equation}
\label{eq:appendix_delta_phase}
    \Delta\phi_j=\phi_j-\phi_{j+1}=(\delta \bq_j - \delta \bq_{j+1}) \cdot \bR/2 =\delta \bG_j\cdot \bR/2.
\end{equation}
We may then rewrite the intralayer potential expressions, Eq.~\eqref{appeq:intralayer_trilayer_decomp}, in a similar fashion by noting that 
\begin{equation}
    \begin{aligned}
       (\bG_j &\pm \delta \bG_j/2) \cdot \brcurs = (\bG_j \pm \delta \bG_j/2) \cdot (\bR + \br)\\
       &= \bG_j \cdot \bR + \bG_j \cdot \br \pm \delta \bG_j \cdot \bR/2 \pm \delta \bG_j \cdot \br/2\\
       &\simeq \bG_j \cdot \bR + \bG_j \cdot \br \pm \Delta \phi_j. 
    \end{aligned}
    \label{appeq:intralayer_adiabatic_expansion}
\end{equation}
Then, noting that $\bG_j \cdot \bR = (\bq_j - \bq_{j+1}) \cdot (n_1 \bL_1 + n_2 \bL_2)$ and using Eq.~\eqref{appeq:qj_dot_R}, we find that 
\begin{equation}
    \begin{aligned}
        \bG_1 \cdot \bR &= \bq_1 \cdot \bR - \bq_2 \cdot \bR = 2\pi n_1\\
        \bG_2 \cdot \bR &= \bq_2 \cdot \bR - \bq_3 \cdot \bR = 2\pi n_2\\
        \bG_3 \cdot \bR &= \bq_3 \cdot \bR - \bq_1 \cdot \bR = -2\pi (n_1 + n_2). 
    \end{aligned}
\end{equation}
As the cosine functions in Eq.~\eqref{appeq:intralayer_trilayer_decomp} are periodic under shifts of $2\pi$, we find that $\bG_j \cdot \bR = \bG_j \cdot (n_1 \bL_1 + n_2 \bL_2)$ merely contributes some integral multiple of $2\pi$ to the arguments of the cosines in Eq.~\eqref{appeq:intralayer_trilayer_decomp}, as we defined $n_1, n_2 \in \mathbb{Z}$. 
And so, the intralayer moir\'e potentials become
\begin{equation}
    \begin{split}
        V_t(\br,\boldsymbol{\phi}) = 2V &\sum_{j=1}^{3}\cos(\bG_j\cdot \br+\psi+\Delta \phi_j),\\
        V_m(\br,\boldsymbol{\phi}) = 2V &\sum_{j=1}^{3}\bigg[\cos(\bG_j\cdot \br-\psi+\Delta\phi_j)\\
        &+\cos(\bG_j\cdot \br+\psi-\Delta\phi_j)\bigg], \\
        V_b(\br,\boldsymbol{\phi}) = 2V &\sum_{j=1}^{3}\cos(\bG_j\cdot \br-\psi-\Delta \phi_j). 
    \end{split}
    \label{appeq:intralayerpotentials}
\end{equation}

We also note that for a given point in the supermoir\'e unit cell, $\mathbf{R}$ can be expressed in terms of the supermoir\'e lattice vectors $\delta \mathbf{L}_i$ using Eqs.~\eqref{eq:appendix_moirelattvec} and~\eqref{eq:appendix_delta_phase} as 
\begin{equation}
    \mathbf{R} = \sum_{j=1,2}\frac{\Delta\phi_j}{\pi}\delta \mathbf{L}_j. 
    \label{eq:slowcoord_appendix}
\end{equation}
We may also explicitly compute 
\begin{equation}
    \begin{aligned}
        \delta \bL_1 &= - \frac{4\pi}{3k_\theta \sqrt{3}} \bL_1 + \frac{8\pi}{3k_\theta \sqrt{3}} \bL_2\\
        \delta \bL_2 &= - \frac{8\pi}{3k_\theta \sqrt{3}} \bL_1 + \frac{4\pi}{3k_\theta \sqrt{3}} \bL_2
    \end{aligned}
\end{equation}
to write $\bR$ in terms of the moir\'e lattice vectors as 
\begin{equation}
    \bR = \frac{4}{3k_\theta \sqrt{3}} \left[ -(\Delta \phi_1 +2 \Delta\phi_2) \, \bL_1 + (2 \Delta \phi_1 + \Delta \phi_2) \, \bL_2  \right]. 
\end{equation}

Thus, we have derived a local continuum Hamiltonian that is parametrically dependent on $\boldsymbol{\phi}$, namely 
\begin{equation}
\label{eq:localHamiltonian_appendix}
    H_K(\br, \boldsymbol{\phi}) =-\frac{\hbar^2 \bk^2}{2m^*}\mathbb 1 + \begin{pmatrix}
    V_t(\br,\boldsymbol{\phi}) &  t_{\boldsymbol{\phi}}(\br)  & 0 \\ 
    t_{\boldsymbol{\phi}}^*(\br) &V_m(\br,\boldsymbol{\phi}) &  t_{-{\boldsymbol{\phi}}}(\br)  \\ 
    0 & t^*_{-\boldsymbol{\phi}}(\br) &V_b(\br,\boldsymbol{\phi})
    \end{pmatrix}. 
\end{equation}
Because $\boldsymbol{\phi}$ varies slowly on the moir\'e length scale, we will fix its values to describe the local stacking configuration for different moir\'e domains within the moir\'e-of-moir\'e supercell. 

\section{Symmetry Properties}

\subsection{High-Symmetry Stacking Configurations}
\label{sect:stacking_phases}
The local Hamiltonian $H_K(\boldsymbol{\phi})$ generally breaks all symmetries except at fine-tuned values of $\boldsymbol{\phi}$ corresponding to high-symmetry stacking configurations. 
If we set all $\phi_j = 0$ (implying $\Delta\phi_j = 0$ via Eq.~\eqref{eq:appendix_delta_phase}), we see that the Hamiltonian comprises two copies of the homobilayer Hamiltonian, one each for the $tm$ and $mb$ bilayers, with identical hoppings between each adjacent layer. 
Namely, there is no real space shift between those two moir\'e structures, and so the $tm$ and $mb$ moir\'e patterns are aligned with each other. 
Therefore, setting $\boldsymbol{\phi} = 0$ corresponds to the high-symmetry configuration $MMM$. 
(We will denote this specific value of $\boldsymbol{\phi}$ by $\boldsymbol{\phi}^{MMM}$.)  

We now wish to find the specific values of the phases for the high-symmetry $MXM/XMX$ stacking configurations, which we similarly denote $\boldsymbol{\phi}^{MXM}$ and $\boldsymbol{\phi}^{XMX}$, respectively. 
The geometry of the supermoir\'e lattice given in Fig.~\ref{fig:appendix_supermoire} allows us to write down $\bR$ vectors for the $MXM$ and $XMX$ stacking configurations. 
Without loss of generality, we select $\bR$ vectors that will yield $\phi_1 = 0$ for ease of computation. 
We are free to do this due to a gauge freedom that allows for a global shift of all phases $\phi_j$~\cite{yuncheng_2023,PhysRevB.109.205411}. 
We thus find that 
\begin{equation}
    \begin{aligned}
        \bR^{MXM} &= 2 \, (\delta \bL_1 - 2\delta \bL_2)/3\\
        &= \frac{2\sqrt{3}}{3} \left( \frac{4\pi}{3k_\theta} \right)^2 \begin{pmatrix}
            0\\
            1
        \end{pmatrix} \\
        \bR^{XMX} &= -2 \, (\delta \bL_1 - 2 \delta \bL_2)/3\\
        &= -\frac{2\sqrt{3}}{3} \left( \frac{4\pi}{3k_\theta} \right)^2 \begin{pmatrix}
            0\\
            1
        \end{pmatrix}. 
    \end{aligned}
    \label{appeq:highsym_R}
\end{equation}

We then compute the phases $\boldsymbol{\phi}^{MXM}$ and $\boldsymbol{\phi}^{XMX}$ for these high-symmetry stacking configurations using Eqs.~\eqref{appeq:highsym_R} and~\eqref{eq:phiphase_appendix} as being
\begin{equation}
    \begin{aligned}
        \delta \bq_1 \cdot \bR^{MXM/XMX}/2 &= 0\\
        \delta \bq_2 \cdot \bR^{MXM/XMX}/2 &= \mp 2\pi/3\\
        \delta \bq_3 \cdot \bR^{MXM/XMX}/2 &= \pm 2\pi/3. 
    \end{aligned}
\end{equation}
As a result, we have explicitly found that 
\begin{equation}
    \boldsymbol{\phi}^{MXM/XMX} = \pm \left[ 0, -\frac{2\pi}{3}, \frac{2\pi}{3} \right]. 
    \label{appeq:high-sym_phases}
\end{equation}

We note that the vectors $\bR^{MXM/XMX}$ given in Eq.~\eqref{appeq:highsym_R} are not generally points on the moir\'e lattice, as required in Eq.~\eqref{eq:app_slow_coord_lincomb}. 
However, we may see that we are still justified in using the phases in Eq.~\eqref{appeq:high-sym_phases} by the following argument. 
For sake of simplicity, we focus on $\bR^{XMX}$. 
First, we use $\bR^{\mathbb{Z}}$ to denote the moir\'e lattice site nearest to $\bR^{XMX}$ and define the off-set $\mathbf{\varepsilon}$ between them via $\bR^{\mathbb{Z}} = \bR^{XMX} + \mathbf{\varepsilon}$. 
This off-set $\mathbf{\varepsilon}$ is therefore a vector on the moir\'e scale. 
Then, as discussed in Appendix~\ref{sect:adiabatic_approximation}, we select the moir\'e lattice site $\bR^{\mathbb{Z}}$ to be the origin of the position vector $\br$ in our local Hamiltonian. 
We then use Eq.~\eqref{appeq:intralayer_adiabatic_expansion} to find 
\begin{equation}
    (\bG_j \pm \delta \bG_j) \cdot \brcurs \simeq \bG_j \cdot \bR^{\mathbb{Z}} + \bG_j \cdot \br \pm \delta \bG_j \cdot \bR^{\mathbb{Z}}/2. 
\end{equation}
We have already shown that the first term $\bG_j \cdot \bR^{\mathbb{Z}}$ merely contributes an integer multiple of $2\pi$ to the argument of the cosine functions in the intralayer potential, and so we drop it moving forward. 
We then continue to find that
\begin{equation}
    \begin{aligned}
        (\bG_j \pm \delta \bG_j) \cdot \brcurs &= \bG_j \cdot \br + \delta \bG_j \cdot (\bR^{XMX} + \mathbf{\varepsilon})/2\\
        &= \bG_j \cdot \br + \Delta \phi_j^{XMX} + \delta \bG_j \cdot \mathbf{\varepsilon}/2, 
    \end{aligned}
\end{equation}
where the second equality is due to the definition of $\Delta \phi_j$ given in Eq.~\eqref{eq:appendix_delta_phase}. 

Now, examining the final term $\delta \bG_j \cdot \mathbf{\varepsilon}/2$, we find that this correction to $\Delta \phi_j^{XMX}$ is of order $\theta \ll 1$, and so we may neglect it entirely. 
Therefore, we have shown that the values of the phases for the high-symmetry stacking configurations given in Eq.~\eqref{appeq:high-sym_phases} are sufficient to describe the local physics of the moir\'e-scale domains, even though the origin from which our local position vector $\br$ is measured is the nearest moir\'e lattice site to the center of the domain, rather than the geometric center of the domain itself. 

\subsection{Symmetries}
\label{sect:localHam_symm}
In this subsection, we discuss the symmetries of the local Hamiltonian $H_K(\boldsymbol{\phi})$ derived above. 
Specifically, we discuss the invariance of the local continuum model for the three moir\'e-scale domains under the three-fold rotational symmetry $\mathcal{C}_{3z}$, $\mathcal{C}_{2y} \mathcal{T}$, and the three-dimensional pseudoinversion symmetry $\mathcal{I}$. 

\subsubsection{MMM stacking}
\label{sect:app_MMM_symm}

The $MMM$ configuration is $\mathcal{C}_{3z}$-symmetric because $V_\ell(\mathcal{C}_{3z} \br,\boldsymbol{\phi}=0)=V_{\ell}(\br,\boldsymbol{\phi}=0)$ and $t_{\boldsymbol{\phi}=0}(\mathcal{C}_{3z}\br)=t_{\boldsymbol{\phi}=0}(\br)$. 
As a result, $H_{\boldsymbol{\phi}=0}(\mathcal{C}_{3z} \br)=H_{\boldsymbol{\phi}=0}(\br)$. 
Furthermore, $V_t(\mathcal{C}_{2y}\br, \boldsymbol{\phi} = 0)=V_{b}(\br, \boldsymbol{\phi} = 0)$ and $V_m(\mathcal{C}_{2y}\br, \boldsymbol{\phi} = 0)=V_m(\br, \boldsymbol{\phi} = 0)$, as $\mathcal{C}_{2y}\bG_2=-\bG_2$ and $\mathcal{C}_{2y}\bG_1=-\bG_3$. 
Note that $\mathcal{C}_{2y}$ swaps the top and bottom layers while preserving the middle layer.
$\mathcal{C}_{2y}$ also sends valley $K$ to $K'$ while acting on the principal moir\'e wavevectors $\bq_j$ as $\mathcal{C}_{2y}\bq_1=\bq_1$ and $\mathcal{C}_{2y}\bq_2=\bq_3$. 
Before proceeding, we introduce the notation 
\begin{equation}\label{standard_tunneling}
    t({\bf r})=t_{\boldsymbol{\phi}=0}(\br)=w\sum_{j=1}^{3} e^{-i{\bf q}_j\cdot\bf r}.
\end{equation}
Thus, we have that
\begin{equation}\begin{split}
    D(\mathcal{C}_{2y})&H^{MMM}_K(\br)D^\dagger(\mathcal{C}_{2y})\\
    &=-\frac{\hbar^2 \bk^2}{2m^*}\mathbb 1 + \begin{pmatrix}
        V_b(\br) &  t^*(\br)  & 0 \\ 
        t(\br) &V_m(\br) &  t^*(\br)  \\ 
        0 & t(\br) &V_t(\br)
    \end{pmatrix} \\
    &= H^{MMM}_{K'}(\mathcal{C}_{2y}\br),
\end{split}\end{equation}
where we omit the dependency on the phase which is set $\boldsymbol{\phi}=0$ everywhere. 
Notice that in the last equation we used $V_{b}(\mathcal{C}_{2y}\br)=V_{t}(\br)$ and $V_{m}(\mathcal{C}_{2y}\br)=V_{m}(\br)$.
Also, the representation of the $\mathcal{C}_{2y}$ symmetry takes the form 
\begin{equation}\label{c2y}
    D(\mathcal{C}_{2y})=i\gamma^x\sigma^y=i\begin{pmatrix}
        0 & 0 & 1 \\ 
        0 & 1 & 0 \\ 
        1 & 0 &0
    \end{pmatrix}\sigma^y,
\end{equation}
with $\sigma^y$ acting in the spin-valley degree of freedom. 
In addition, the time-reversal symmetry reads $\mathcal T=i\sigma^y \mathcal K$, where $\mathcal K$ is complex conjugation. 
Thus, we have 
\begin{equation}
    \mathcal T D(\mathcal{C}_{2y}) H^{MMM}_K(\br) D^\dagger (\mathcal{C}_{2y})\mathcal T^\dagger=H^{MMM}_K(\mathcal{C}_{2y}\br).
\end{equation}

Similarly to twisted homobilayer there is an additional three-dimensional pseudoinversion symmetry with representation $D(\mathcal{I})=\gamma^x$, as defined in Eq.~\eqref{eq:pseudoinv_oper} of the main text, 
\begin{equation}
    D(\mathcal{I}) = \gamma^x = \begin{pmatrix}
        0 & 0 & 1\\
        0 & 1 & 0\\
        1 & 0 & 0
    \end{pmatrix}, 
    \label{eq:appendix_pseudoinv}
\end{equation}
which acts as 
\begin{equation}\begin{split}
    \gamma^xH^{MMM}_K(\br)\gamma^x&=-\frac{\hbar^2 \bk^2}{2m^*}\mathbb 1 + \begin{pmatrix}
 V_b(\br) &  t^*(\br)  & 0 \\ 
        t(\br) &V_m(\br) &  t^*(\br)  \\ 
        0 & t(\br) &V_t(\br)
    \end{pmatrix}\\
    &=H^{MMM}_K(-\br).
\end{split}\end{equation}
This latter symmetry is broken if we introduce higher-order wavevector modulations. 

\subsubsection{XMX stacking}

In Appendix~\ref{sect:stacking_phases}, we found that $\boldsymbol{\phi}^{XMX} = [0, 2\pi/3,-2\pi/3]$ and $\boldsymbol{\Delta \phi}^{XMX} = [-2\pi/3, 4\pi/3, -2\pi/3]$. 
With the aim of simplifying the notation we introduce the tunneling function 
\begin{equation}
\begin{aligned}
  t_{tm}^{MXM}(\br) =  t_{\boldsymbol{\phi}^{MXM}}(\br) &\equiv t_{\omega}(\br)=w\sum_{j=1}^{3}e^{-i\bq_j\cdot \br}\omega^{j-1}\\
  \implies t_{\boldsymbol{\phi}^{XMX}} (\br) &= t_{\omega^*} (\br),  
\end{aligned}
\end{equation}
where $\omega = e^{2\pi i/3}$. 

$\mathcal{C}_{3z}$ acts on the tunnelings $t_{\omega}(\br)$ and $t_{\omega^*}(\br)$ as 
\begin{equation}
\begin{split}
    &t_{\omega}(\mathcal{C}_{3z}\br)= \omega t_{\omega}(\br),\\
    &t_{\omega^*}(\mathcal{C}_{3z}\br)= \omega^* t_{\omega^*}(\br).
\end{split}
\end{equation}
We also note that $t_{\omega}(\br)=t_{\omega^*}(\mathcal{C}_{2y}\br)$.

For the intralayer potentials we have
\begin{equation}
    \begin{split}
         V_{t}(\br,\boldsymbol{\phi}^{XMX}) \equiv V_{-}(\br) = 2V&\sum_{j=1}^{3}\cos(\bG_j\cdot \br+\psi-2\pi/3),\\
        V_{m}(\br,\boldsymbol{\phi}^{XMX}) \equiv V_0(\br) = 2V&\sum_{j=1}^{3}\bigg[\cos(\bG_j\cdot \br-\psi-2\pi/3)\\
        &+\cos(\bG_j\cdot \br+\psi+2\pi/3)\bigg],\\
         V_b(\br,\boldsymbol{\phi}^{XMX}) \equiv V_{+}(\br) = 2V&\sum_{j=1}^{3}\cos(\bG_j\cdot \br - \psi + 2\pi/3), 
    \end{split} 
\end{equation}
where we used $\cos{(\bG_2 \cdot \br + \psi \pm 4\pi/3)} = \cos{(\bG_2 \cdot \br + \psi \mp 2\pi/3)}$ for the $j=2$ term. 
Using this notation, the Hamiltonian for the $XMX$ stacking is 
\begin{equation}
    H^{XMX}_K(\br) = - \frac{\hbar^2 \bk^2}{2m^*} \mathbb{1} + \begin{pmatrix}
        V_-(\br) & t_{\omega^*}(\br) & 0\\
        t^*_{\omega^*} (\br) & V_0 (\br) & t_\omega(\br) \\
        0 & t^*_{\omega}(\br) & V_+(\br)
    \end{pmatrix}. 
    \label{eq:appendix_xmxHamiltonian}
\end{equation}
Under a threefold rotation, we find $V_\ell(\mathcal{C}_{3z} \br)=V_\ell(\br)$. 
Thus, we obtain 
\begin{equation}
    \begin{split}
    H^{XMX}_K(\mathcal{C}_{3z}\br)&=-\frac{\hbar^2 \bk^2}{2m^*}\mathbb 1+ \begin{pmatrix}
    V_-(\br) &  \omega^* t_{\omega^*}(\br)  & 0 \\ 
        \omega t^*_{\omega^*}(\br) &V_0(\br) & \omega t_{\omega}(\br)  \\ 
        0 & \omega^* t^*_{\omega}(\br) &V_+(\br)
    \end{pmatrix} \\
    &= D(\mathcal{C}_{3z}) H^{XMX}_K(\br) D^\dagger (\mathcal{C}_{3z}),
    \end{split}
\end{equation}
where, as in Eq.~\eqref{eq:c3xmx} of the main text, the representation of the $\mathcal{C}_{3z}$ symmetry for the $XMX$ stacking is 
\begin{equation}
    D^{XMX}(\mathcal{C}_{3z})=\begin{pmatrix}
        \omega^* & 0 & 0 \\ 
        0 & 1 & 0 \\ 
        0 & 0 & \omega^*
    \end{pmatrix}.\label{eq:c3xmx_appendix}
\end{equation}

The approximate three-dimensional inversion $\mathcal{I}$ is a symmetry for the model since $\gamma^x H^{XMX}_K(\br)\gamma^x=H^{XMX}_K(-\br)$, just as with the $MMM$ stacking case. 
Furthermore, the symmetry $\mathcal{C}_{2y}$ (Eq.~\eqref{c2y}) acts on the Hamiltonian as 
\begin{equation}
    \begin{split}
    D&(\mathcal{C}_{2y})H^{XMX}_{K}(\br)D^\dagger (\mathcal{C}_{2y})\\
    &= -\frac{\hbar^2 \bk^2}{2m^*}\mathbb 1
    + \begin{pmatrix}
    V_-(\mathcal{C}_{2y}\br) &  t^*_{\omega^*}(\mathcal{C}_{2y}\br)  & 0 \\ 
        t_{\omega^*}(\mathcal{C}_{2y}\br) & V_0(\mathcal{C}_{2y}\br)  &  t^*_{\omega}(\mathcal{C}_{2y}\br)  \\ 
        0 & t_{\omega}(\mathcal{C}_{2y}\br) & V_+(\mathcal{C}_{2y}\br) 
    \end{pmatrix}\\
    &=H^{*XMX}_{K'}(\mathcal{C}_{2y}\br).
    \end{split}
\end{equation}
As a result, we also have 
\begin{equation}
    \mathcal T D(\mathcal{C}_{2y}) H^{XMX}_K(\br) D^\dagger (\mathcal{C}_{2y})\mathcal T^\dagger = H^{XMX}_{K}(\mathcal{C}_{2y}\br).
\end{equation}

\subsubsection{MXM stacking}

The $MXM$ stacking configuration is characterized by $\boldsymbol{\phi}^{MXM}=-\boldsymbol{\phi}^{XMX}$ and $\boldsymbol{\Delta\phi}^{MXM}=-\boldsymbol{\Delta\phi}^{XMX}$. 
Thus, the Hamiltonian for this stacking reads 
\begin{equation}
     H^{MXM}_K(\br)=-\frac{\hbar^2\bk^2}{2m^*}\mathbb 1+ \begin{pmatrix}
    \tilde V_-(\br) &  t_{\omega}(\br)  & 0 \\ 
        t^*_{\omega}(\br) &\tilde V_0(\br) &  t_{\omega^*}(\br)  \\ 
        0 &  t^*_{\omega^*}(\br) &\tilde V_+(\br)
    \end{pmatrix},
\end{equation}
where
\begin{equation}
    \begin{split}
         \tilde V_{-}(\br) = 2V&\sum_{j=1}^{3}\cos(\bG_j\cdot \br+\psi+2\pi/3),\\
        \tilde V_0(\br) = 2V&\sum_{j=1}^{3}\bigg[\cos(\bG_j\cdot \br-\psi+2\pi/3)\\
        &+\cos(\bG_j\cdot \br +\psi-2\pi/3)\bigg],\\
         \tilde V_{+}(\br) = 2V&\sum_{j=1}^{3}\cos(\bG_j\cdot \br - \psi -  2\pi/3).
    \end{split} 
\end{equation}
In this case, the representation of the $\mathcal C_{3z}$ operator is different, namely, 
\begin{equation}
    D^{MXM}(\mathcal{C}_{3z})=\begin{pmatrix}
        \omega & 0 & 0 \\ 
        0 & 1 & 0 \\ 
        0 & 0 & \omega
    \end{pmatrix}, 
    \label{eq:c3mxm_appendix}
\end{equation}
where the opposite angular momentum compared to the $XMX$ case originates from the opposite phase in the interlayer tunneling. 

We notice that sending $\psi\to-\psi$ and $\br\to -\br$ allows us to obtain the $XMX$ stacking for the $K'$ valley via 
\begin{equation}\begin{split}    
     H^{MXM}_K\to&-\frac{\hbar^2\bk^2}{2m^*}\mathbb 1+ \begin{pmatrix}
    V_-(\br) &  t_{\omega}(-\br)  & 0 \\ 
        t^*_{\omega}(-\br) & V_0(\br) &  t_{\omega^*}(-\br)  \\ 
        0 &  t^*_{\omega^*}(-\br) & V_+(\br)
    \end{pmatrix}\\
    &= -\frac{\hbar^2\bk^2}{2m^*}\mathbb 1+ \begin{pmatrix}
    V_-(\br) &  t^*_{\omega^*}(\br)  & 0 \\ 
        t_{\omega^*}(\br) & V_0(\br) &  t^*_{\omega}(\br)  \\ 
        0 &  t_{\omega}(\br) & V_+(\br)
    \end{pmatrix}\\
    &=H^{XMX}_{K'}.
\end{split}
\label{appeq:XMX_MXM_symmrelation}
\end{equation}
Therefore, we focus on phase diagrams in the $XMX$ stacking in the main text, knowing that we can deduce the equivalent results for the $MXM$ stacking using the relation above. 

\subsection{Pseudoinversion when $D \neq 0$}
\label{sect:pseudoinv_at_Dneq0}

The application of an out-of-plane displacement field $D$ adds a term, Eq.~\eqref{eq:displHamil}, 
\begin{equation}
    H_K^\text{disp}(D) = \text{diag}[+D,\,0,\,-D]
    \label{appeq:displHamil}
\end{equation}
to the local continuum Hamiltonian, Eq.~\eqref{eq:localHamiltonian_appendix}. 
This additional term breaks the pseudoinversion symmetry given in Eq.~\eqref{eq:appendix_pseudoinv}, as $\gamma^x$ performs a change of basis that permutes the top and bottom layer while leaving the middle one invariant. 

However, taking $\br \rightarrow -\br$ and $D \rightarrow -D$ in addition to applying the pseudoinversion operator is a symmetry that leaves the Hamiltonian invariant. 
To see this, we first apply the pseudoinversion symmetry to the Hamiltonian of Eq.~\eqref{eq:localHamiltonian_appendix} for some stacking configuration parametrized by $\boldsymbol{\phi}$ with $D \neq 0$
\begin{equation}
    \begin{aligned}
        \gamma^x &\left[ H_K(\br, \boldsymbol{\phi}) + H_K^\text{disp}(D) \right]\gamma^x = -\frac{\hbar^2 \bk^2}{2m^*} \mathbb{1}\\
        &+ \begin{pmatrix}
            V_b(\br, \boldsymbol{\phi}) - D & t_{-\boldsymbol{\phi}}^* (\br) & 0\\
            t_{-\boldsymbol{\phi}}^*(\br) & V_m(\br, \boldsymbol{\phi}) & t_{\boldsymbol{\phi}} (\br)\\
            0 & t_{\boldsymbol{\phi}}(\br) & V_t(\br, \boldsymbol{\phi}) + D
        \end{pmatrix}, 
    \end{aligned}
\end{equation}
showing that the displacement field term in Eq.~\eqref{appeq:displHamil} is odd under pseudoinversion. 
Then, we take $\br \rightarrow -\br$ and find that the top/bottom intralayer potentials given in Eq.~\eqref{appeq:intralayerpotentials} transform as
\begin{equation}
    \begin{aligned}
        V^{t/b}(-\br, \boldsymbol{\phi}) &= 2V \sum\limits_{j=1}^3 \cos{(\bG_j \cdot(-\br) \pm \psi \pm \Delta\phi_j)}\\
        &= 2V \sum\limits_{j=1}^3 \cos{(-\bG_j \cdot \br \pm \psi \pm \Delta\phi_j)}\\
        &= 2V \sum\limits_{j=1}^3 \cos{(\bG_j \cdot \br \mp \psi \mp \Delta \phi_j)}\\
        &= V^{b/t}(\br, \boldsymbol{\phi}). 
    \end{aligned}
\end{equation}
We may use the identity $t_{\boldsymbol{\phi}} (\br) = t_{-\boldsymbol{\phi}}^*(-\br)$ given in Eq.~\eqref{appeq:interlayer_minusr_identity} to see that the interlayer terms also remain invariant. 

Thus, we have found that
\begin{equation}
    \gamma^x \left[ H_K(-\br, \boldsymbol{\phi}) + H_K^\text{disp} (D) \right] \gamma^x = H_K(\br, \boldsymbol{\phi}) + H_K^\text{disp}(-D). 
\end{equation}
As neither the change of basis due to $\gamma^x$ nor taking $\br \rightarrow -\br$ changes the Chern number, we find that taking $D \rightarrow -D$ will leave the Chern number invariant as well. 
Therefore, taking $D \rightarrow -D$ will not change the Chern numbers presented in the phase diagram as a function of $D$ vs. $\theta$, Fig.~\ref{fig:chernphases}(b). 

\subsection{Pseudoinversion Eigenvalues}
\label{sect:pseudoinv_eigs}
In Sec.~\ref{sect:SymmAtLargeAngle} of the main text, we presented the results of a perturbation theory calculation to find the eigenvalues of the pseudoinversion symmetry operator in the large angle limit for the $XMX$-stacked MoTe$_2$ homotrilayer. 
Here, we present the details of that calculation. 

In the large twist angle limit where the intralayer dispersions dominate over the moir\'e potentials (typically above 6.8$^\circ$), we can derive a rigorous perturbative result. 
Here, the band structure is primarily set by the kinetic term
\begin{equation}
\label{eq:appendix_kineticHamiltonian}
    H_K^{\rm kin}(\bk) = \begin{pmatrix}
    -\frac{\hbar^2 (\bk-\bq_1)^2}{2m^*} &  0  & 0 \\ 
    0 & -\frac{\hbar^2 \bk^2}{2m^*} & 0  \\ 
    0 & 0 & -\frac{\hbar^2 ( \bk+ \bq_1)^2}{2m^*}
    \end{pmatrix}.
\end{equation}
dominates the band structure. 
The first band crossing at $M$ occurs when the parabolic dispersions centered at $K$ and $K'$ on the top and bottom layers respectively intersect. 
However, the local Hamiltonian Eq.~\eqref{eq:appendix_xmxHamiltonian} only directly couples parabolas from adjacent layers, and so the degenerate perturbation theory must be taken to second order. 
The second-order process occurs via the crossing of two parabolic bands originating from the middle layer due to $\Gamma$ points in neighboring moir\'e reciprocal shells. 

We begin by introducing the eigenstates of the kinetic energy in the plane wave basis, as 
\begin{equation}
    \braket{\br, \ell}{\psi_{\bk,\bG,\ell}}=\frac{e^{i(\bk-s_\ell \bq_1+\bG)\cdot \br}}{\sqrt{A}}, 
\end{equation}
with energies $\epsilon_{\bk,\bG,\ell}=-\hbar^2|\bk-s_\ell \bq_1+\bG|^2/2m$. We have introduced $s_\ell = 1, 0, -1$ respectively for the layers $\ell = t,m,b$. 
As we are considering the pseudoinversion symmetry, we focus on the states at one of its high-symmetry points $M_1$, shown in Fig.~\ref{fig:helicalgeometry}(b) of the main text, and defined such that $\bk_{M_1}=\bG_1/2 = (\bq_1 - \bq_2)/2$. 
Then, denoting a plane wave eigenstate that originates from $\bp$ by $\psi_\bp(\br)$, we find two degenerate plane waves 
\begin{align}
    \psi_K(\br) &\equiv \braket{\br,t}{\psi_{M_1,0,t}}=\frac{e^{i(\bk_{M_1}-\bq_1)\cdot \br}}{\sqrt{A}} ,\\
    \psi_{K'}(\br) &\equiv \braket{\br,b}{\psi_{M_1,-\bG_1,b}} =\frac{e^{i(\bk_{M_1}+\bq_1-\bG_1)\cdot \br}}{\sqrt{A}},
\end{align}
each with energy 
\begin{equation}
    \epsilon_{M_1}=-\frac{\hbar^2}{2m^*}\frac{|\bq_1+\bq_2|^2}{4}=-\frac{\hbar^2}{2m^*a^2_M}\left(\frac{2\pi}{3}\right)^2.
\end{equation} 

As noted above, the interlayer potential $t_{\omega/\omega^*}(\br)$ couples adjacent layers, so the two plane waves are decoupled to first order in $w$. 
The matrix element mixing $\psi_{M_1,0,t}$ and $\psi_{M_1,-\bG_1,b}$ originates from second-order processes involving states belonging to the middle layer and originating from $\Gamma$, and so 
\begin{align}
    \psi_\Gamma(\br) &\equiv \braket{\br,m}{\psi_{M_1,0,m}} =\frac{e^{i \bk_{M_1}\cdot \br}}{\sqrt A},\\
    \psi_{\Gamma+\bG_1}(\br)&\equiv \braket{\br,m}{\psi_{M_1,-\bG_1,m}} =\frac{e^{i(\bk_{M_1}-\bG_1)\cdot \br}}{\sqrt{A}},
 \end{align}
with energy
\begin{equation}
    \epsilon'_{M_{1}}= -\frac{\hbar^2}{2m^*}|\bk_{M_1}|^2=- \frac{\hbar^2}{2m^*a^2_M}\frac{4\pi^2}{3}.
\end{equation}
The two sets of plane waves are separated by a nonzero energy gap
\begin{equation}
E_{\rm gap}=\epsilon_{M_1}-\epsilon'_{M_1}=\frac{8\pi^2}{9}\frac{\hbar^2}{2m^* a^2_M}.
\end{equation}
To determine the low-energy Hamiltonian in the two-fold degenerate manifold, we now compute the matrix elements connecting the different states at $M_1$ to second order in $w/E_{\rm gap}$. 
Denoting the matrix elements of the interlayer potential that couple states $\ket{\psi_{\bp}}$ and $\ket{\psi_{\bp'}}$ by $T_{\bp,  \bp'}$, we find that 
\begin{align}
    \begin{split}
    T_{K, \Gamma} &= \int t_{\omega^*}(\br)~\psi^*_K(\br)~\psi_\Gamma (\br)~d^2\br=w, \\ 
    T_{K, \Gamma+\bG_1} &= \int t_{\omega^*}(\br)~\psi^*_{K}(\br)~\psi_{\Gamma+\bm G_1}(\br)~d^2\br=w\omega^*,\\  
    T_{\Gamma, K'} &= \int t_{\omega}(\br)~\psi^*_{\Gamma}(\br)~\psi_{K'}(\br)~d^2\br =w\omega,\\ 
    T_{\Gamma+\bG_1, K'} &=\int t_{\omega}(\br)~\psi^*_{\Gamma+\bm G_1}(\br)~\psi_{K'}(\br)~d^2\br =w.
    \end{split}
\end{align}
Then, the second-order perturbative correction within the subspace of kinetic energy eigenstates at $M_1$, denoted $[ \ket{\psi_K}, \ket{\psi_{K'}}]$, is given by Eq.~\eqref{eq:perturbative_hamiltonian}, 
\begin{equation}
    \left[\delta H_{w}\right]_{ij} =\sum_{l=\Gamma, \Gamma + \bG_1}\frac{T_{il}T_{lj}}{E_{\rm gap}}=\left[\frac{2w^2}{E_{\rm gap}}+2\text{ Re }\omega\frac{w^2}{E_{\rm gap}}\sigma^x\right]_{ij}.
\end{equation}
Then, as stated in the main text, this perturbation selects the topmost eigenstate with energy $\epsilon_{M_1}+3w^2/E_{\rm gap}$ to be the linear combination $\ket{-}=(\ket{\psi_K}-\ket{\psi_{K'}})/\sqrt{2}$, which is odd under inversion. 
This allows us to conclude that in this large angle limit, the pseudoinversion eigenvalue is $-1$. 

\section{Additional Numerical Results}\label{app:additional_numerical}

\subsection{DFT-Determined Parameter Validity for Bilayer TMDs}
\label{sect:dft_prms_bilayer}

\begin{figure}
    \centering
    \includegraphics[width=0.98\linewidth]{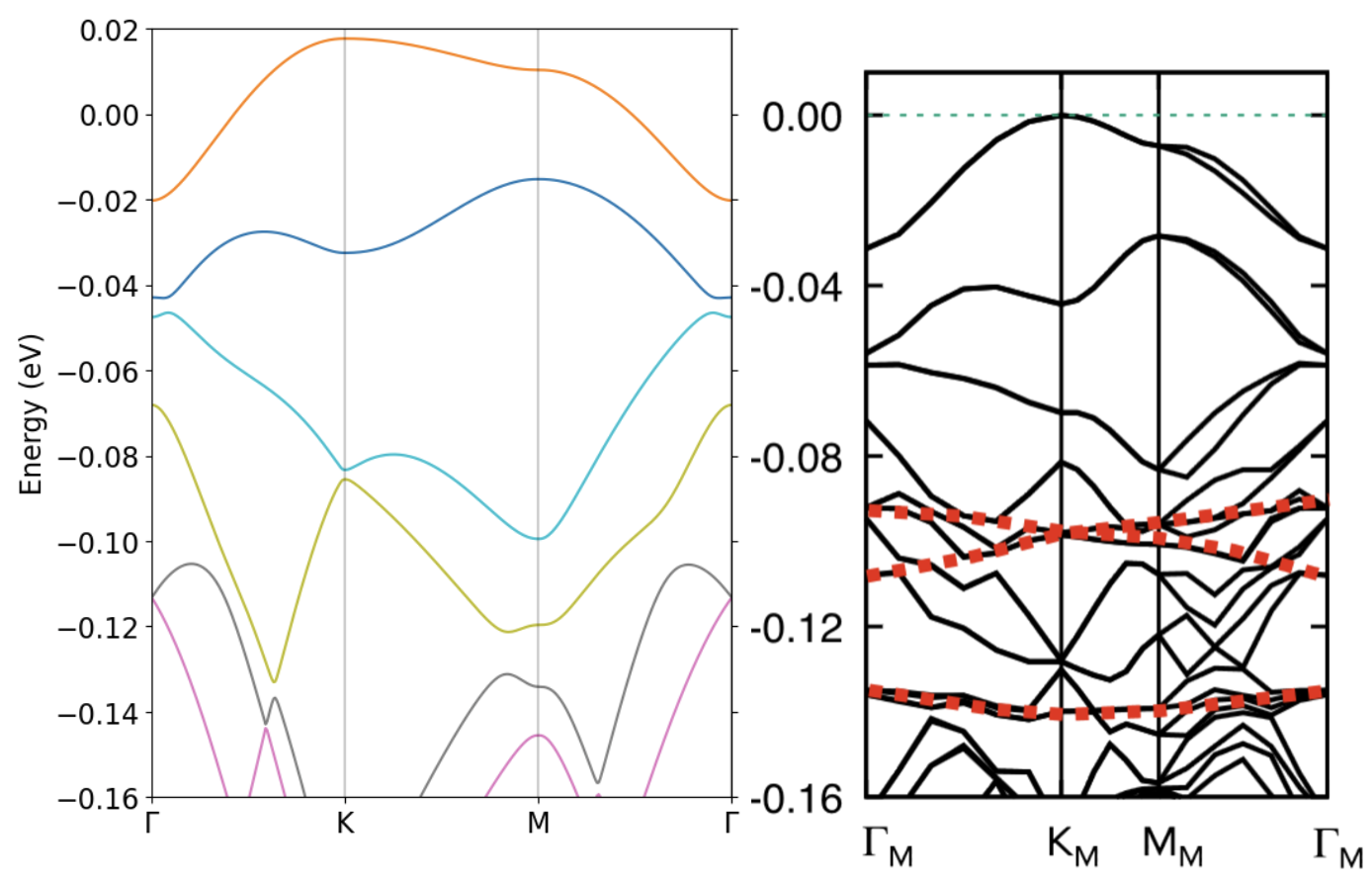}
    \caption{\textbf{Left:} Continuum model band structure of bilayer MoTe$_2$ at $\theta = 5.09^\circ$ using $V = 16.5$ meV, $\psi = -105.9^\circ, w = -18.8$ meV. \textbf{Right:} DFT calculation of band structure published in Ref.~\cite{Jia-PhysRevB.109.205121} for bilayer MoTe$_2$ at identical twist angle. We note that the doubled bands in the DFT calculation are due to the effects of spin-orbit coupling, while our continuum model band structure was computed while suppressing the spin degree of freedom. }
    \label{fig:dftvsctm}
\end{figure}

Though the TMD homobilayer continuum model parameters computed via DFT in Refs.~\cite{Jia-PhysRevB.109.205121,wang2024fractional,reddy2023,wu2019topological} were obtained from an array of twist angles $\theta = 0^\circ, 3.89^\circ, 4.4^\circ$, they are clustered together on the phase diagram shown in Fig.~\ref{fig:chernphases}(a). 
We show in Fig.~\ref{fig:dftvsctm} that the continuum model parameters obtained in Ref.~\cite{Jia-PhysRevB.109.205121} for $\theta = 3.89^\circ$ still provide reasonable results for larger twist angles for the topmost valence bands in the bilayer case by using the parameters to compute the continuum model band structure at $\theta = 5.09^\circ$ and comparing it with the DFT-obtained band structure from the same reference. 

\subsection{MXM Stacking Configuration}\label{sect:addl_mxm}
\begin{figure}[t!]
    \centering
    \includegraphics[width=0.98\linewidth]{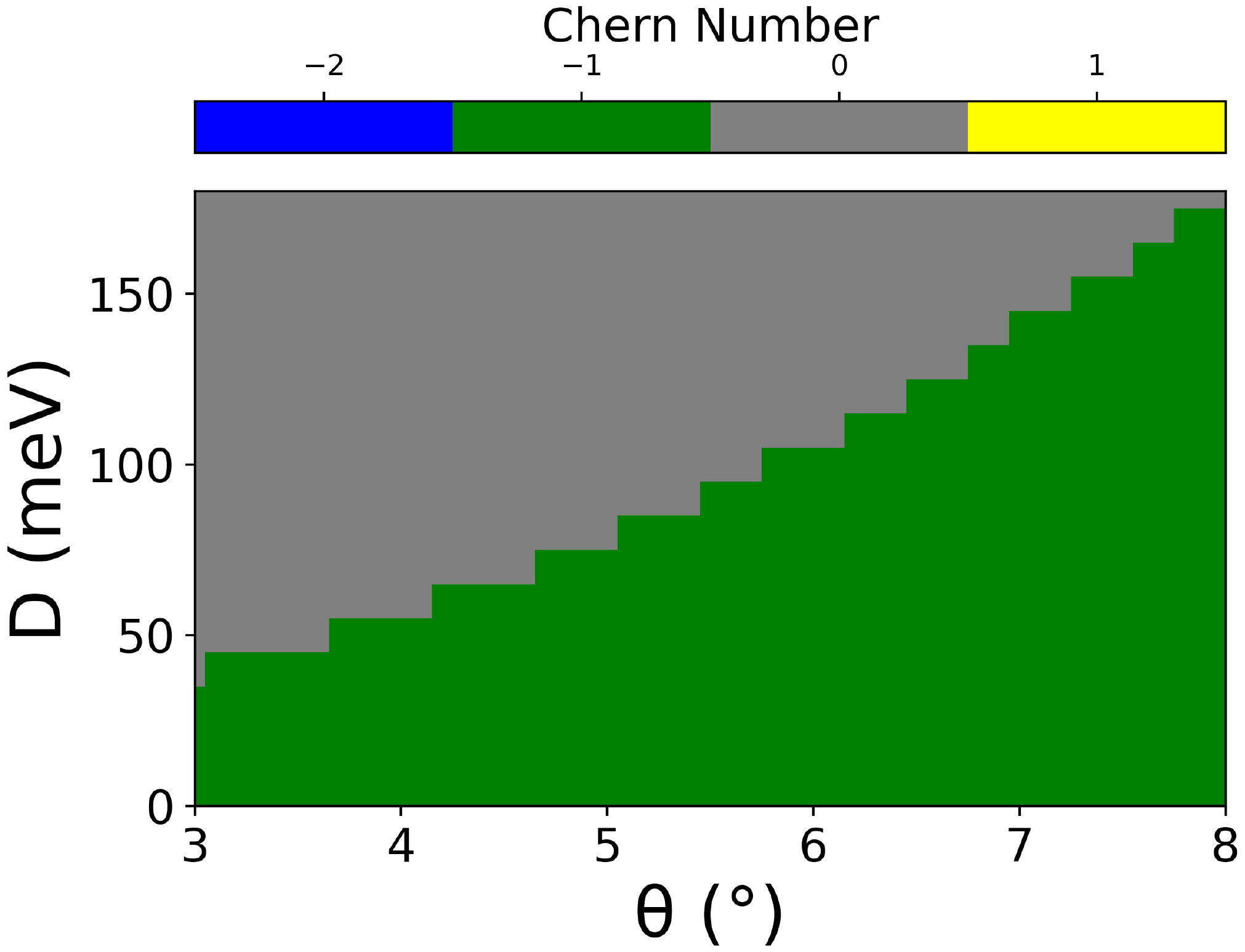}
    \caption{Chern number phase diagram of topmost valence band as a function of external displacement field $D$ vs. twist angle $\theta$ for $MXM$-stacked helical trilayer MoTe$_2$}
    \label{fig:MXM_chernphases}
\end{figure}
As shown in Eq.~\eqref{appeq:XMX_MXM_symmrelation}, taking $\psi \rightarrow -\psi$ leads to $H_K^{XMX} (\br) \rightarrow H_{K'}^{MXM}(-\br) = {H_K^*}^{MXM}(-\br)$. 
As a result, we infer from Fig.~\ref{fig:chernphases}(a) of the main text that at $D = 0$, the Chern number for the $MXM$ stacking using the same realistic continuum model parameters $(V,~\psi,~w,~m^*) = (16.5\text{ meV},-100^\circ,-18.8\text{ meV},~0.60m_e)$ will be $C=-1$. 
For sake of completeness, we provide a Chern number phase diagram of the topmost valence band of the $MXM$ stacking as a function of $D$ vs. $\theta$ in Fig.~\ref{fig:MXM_chernphases}. 
We find that the Chern number is indeed $C=-1$ across a wide range of twist angles for the $MXM$-stacked domain when $D = 0$. 
In addition, at a sufficiently strong $D > 0$, the topmost band is trivialized to become a $C = 0$ band regardless of the twist angle $\theta$, as we expect. Moreover, we do not find any $|C|=2$ region irrespective of the twist angle or displacement field.

\subsection{MMM Stacking Configuration}
\label{sect:mmm-stacked_numerics}

\begin{figure}
    \centering
    \includegraphics[width=0.98\linewidth]{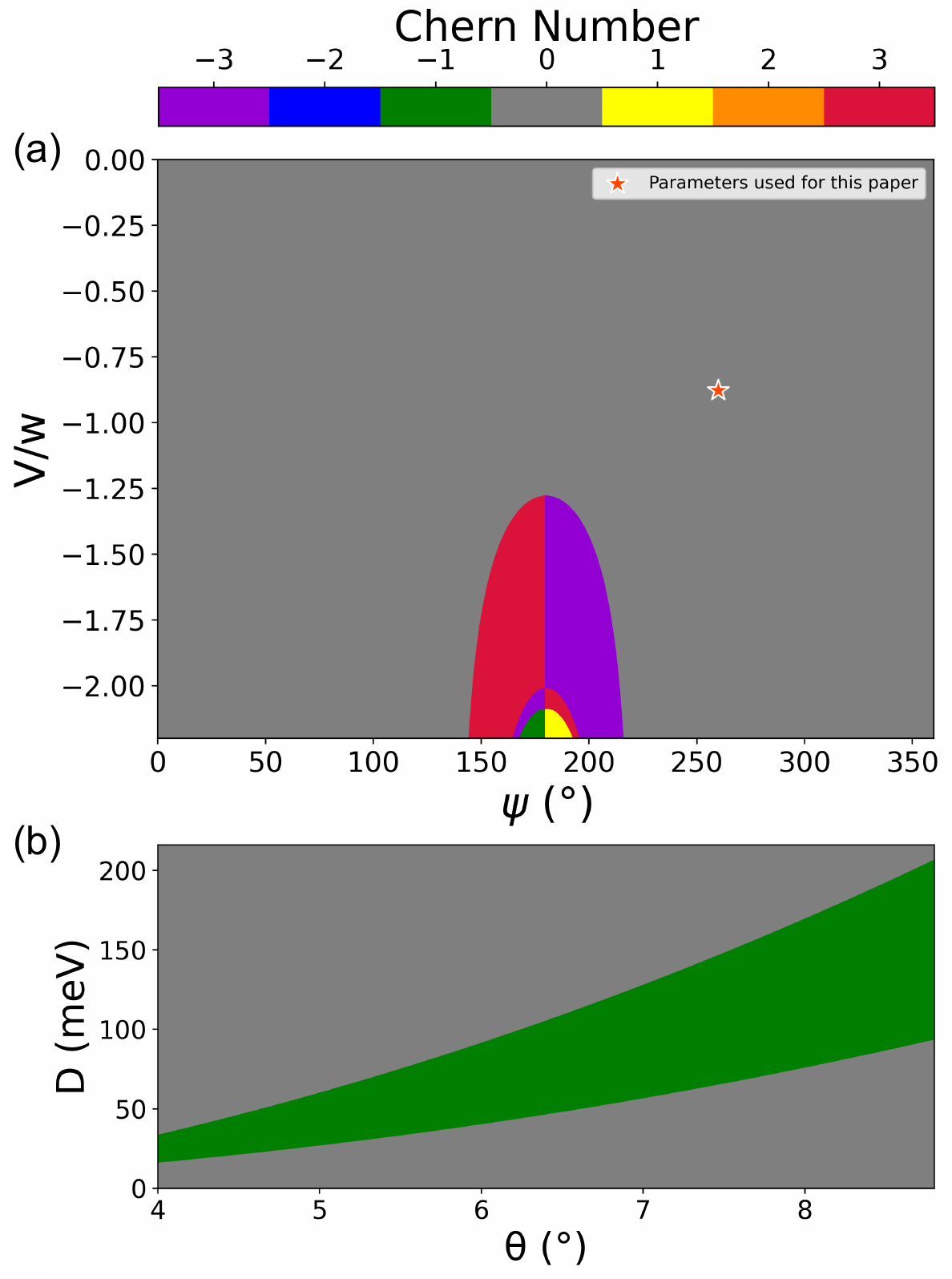}
    \caption{\textbf{(a)} Chern number phase diagram for the topmost valence band of $MMM$-stacked helical trilayer MoTe$_2$ at twist angle $\theta = 5.6^\circ$ as a function of varying continuum model parameters $V/w$ vs. $\psi$ with fixed $w = -18.8$ meV and $m^* = 0.60m_e$. 
    The orange star denotes the continuum model parameters that we choose for the figures in this paper, i.e. $(V, \psi, w, m^*) = (16.5\text{ meV}, -100^\circ, -18.8\text{ meV}, 0.60m_e)$, which are close to those obtained via DFT calculations, as shown in Fig.~\ref{fig:chernphases}(a) of the main text.
    The topmost valence band in this stacking is topologically trivial for a wide range of continuum model parameters, except within a dome centered about $\psi = 180^\circ$ at $\lvert V/w\rvert \gtrsim 1.25$. 
    \textbf{(b)} Chern number phase diagram for the topmost valence band of $MMM$ stacking as a function of varying twist angle and external displacement field $D$. }
    \label{fig:mmm_chernphases}
\end{figure}

In Fig.~\ref{fig:mmm_chernphases}(a), we present a phase diagram of Chern numbers of the topmost valence bands for the $MMM$ stacking configuration as a function of the continuum model parameters $V/w$ vs.~$\psi$ for $\theta = 5.6^\circ$, with fixed $w = -18.8\text{ meV}$ and $m^* = 0.60m_e$. 
We find that the topmost valence band is topologically trivial for most values of $(\psi,\,V/w)$, with the exception of a dome centered at $\psi = 180^\circ$ for large values of $\lvert V/w \rvert$ ($V/w \lesssim -1.25$). 
Interestingly, we find a large region with $\lvert C \rvert = 3$ and a smaller region with $\lvert C \rvert = 1$. 
Note that the phase diagram is symmetric about the $\psi = 180^\circ$ line. 
This is because taking $\psi \rightarrow -\psi$ leads to $H^{MMM}_{K}(\br) \rightarrow {H^*_K}^{MMM}(-\br)$, thus switching the sign of the Chern number. 

We do not focus on $MMM$ stacking in the main text as this paper is primarily interested in bands with $\lvert C \rvert > 1$, and the regions of the phase diagram in which the topmost valence band has a $\lvert C \rvert > 1$ are far from the DFT-obtained parameters for twisted homobilayer MoTe$_2$. 
The orange star in Fig.~\ref{fig:mmm_chernphases}(a) marks the continuum model parameters $(V, \psi, w, m^*) = (16.5\text{ meV}, -100^\circ, -18.8\text{ meV}, 0.60m_e)$ that we use throughout this paper. 
This set of parameters was selected because of its proximity to the parameters computed using DFT calculations for the MoTe$_2$ homobilayer, as shown in Fig.~\ref{fig:chernphases}. 
As can be seen in Fig.~\ref{fig:mmm_chernphases}(b), tuning both the twist angle and the displacement field using these parameters does not lead to any region with $\lvert C \rvert > 1$. More precisely, as $D$ increases for a given twist angle, the band acquires a $C=-1$ character, but will be trivialized again at a sufficiently strong $D$. 
The transition from the $C = 0$ to $C = -1$ is marked by a gap closure at the $K'$ point, while the transition back to a trivial band is marked by a gap closure at $\Gamma$. 

In addition to the absence of a topmost band with $\lvert C \rvert > 1$ for the MoTe$_2$ homobilayer model parameters, the band gap between the two highest-energy valence bands is very small in the $\lvert C \rvert = 3$ region of Fig.~\ref{fig:mmm_chernphases} (at most $2\text{ meV}$) as compared to the gap that we can obtain for $\lvert C \rvert = 2$ bands in the $XMX/MXM$ stacking (up to $11 \text{ meV}$). 
Moreover, we observe that even if we focus on the optimal parameters for the $\lvert C \rvert = 3$ band, the quantum metric of the topmost band is highly concentrated, as we will show in the following subsection Appendix~\ref{app:QG_results}. 

\subsection{Quantum Geometry Results}\label{app:QG_results}

\begin{figure}
    \centering
    \includegraphics[width=0.98\linewidth]{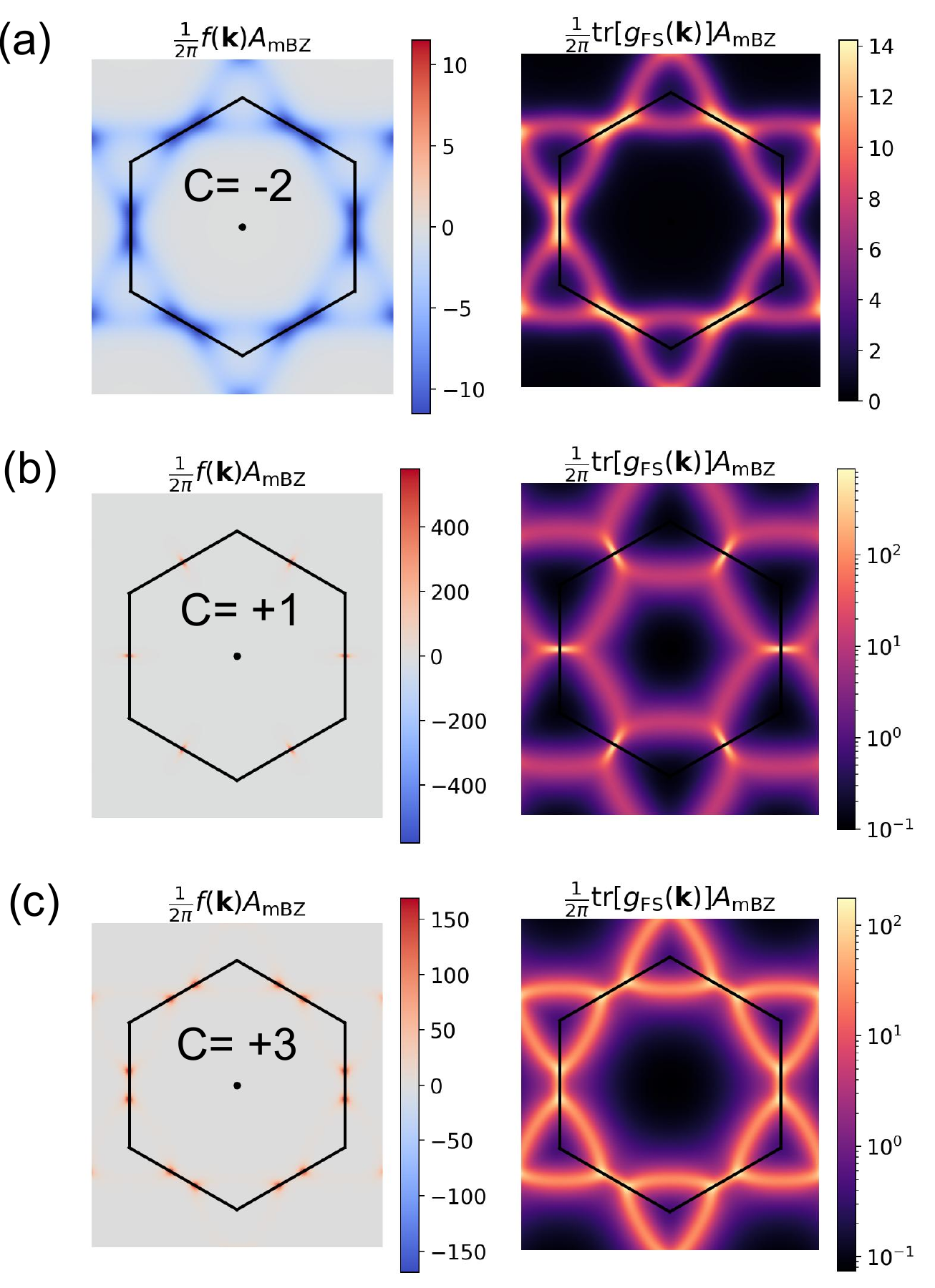}
    \caption{Berry curvature $f(k)$ and trace of Fubini-Study metric $\text{tr}[g_\text{FS}(\bm{k})]$ for the top valence band. 
    The violation of the trace condition $\bar{T}$ is defined in Eq.~\eqref{appeq:trace_viol}.
    \textbf{(a)} $C=-2$ band for $XMX$ stacking at $\theta=5.6^\circ$, with $(V, \psi, w, m^*) = (16.5\text{ meV}, -100^\circ, -18.8\text{ meV}, 0.60m_e)$, and $\bar{T}=1.3$. 
    \textbf{(b)} $C=1$ band for $XMX$ stacking at $\theta=7.4^\circ$, with $(V,~\psi,~w,~m^*) = (16.5\text{ meV}, -100^\circ, -18.8\text{ meV}, 0.60m_e)$, and $\bar{T}=6.1$. 
    \textbf{(c)} $C=3$ band for $MMM$ stacking at $\theta=5.6^\circ$, with $(V, \psi, w, m^*) = (39.48\text{ meV}, -211^\circ, -18.8\text{ meV}, 0.60m_e)$, and $\bar{T}=4.4$. Note the logarithmic color scale for $\text{tr}[g_\text{FS}(\bm{k})]$ for (b) and (c).
    Black hexagon delineates the mBZ boundary, and black dot corresponds to $\Gamma$. 
     Quantities are computed over a $120\times120$ mesh in the mBZ.}
    \label{fig:BC_QG}
\end{figure}

In Fig.~\ref{fig:BC_QG}, we plot the distribution of the Berry curvature $f(\bm{k})$ and the trace of the Fubini-Study metric $\text{tr}[g_\text{FS}(\bm{k})]$~\cite{Roy-PhysRevB.90.165139} of the top valence band. 
We define the violation of the trace condition as 
\begin{equation}
    \bar{T}=\frac{A_\text{mBZ}}{2\pi}\int d\bm{k}\int\left(\text{tr}[g_\text{FS}(\bm{k})]-|f(\bm{k})|\right).
    \label{appeq:trace_viol}
\end{equation}
Note that we have the constraint $\text{tr}[g_\text{FS}(\bm{k})]\geq|f(\bm{k})|$~\cite{Roy-PhysRevB.90.165139}. Fig.~\ref{fig:BC_QG}(a) shows that for a representative $C=-2$ band in $XMX$-stacking, the trace violation $\bar{T}=1.3$ is low. 
Fig.~\ref{fig:BC_QG}(b) shows the corresponding results for the $C=1$ band at a twist angle $\theta=7.4^\circ$ just above the $C=-2\rightarrow 1$ topological transition. 
Since that transition is mediated by gap closures at the $M$ points, we find that the quantum geometry is extremely peaked in those regions. 
In Fig.~\ref{fig:BC_QG}(c), we consider a representative $C=3$ band for $MMM$-stacking. 
Note that the parameters $V=39.48\,$meV and $\psi=-211^\circ$ are far from the expected physical values (see Fig.~\ref{fig:chernphases}). 
We find a significant trace condition violation $\bar{T}=4.4$. 
Note that in all cases, $\text{tr}[g_\text{FS}(\bm{k})]$ and $|f(\bm{k})|$ are relatively small and smooth at $\Gamma$.

\subsection{Taking $V_o \neq V_m$, $\psi_o \neq \psi_m$}
\label{sect:addl_ctm_prms}
\begin{figure}
    \centering
    \includegraphics[width=0.98\linewidth]{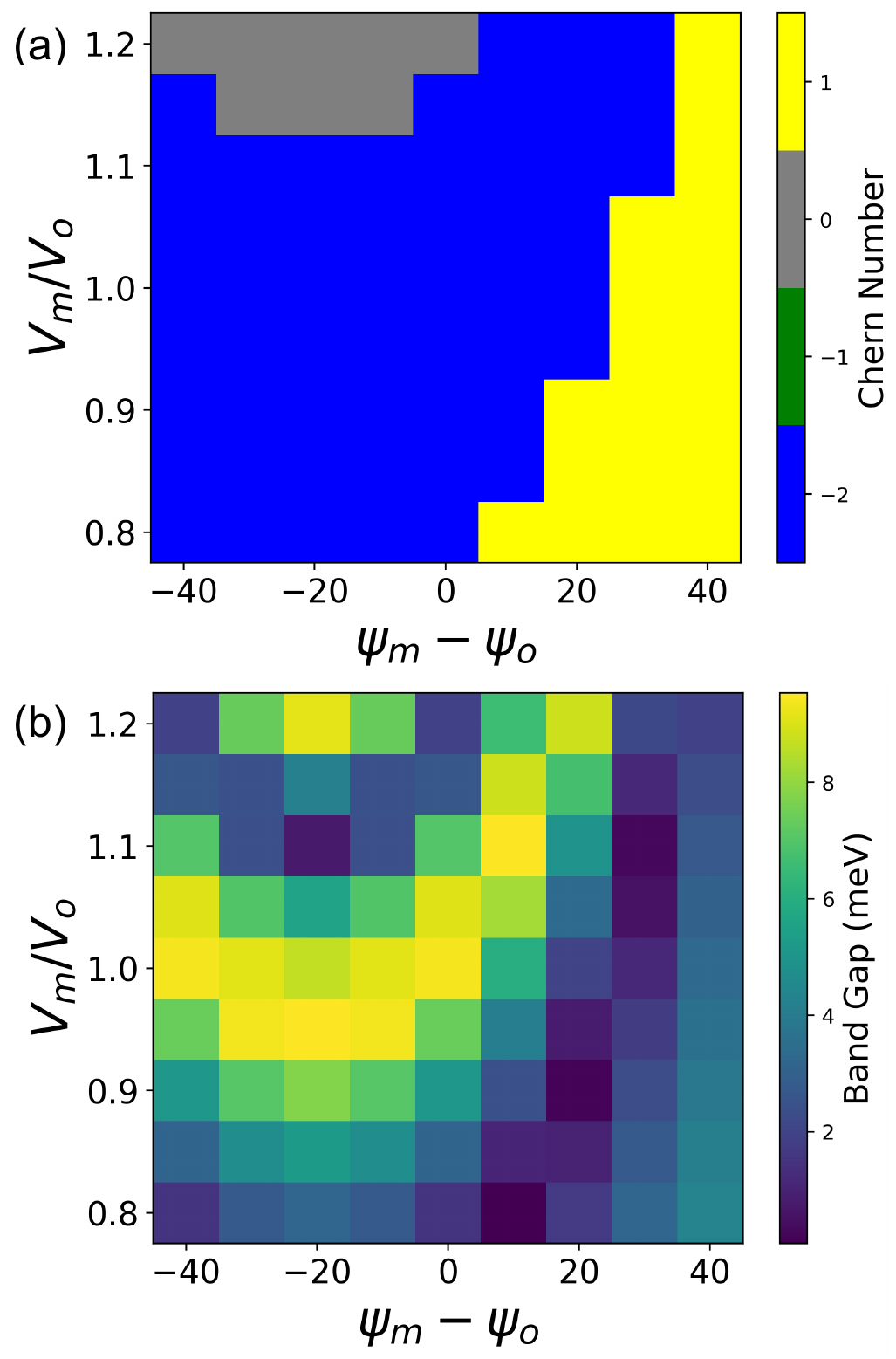}
    \caption{\textbf{(a)} Chern number of the topmost valence band when detuning the intralayer parameters of the middle layer, as a function of $V_m/V_o$ vs. $\psi_m - \psi_o$, at twist angle $\theta = 5.6^\circ$ and using $(V_o, \psi_o, w, m^*) = (16.5\text{ meV}, -100^\circ, -18.8\text{ meV}, 0.60m_e)$. 
    \textbf{(b)} Direct band gap between the top two valence bands as a function of $V_m/V_o$ vs. $\psi_m - \psi_o$. }
    \label{fig:vary_mid_prms}
\end{figure}

In our derivation of the local moir\'e-scale Hamiltonian presented in Appendix~\ref{sect:ctm_mod_prms}, we used the symmetries of the helically twisted MoTe$_2$ homotrilayer to constrain the number of continuum model parameters in our model. 
In doing so, we showed that though the symmetries relate the intralayer parameters $V$ and $\psi$ of the outer layers to each other, no symmetry constraint exists to relate the intralayer parameters of the outer layers to those of the middle layer. 
Denoting the parameters of the outer layers as $(V_o, \psi_o)$ and those of the middle layers as $(V_m, \psi_m)$, we present a phase diagram of the Chern number of the topmost valence band as a function of $V_m/V_o$ vs.~$\psi_m - \psi_o$ in Fig.~\ref{fig:vary_mid_prms}(a). 
We also show the direct band gap between the first two valence bands in Fig.~\ref{fig:vary_mid_prms}(b). 

We find that the $C = -2$ phase persists even when either $V_m$ or $\psi_m$ are tuned away from their respective outer layer values. 
As a result, we conclude that though the Hamiltonian, Eq.~\eqref{eq:Hamiltonian_1}, that we study in the main text is not fully constrained by the symmetries of the system, the results that we present should remain valid in the case that the continuum model parameters of the middle layer are not equal to those of the outer layers. 

\section{Further Results on Wannier Orbitals}

\subsection{Wannierization procedure}
\label{sect:Wannierisation_method}

In order to construct the Wannier orbitals from the eigenstates of the topmost continuum model bands, we begin with the Bloch states given in Eq.~\eqref{eq:Bloch_States} 
\begin{equation}
    \psi_{\bm k}^{n}({\bm r})=u^n_{{\bm k}}({\bm r})e^{i{\bm k}\cdot {\bm r}}=\sum_{\bm G}z^{n}_{{\bm k},{\bm G}}\,e^{i{({\bm k}+ {\bm G})}\cdot {\bm r}},
    \label{eq:appendix_Bloch_States}
\end{equation}
where $n$ denotes the band and $u_\bk^n (\br)$ is the periodic component of the Bloch wave function. 
As discussed in the main text, the second equality results from a Fourier expansion, and the Fourier coefficients $z_{\bk, \bG}^n$ are computed by diagonalizing the continuum model.  
Then, we construct a unitary transformation $U_\Pi(\bk)$ at each $\bk$ in the mBZ that performs a change of basis from that of the basis continuum model eigenstates to the basis of the desired Wannier orbitals (see  Eq.~\eqref{eq:quasiBloch})
\begin{equation}
    \tilde{\psi}_\bk^\alpha(\bk) = \sum\limits_m U_\Pi^{m\alpha} (\bk) \psi_\bk^m(\br), 
    \label{appeq:quasiBloch}
\end{equation}
where $\psi^n_{\bm k}=(\psi_{\bm k}^{n,t},\psi_{\bm k}^{n,m},\psi_{\bm k}^{n,b})^T$ is a three-component spinor. 
The unitary transformation $U_\Pi(\bk)$ is constructed by diagonalizing the layer polarization operator in layer space. 
For the trilayer system that we study, the layer polarization operator is given by the matrix $\Lambda=\,$diag$\,(1,0,-1)$, which yields a Wannier orbital construction that respects the $\mathcal{C}_{2y} \mathcal{T}$ symmetry of the model. 

The reason that this construction yields well-localized orbitals may be understood in the large angle limit. 
In that limit, where the moir\'e potential is weak, the band structure consists of three decoupled parabolas in $\bk$-space, each of which originate from a monolayer, meaning that these parabolic dispersions are layer-polarized. 
Though this is no longer true at the smaller twist angles we consider where the moir\'e potential is stronger, the continuum model eigenstates are still adiabatically connected to those in the large angle limit. 
Therefore, transforming the eigenstates into a basis in which the layer polarization operator is diagonal localizes the eigenstates to each layer in real space as much as possible. 
This also explains why the procedure does not work in the small angle limit, where the moir\'e potential is too strong for the same change-of-basis to work to localize the wave functions into exponentially localized Wannier orbitals. 

Having established the relevance of the layer polarization operator $\Lambda$, the problem reduces to diagonalizing the matrix
\begin{align}
    \Pi_{\bk}^{mn}=\braket{\psi_{\bk}^m|\Lambda}{\psi_{\bk}^n},
\end{align}
such that $U_{\Pi}^{\dagger}({\bm k})\,\Pi_{\bm k}\, U_{\Pi}({\bm k})=\text{diag}(\lambda_1(
\bk), \lambda_2 (\bk), \lambda_3 (\bk))$. 
We note that for this Wannierization procedure to work, the eigenvalues $\lambda_j (\bk)$ must be well-separated from each other at every $\bk$ in the BZ. 
Here we have introduced the notation
\begin{equation}
\begin{aligned}
    \braket{\psi_{\bm k}^m}{\psi_{\bm k}^n} &= \int \sum\limits_{\ell, \bG, \bG'} (z_{\bk, \bG'}^{m, \ell} )^* z^{n, \ell}_{\bk, \bG} e^{i (\bG - \bG') \cdot \br} \, d^2\br \\
    &= \sum\limits_{\ell, \bG, \bG'} (z_{\bk, \bG'}^{m, \ell} )^* z^{n, \ell}_{\bk, \bG} \int e^{i (\bG - \bG') \cdot \br} \, d^2 \br \\
    &= \sum\limits_{\ell, \bG, \bG'} (z_{\bk, \bG'}^{m, \ell} )^* z^{n, \ell}_{\bk, \bG} \delta_{\bG, \bG'} \\
    &=\sum_{\ell, \,{\bm G}}(z^{m,\ell}_{{\bm k},{\bm G}})^*\, z^{n,\ell}_{{\bm k},{\bm G}},\qquad \ell=t,m,b.
\end{aligned}
\end{equation}

In addition, we provide some additional details of the gauge-fixing step of the Wannierization. 
The gauge fixing is performed on the quasi-Bloch states given in Eq.~\eqref{appeq:quasiBloch}, and can be achieved by 
multiplying the vector of quasi-Bloch states by a diagonal matrix
\begin{align}
    \begin{pmatrix}
        \phi_{\bm k}^{H_1}\\
        \phi_{\bm k}^{T}\\
        \phi_{\bm k}^{H_2}
    \end{pmatrix}=
    \begin{pmatrix}
        e^{-i\,\phi^{H_1}}&0&0\\
        0&e^{-i\,\phi^{T}}&0\\
        0&0&e^{-i\,\phi^{H_2}}
    \end{pmatrix} \begin{pmatrix}
        \tilde{\psi}_{\bm k}^{H_1}\\
        \tilde{\psi}_{\bm k}^{T}\\
        \tilde{\psi}_{\bm k}^{H_2}
    \end{pmatrix},
\end{align}
whose elements are given by
\begin{align}
    \phi^{H_1}&=\text{arg}\left(\sum_{\bm G}\tilde{z}_{{\bm k},{\bm G},b}^{H_1}\,e^{i({\bm k}+{\bm G})\cdot{\bm R_{H_1}}}\right),\\
    \phi^{T}&=\text{arg}\left(\sum_{\bm G}\tilde{z}_{{\bm k},{\bm G},m}^{T}\,e^{i({\bm k}+{\bm G})\cdot{\bm R_{T}}}\right),\\
    \phi^{H_2}&=\text{arg}\left(\sum_{\bm G}\tilde{z}_{{\bm k},{\bm G},t}^{H_2}\,e^{i({\bm k}+{\bm G})\cdot{\bm R_{H_2}}}\right).
\end{align}
As in the main text, the orbital indices $H_1,~T,~H_2$ denote the hexagonal and triangular lattice sites. 
This diagonal matrix is the $U_\varphi(\bk)$ used in Eq.~\eqref{eq:gauge_quasiBloch} of the main text, and its elements are the phases at the center of each orbital that ensures that $\tilde{\psi}_\bk$ is real and positive there, in accordance with Eq.~\eqref{eq:appendix_Bloch_States}. 
Then, the Wannier orbital centered at $\alpha$, $W^\alpha(\br)$, is given by Eq.~\eqref{eq:Wannier_orbitals} of the main text
\begin{equation}
    \label{appeq:Wannier_orbitals}
    W^\alpha(\br) = \frac{1}{\sqrt{N}} \sum\limits_\bk \varphi^\alpha_\bk(\br), 
\end{equation}
where $N$ is the number of moir\'e unit cells, and may be written in terms of its layer components $W^\alpha(\br) = (W_t^\alpha(\br), W_m^\alpha(\br), W_b^\alpha(\br))^T$. 

\begin{figure}[t!]
    \centering
    \includegraphics[width=0.98\linewidth]{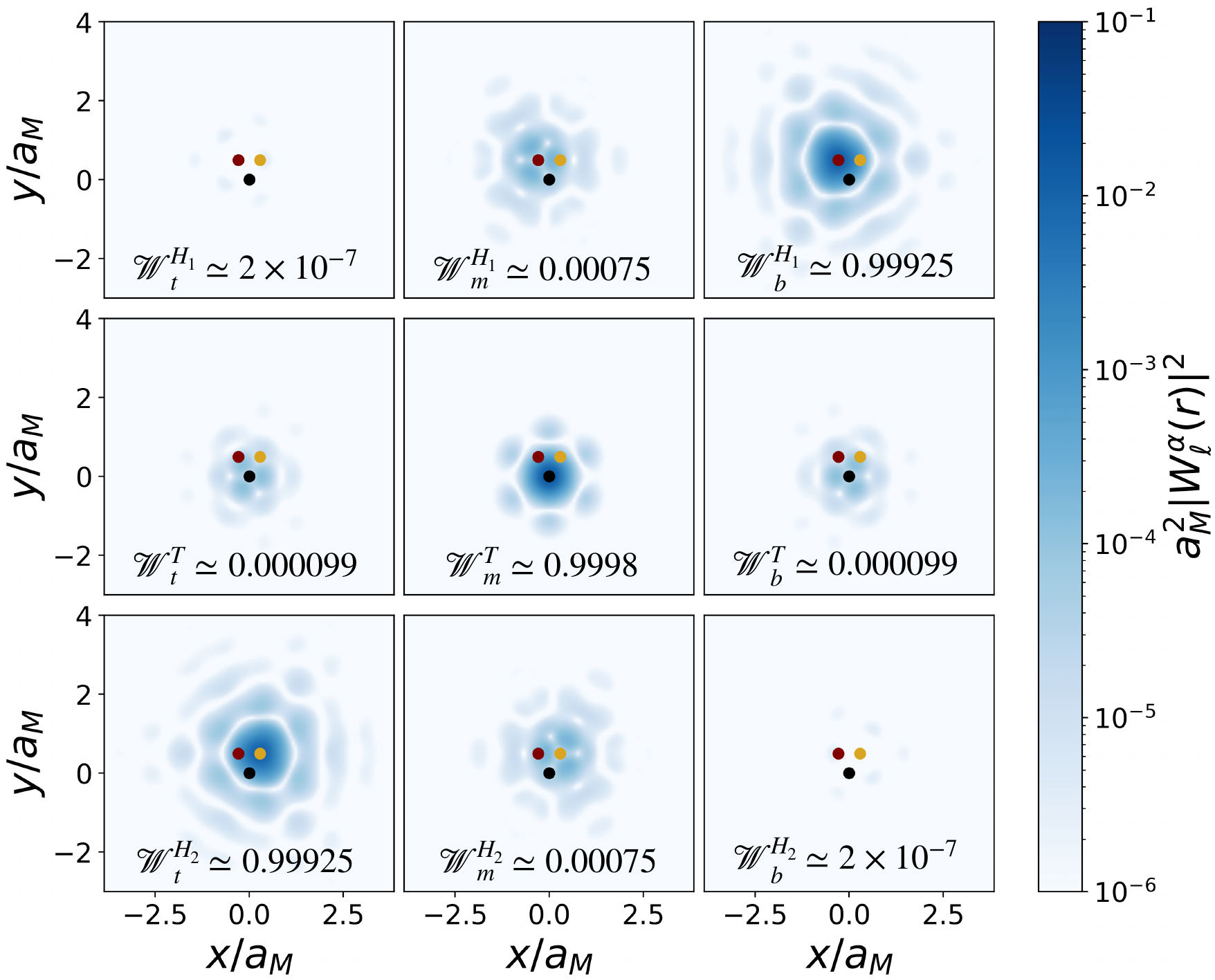}
    \caption{Squared modulus of Wannier orbitals centered at $\alpha$, $W^\alpha(\br)$, for each layer $\ell \in \{t,m,b\}$ at $\theta = 5.6^\circ$ plotted on a logarithmic color scale in real space, with a cut-off of $10^{-6}$. 
    Each plot provides the weight $\mathcal{W}^\alpha_\ell$ as defined in Eq.~\eqref{appeq:wannierorbital_weights}. 
    The maroon, black, and yellow dots mark the $\alpha \in \{ H_1, T, H_2 \}$ sites, respectively. 
    Real space coordinates are measured in units of $a_M$, the length of a moir\'e real space lattice vector. }
    \label{fig:wannierorbs_theta5.6}
\end{figure}

\subsection{Exponential Localization of Wannier Orbitals}
\label{sect:wannier_localisation}

In Fig.~\ref{fig:wannierorbs_theta5.6}, we plot the squared modulus of the Wannier orbitals for twist angle $\theta = 5.6^\circ$ on a logarithmic scale, with labeled weights of each orbital on a given layer, 
\begin{equation}
    \label{appeq:wannierorbital_weights}
    \mathcal{W}^\alpha_\ell = \int \lvert W^\alpha_\ell (\br) \rvert^2 \, d^2\br, 
\end{equation}
as given in Eq.~\eqref{eq:wannierorbital_weights} of the main text. 
Comparison between this figure and Fig.~\ref{fig:Wannier_theta3} of the main text (depicting the analogous plots for $\theta = 3.0^\circ$) shows that the Wannier orbitals are extended farther away in real space from their respective centers for larger twist angles. 
This contrast also shows that the Wannier orbitals are slightly better layer-polarized at the larger twist angle as well. 

Despite these differences in the Wannier orbitals with varying $\theta$, both sets of Wannier functions are still exponentially localized. 
\begin{figure*}[t!]
    \centering
    \includegraphics[width=0.98\linewidth]{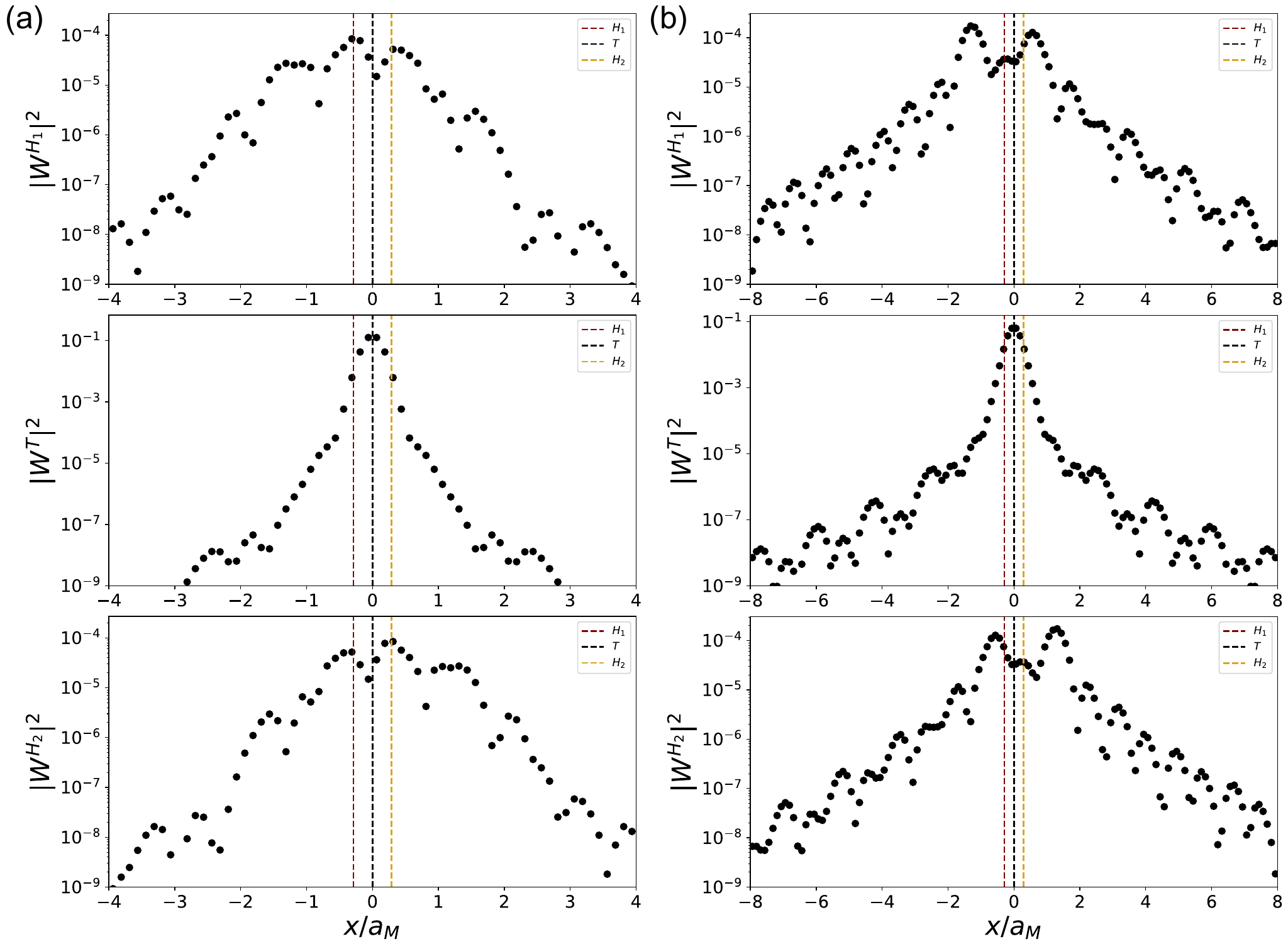}
    \caption{Squared modulus of Wannier orbitals $W^\alpha(\br)$ plotted on a logarithmic scale across a horizontal line cut for \textbf{(a)} $\theta = 3.0^\circ$ and \textbf{(b)} $\theta = 5.6^\circ$. The top, middle and bottom rows correspond to $\alpha=H_1,T,H_2$ orbitals respectively.
    For each orbital, the line cut is taken at a fixed $y$-coordinate corresponding to its Wannier center.  
    Dashed lines indicate the $x$-coordinate of the Wannier center corresponding to $\alpha = H_1$ (maroon), $T$ (black), and $H_2$ (yellow). }
    \label{fig:logwannier_theta}
\end{figure*}
We show this by plotting the squared modulus of each Wannier orbital on a logarithmic scale as a function of $x/a_M$ across a horizontal line cut through each orbital's stacking site center $\alpha$ for $\theta=3^\circ$ and $\theta=5.6^\circ$ in Fig.~\ref{fig:logwannier_theta}. 

\subsection{Tight-Binding vs. Continuum Model}
\label{sect:wannier_vs_ctm}

We demonstrate that the tight-binding model construction provided in Sec.~\ref{sect:tight-binding_model} of the main text yields qualitatively similar band structures to the continuum model. 
\begin{figure*}[t!]
    \centering
    \includegraphics[width=0.98\linewidth]{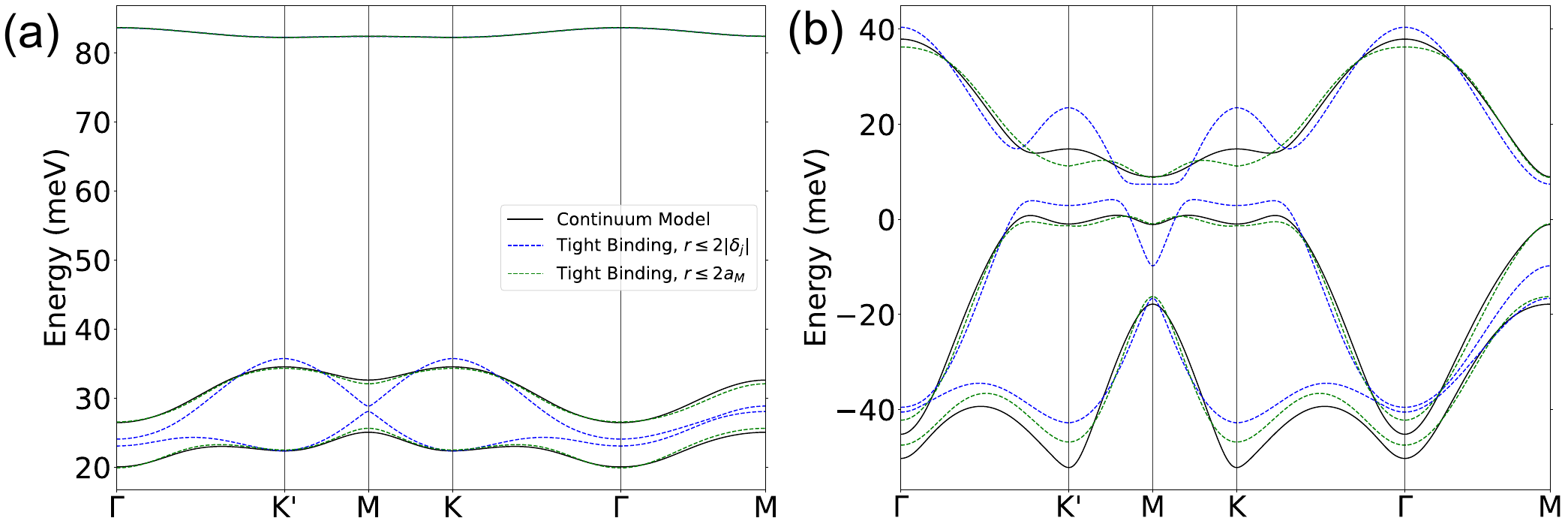}
    \caption{Comparison of continuum model (solid black) band structures with those of the tight-binding model using cut-offs of $|\br| \leq 2|\boldsymbol{\delta}_j|$ (dashed blue) and $|\br| \leq 2a_M$ (dashed green) for the topmost three valence bands at twist angles \textbf{(a)} $\theta = 3.0^\circ$ and \textbf{(b)} $\theta = 5.6^\circ$. }
    \label{fig:TBvsCtmBands}
\end{figure*}
Fig.~\ref{fig:TBvsCtmBands} compares the band structures from our tight-binding model with those of the continuum model. 
In plotting the band structures of the tight-binding model, we cut off the hoppings such that we only include parameters where the hopping distance is $|\br| \leq 2|\boldsymbol{\delta}_j|$ (dashed blue lines) or $\lvert \br \rvert \leq 2a_M$ (dashed green lines). 

\clearpage

\end{document}